\documentclass[brazil,12pt,a4paper,openany]{book}
\usepackage{anysize}
\usepackage{graphicx}
\usepackage[latin1]{inputenc}
\usepackage[portuges,brazilian]{babel}
\usepackage{fancyhdr}
\usepackage{ifthen}
\usepackage{array}
\usepackage{hyperref}

\marginsize{30mm}{30mm}{30mm}{30mm}

\begin{document}

\setlength{\abovecaptionskip}{0.0cm}
\setlength{\belowcaptionskip}{0.0cm}

\def\cqd{\vrule height7.0pt width6.0pt depth0pt} 

\setlength{\baselineskip}{18pt}

\newcommand{\titulo}
{
    Dualidades Eletromagnéticas no Espaço-Tempo Não-Comutativo e Formalismos Simpléticos
}
\newcommand{\capitulo}[1]
{
    \chapter{#1}
}
\newcommand{\secao}[1] { \section{#1} }
\newcommand{\subsecao}[1] { \subsection{#1} }
\def\[{\left\lbrack}
\def\]{\right\rbrack}
\def\hb{\hfill\break}
\def\({\left(}
\def\){\right)}
\def\< {\left \langle}
\def\> {\right \rangle}

\newcommand{\be}{\begin{equation}}
\newcommand{\ee}{\end{equation}}
\newcommand{\ea}{\end{eqnarray}}
\newcommand{\ba}{\begin{eqnarray}}
\newcommand{\lan}{\langle}
\newcommand{\la}{\langle}
\newcommand{\ran}{\rangle}
\newcommand{\asp}{\textquotedblleft}  
\newcommand{\vx}{{\vec{x}}}
\newcommand{\vy}{{\vec{y}}}
\newcommand{\vep}{{\varepsilon}}
\newcommand{\ep}{{\epsilon}}
\newcommand{\tal}{{\tilde \alpha}}
\newcommand{\tbe}{{\tilde \beta}}
\newcommand{\cl}{{\cal L}}
\newcommand{\cg}{{\cal G}}
\newcommand{\cd}{{\cal D}}
\newcommand{\dirac}{{\delta(\vx - \vy)}}
\newcommand{\lnm}{{\cal L}_{\mbox{\tiny \it MCS}_\theta}}
\newcommand{\lns}{{\cal L}_{\mbox{\tiny \it AD}_\theta}}
\newcommand{\str}{{^{\star}}}
\newcommand{\fdu}{{^{\star}F}}
\newcommand{\adu}{{^{\star}A}}
\newcommand{\jdu}{{^{\star}J}}
\newcommand{\ddu}{{^{\star}d}}
\newcommand{\tdu}{{^{\star}\theta}}
\newcommand{\hfu}{\hat F^{\mu \nu}}
\newcommand{\hfd}{\hat F_{\mu \nu}}
\newcommand{\tf}{\tilde F}
\newcommand{\tfmu}{\tilde F^\mu}
\newcommand{\tfmd}{\tilde F_\mu}
\newcommand{\tfnu}{\tilde F^\nu}
\newcommand{\tfnd}{\tilde F_\nu}
\newcommand{\tflu}{\tilde F^\lambda}
\newcommand{\tfld}{\tilde F_\lambda}
\newcommand{\tth}{{\tilde \theta}}
\newcommand{\fmnu}{F^{\mu \nu}}
\newcommand{\fmnd}{F_{\mu \nu}}
\newcommand{\hamu}{\hat A^\mu}
\newcommand{\hamd}{\hat A_\mu}
\newcommand{\hanu}{\hat A^\nu}
\newcommand{\hand}{\hat A_\nu}
\newcommand{\halu}{\hat A^\lambda}
\newcommand{\hald}{\hat A_\lambda}
\newcommand{\amu}{A^\mu}
\newcommand{\amd}{A_\mu}
\newcommand{\auu}{A^\nu}
\newcommand{\aud}{A_\nu}
\newcommand{\alu}{A^\lambda}
\newcommand{\ald}{A_\lambda}
\newcommand{\epu}{{\epsilon^{\mu \nu \lambda}}}
\newcommand{\epd}{{\epsilon_{\mu \nu \lambda}}}
\newcommand{\prt}{{\partial}}
\newcommand{\diag}{\mbox{diag}}
\newcommand{\tr}{\mbox{tr}}
\newcommand{\Tr}{\mbox{Tr}}
\newcommand{\tht}{\tilde \theta}
\newcommand{\real}{{\rm I\!R}}
\newcommand{\nat}{{\rm I \! N}} 
\newcommand{\euc}{{\rm I\!E}} 
\newcommand{\comp} {{\,  \rm \mbox{\sffamily  I} \hspace{-5.5pt} C}}
\newcommand{\id}{{\rm 1 \! \hspace{-1.3pt}   I}}
\newcommand{\lra}{\leftrightarrow}
\newcommand{\sen}{\mbox{sen}}


\newcommand{\bq}{\begin{eqnarray}}
\newcommand{\eq}{\end{eqnarray}}
\newcommand{\ao}{\~ao}
\newcommand{\oes}{\~oes}
\newcommand{\cao}{\c c\~ao}
\newcommand{\coes}{\c c\~oes}
\newcommand{\bqn}{\begin{eqnarray*}}
\newcommand{\eqn}{\end{eqnarray*}}
\newcommand{\0}{{(0)}}
\newcommand{\1}{{(1)}}
\newcommand{\2}{{(2)}}
\newcommand{\n}{{(n)}}
\newcommand{\sx}{\stackrel{x}}
\newcommand{\sy}{\stackrel{y}}

\pagestyle{fancy}
\fancyhf{}
\lhead{\rightmark}
\rhead{\thepage}

\fancypagestyle{plain}
{
	\fancyhf{}
	\lhead{\rightmark}
	\rhead{\thepage}
}


\title{ \vspace*{3in} {\Large	 \titulo \vspace{-.3in}}}

\author{  \Large Davi Cabral Rodrigues  \\[.3in]  \large Cl\'ovis  José Wotzasek \\[.1in]  \large Wilson Oliveira}

\date{}

\maketitle

\thispagestyle{empty}

\pagenumbering{roman}

\noindent
\includegraphics[width=0.10\textwidth]{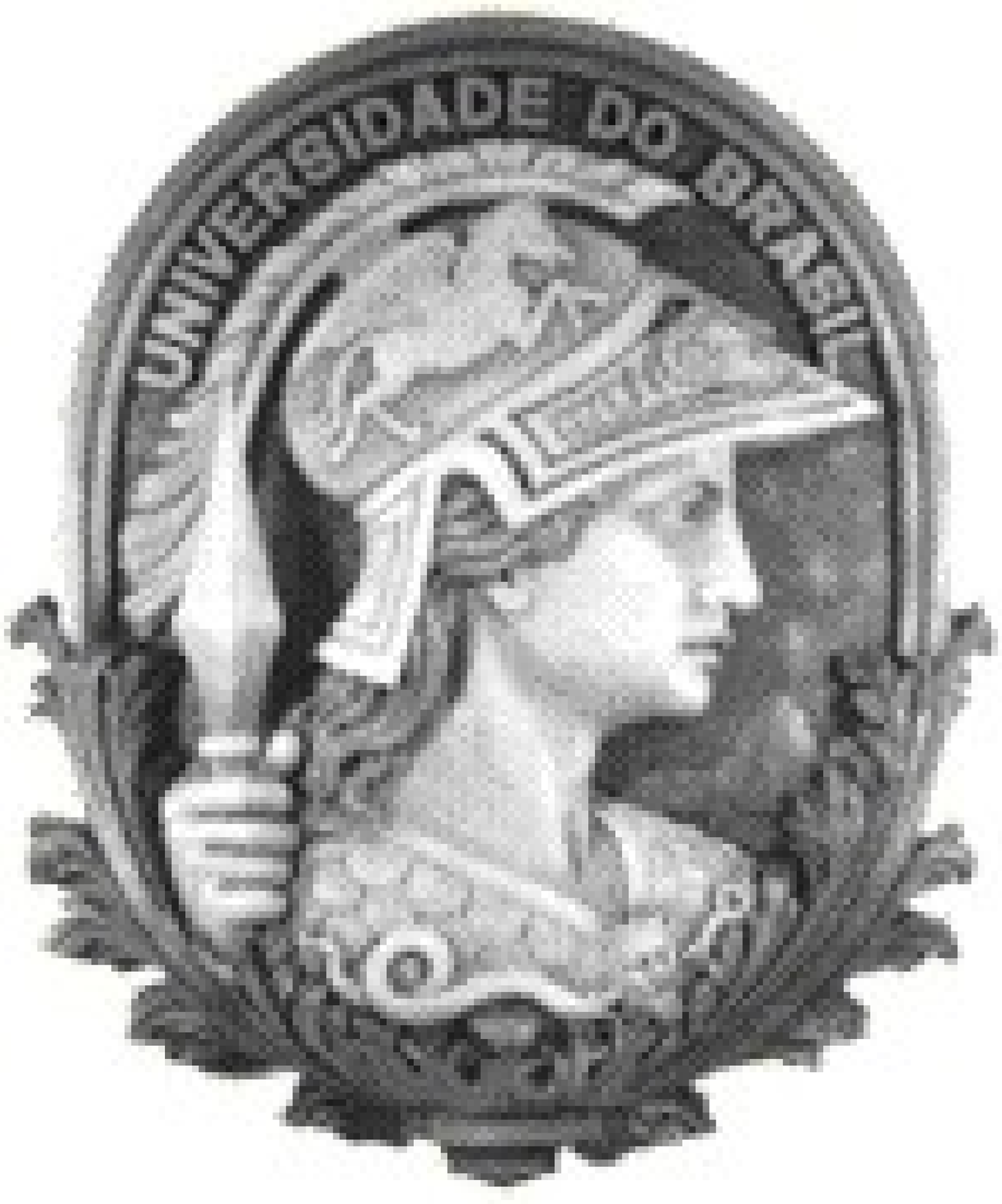}

\noindent
{\normalsize UFRJ}

\begin{center}

	{\large \titulo} \\[.5in]

	Davi Cabral Rodrigues \\[.8in]

\end{center}

\begin{flushright}

	\parbox{8cm}{Tese de Doutorado apresentada ao Programa de Pós-graduação em Física, Instituto de Física, da Universidade Federal do Rio de Janeiro, como parte dos requisitos necessários à obtenção do título de Doutor em Ciências (Física).\\[.4in]
	Orientador: Clóvis José Wotzasek \\
	Co-orientador: Wilson Oliveira} \\[2in]
\end{flushright}

\begin{center}
	Rio de Janeiro \\
	Setembro de 2006

\end{center}

\pagebreak


\thispagestyle{empty}

\begin{center}

{\Large \bf \titulo\\[5pt]}

{ Davi Cabral Rodrigues}

{ Orientador: Clóvis José Wotzasek}

{ Co-orientador: Wilson Oliveira}

\end{center}

\vspace*{5pt}

\begin{center}

Tese de doutorado submetida ao Programa de Pós-graduação em Física, Instituto de Física, da Universidade Federal do Rio de Janeiro - UFRJ, como parte dos requisitos necessários à obtenção do título de Doutor em Física. 

\end{center}


Aprovada por:\\

\begin{flushright}
\begin{tabular}{c}\hline

Dr. Clóvis José Wotzasek \\(Orientador)\\[30pt]\hline

Dr. Wilson Oliveira \\(Co-orientador)\\[30pt]\hline

Dr. Adilson José da Silva\\[30pt]\hline 

Dr. Henrique Boschi Filho\\[30pt]\hline

Dr. Luis Esteban Oxman\\[30pt]\hline

Dr. Nelson Ricardo de Freitas Braga\\[27pt]



\end{tabular}
\end{flushright}

\begin{center}
	{\normalsize Rio de Janeiro, setembro de 2006.}
\end{center} 

\pagebreak


\vspace*{1in}
\begin{center}

\framebox[15cm]{\parbox{15cm}{\vspace{1truecm}
R696\hspace{1cm}    Rodrigues, Davi Cabral\\
\mbox{\hspace{2cm}} Dualidades Eletromagnéticas no Espaço-Tempo Não-Comutativo e Formalismos Simpléticos - Rio de Janeiro: UFRJ/IF, 2006.\\
\mbox{\hspace{2cm}} x, 151f.: il.; 31 cm.\\
\mbox{\hspace{2cm}} Orientador: Clóvis José Wotzasek\\
\mbox{\hspace{2cm}} Tese (doutorado) - UFRJ / Instituto de Física / Programa de Pós-graduação em Física, 2006.\\
\mbox{\hspace{2cm}} Referências Bibliográficas: f. 137-151.\\
\mbox{\hspace{2cm}} 1. Não-comutatividade. 2. Dualidade. 3. Formalismos simpléticos. 4. Mapa de Seiberg-Witten. 5. Eletromagnetismo. 6. Massa topológica.  I. Wotzasek, Clóvis José. II. Universidade Federal do Rio de Janeiro, Instituto de Física, Programa de Pós-graduação em Física. III. Dualidades Eletromagnéticas no Espaço-Tempo Não-Comutativo e Formalismos Simpléticos.
}}

\end{center}


\newpage

\noindent

\vspace*{30pt}

\begin{center}

{\Large \bf Resumo} \\[.2in]

{\Large \titulo} \\[.2in]

Davi Cabral Rodrigues \\[.05in]
Clóvis José Wotzasek \\
Wilson Oliveira \\[.2in]

\end{center}

Resumo da Tese de Doutorado submetida ao Programa de Pós-Graduação em Física. Instituto de Física, da Universidade Federal do Rio de Janeiro - UFRJ, como parte dos requisitos necessários à obtenção do título de Doutor em Ciências (Física). \\[.5in]

O estudo de teorias de espaço-tempo não-comutativo tem atraído muita atenção nos últimos anos, o que se deve essencialmente a duas razões: i) teorias desse tipo aparecem como casos limites de modelos físicos \emph{a priori} comutativos, como é o caso de teoria de cordas no limite de Seiberg-Witten e o de uma partícula no mais baixo nível de Landau; ii) por si mesmas são teorias de interesse físico, pois possuem propriedades bem particulares próximas a várias expectativas, de gravitação quântica a efeito Hall quântico fracionário.
 
Apresentamos nesta tese uma recente proposta de \emph{gauge embedding}  baseada no formalismo simplético de sistemas vinculados e analisamos  extensões de dualidades eletromagnéticas clássicas para o espaço-tempo não-comutativo em 3D e em 4D. A análise dessas dualidades no espaço tempo não-comutativo é importante pois dualidades estabelecem equivalências não-triviais de formulações de uma mesma teoria, levando à descoberta de novas expressões e novas propriedades físicas. Em particular, é sabido que em 4D no limite de campos lentamente variantes (CLV), o parâmetro da não-comutatividade $\theta$ é transformado em seu dual Hodge $\str \theta$ via dualidade eletromagnética, essa transformação conecta não-comutatividade no espaço somente com não-comutatividade entre tempo e espaço. Nesta tese estendemos essa dualidade para 3D, avaliamos a necessidade do limite de CLV em 4D e em 3D, estudamos o caso tridimensional com massa topológica e estabelecemos uma extensão não-comutativa para o modelo autodual, esclarecendo certo conflito existente na literatura.

Também aqui apresentamos o desenvolvimento de uma recente técnica de \emph{gauge embedding}, baseada no tratamento simplético de sistemas vinculados que já foi aplicada em alguns modelos, tanto comutativos quanto não-comutativos, tendo conseguido reproduzir com sucesso resultados obtidos através de outros métodos. Consideramos que esse recente método é promissor e esperamos que seu desenvolvimento conduza a soluções de problemas físicos atuais. \\[.3in]

\noindent
Palvras-chave: não-comutatividade, dualidade, formalismos simpléticos, mapa de Seiberg-Witten, eletromagnetismo, massa topológica. \\[.5in]

\begin{center}
	Rio de Janeiro\\
	Setembro de 2006

\end{center}



\newpage

\noindent

\vspace*{30pt}

\begin{center}

{\Large \bf \emph{Abstract}} \\[.2in]

{\Large Electromagnetic Dualities on Noncommutative Space-Time and Symplectic Formalisms} \\[.2in]

Davi Cabral Rodrigues \\[.05in]
Clóvis José Wotzasek \\
Wilson Oliveira \\[.2in]

\end{center}

\emph{Abstract} da Tese de Doutorado submetida ao Programa de Pós-Graduação em Física. Instituto de Física, da Universidade Federal do Rio de Janeiro - UFRJ, como parte dos requisitos necessários à obtenção do título de Doutor em Ciências (Física). \\[.5in]

The study of noncommutative space-time theories has being attracting much attention in the last years, this is mainly due to two reasons: i) these theories appear as limiting cases of physical models which are \emph{a priori} commutative, e.g., strings in the Seiberg-Witten limit and a charged  particle in the Lowest Landau Level; ii) by themselves these theories are of physical interest, since they have very particular properties which are close to many expectations, from quantum gravity to fractionary quantum Hall effect. 

In this thesis we present a recent proposal for gauge embedding inspired in the symplectic formalism for constrained systems and we analyze extensions of classical electromagnetic dualities to the noncommutative 3D and 4D space-times. This investigation is important since dualities establish nontrivial equivalences between different formulations of the same physical theory, leading to the discovery of new expressions and new physical properties. In particular, it is known that in the slowly varying fields (SVF) limit, the noncomutativity parameter $\theta$ becomes its Hodge dual $\str \theta$ through the  4D electromagnetic duality, this transformation connects space noncommutativity with noncommutativity between space and time. In this thesis we extend this duality to the 3D space-time, evaluate the necessity of the SVF limit in 4D and in 3D,  study the 3D case with topological mass and establish a noncommutative extension to the the selfdual model, clarifying certain conflict found in the literature.

We also present here the development of a recent technique of gauge embedding, inspired in the symplectic handling of constraints which has already been applied in a number of models, both commutative and noncommutative, and success has been achieved in reproducing results obtained by other methods. We consider that this recent method is promising and we hope that its development will be helpful to the solution of current physical problems.\\[.3in]


\begin{center}
	Rio de Janeiro\\
	Setembro de 2006

\end{center}















\newpage

\noindent

\vspace*{10pt}

\begin{center}

{\LARGE\bf Agradecimentos}
\end{center}

\vspace*{10pt}

\noindent
$\diamond$ Em especial agradeço ao meu orientador, Prof. Clóvis Wotzasek. Agradeço a ele  pelos seus diversos ensinamentos  (técnicos ou não), por acreditar no meu potencial e me estimular, por mostrar que trabalho e diversão em física devem se misturar, pela ajuda nas questões referentes ao pós-doutorado, além das várias caronas para Niterói.

\noindent
$\diamond$ Prof. Wilson Oliveira, meu co-orientador no doutorado e orientador de graduação e  mestrado na UFJF. Agradeço a ele pela colaboração nos trabalhos sobre método simplético, pela oportunidade de apresentar o colóquio sobre dualidades eletromagnéticas não-comutativas na  UFJF e por sempre ter  me auxiliado no que precisei.

\noindent
$\diamond$ Agradeço a Clifford Neves, meu co-orientador no mestrado, pela fundamental colaboração nos trabalhos sobre método simplético, pela grande paciência em aturar minhas incisivas perguntas e pelas ótimas discussões.

\noindent
$\diamond$ Agradeço aos meus colaboradores e colegas Jorge Noronha e Marcelo Guimarães, aprendemos muito em conjunto. Em especial agradeço ao Marcelo por ter-me indicado várias referências e pelos comentários sobre esta tese.

\noindent
$\diamond$ Agradeço ao Prof. Victor Rivelles pela discussão sobre o mapa de Seiberg-Witten e pela oportunidade de apresentar um colóquio na USP sobre dualidades eletromagnéticas não-comutativas.

\noindent
$\diamond$ Agradeço ao Prof. Sergio Cacciatori e colaboradores pela atenção ao nosso problema e pela coordial correspondência.

\noindent
$\diamond$ Agradeço ao meu amigo e consultor de matemática e informática Prof. Orestes Piermatei Filho.

\noindent
$\diamond$ Para esta tese foi também importante o apoio, estímulo e compreensão de meus pais, avós e de minha irmã Ana, sou muito grato a eles. Agradeço também aos meus amigos que pouco entendem do que eu faço, mas estão sempre prontos para me ajudar no que eu precisar, conversar ou ouvir alguma estranha música.

\noindent
$\diamond$ Agradeço ao CNPq pela bolsa de doutorado de ago/2002 a fev/2005 e à FAPERJ pela bolsa do programa  \emph{bolsa nota 10} de mar/2005 a jul/2006.


\tableofcontents




\markboth{\titulo}{\titulo}


\setlength{\baselineskip} {22 pt}

\capitulo{Introdução}

\pagenumbering{arabic}

Desde as origens da ciência moderna sempre coube aos cientistas o trabalho de elaboração de teorias e verificação dessas através da experiência. Com o passar do tempo, mais fenômenos vieram a ser explicados pela ciência e com maior precisão, paralelamente as teorias foram se tornando mais complexas, exigindo cada vez mais trabalho teórico. No século XVII, um cidadão interessado era capaz de ler os últimos tratados científicos de seu tempo, como o \emph{Diálogos Sobre as Duas Novas Ciências} (1638) de Galileu Galilei; entretanto, a partir do século XX, mesmo os conceitos mais básicos da física dificilmente são compreendidos por não-cientistas\footnote{O filósofo K. Popper além de interesse pela física teve contato com grandes físicos de seu tempo e argumentou que uma interpretação da mecânica quântica deveria caber à filosofia. Todavia, ao apresentar sua interpretação em um capítulo de \emph{A lógica da pesquisa científica} (1934), se aventurou além do que poderia e incorreu em erros básicos, como ele mesmo admite em edições subseqüentes.}. Os livros atuais voltados para o público leigo, antes de descreverem superficialmente as pesquisas atuais, precisam introduzir os fundamentos da física que foram estabelecidos no início do século passado! Esse alto grau de distanciamento da ciência contemporânea pode, por outo lado, ser visto como um indício de que a ciência estaria realmente rumando de acordo com seus princípios de buscar pelos fundamentos da Natureza a despeito das crenças e apelos estéticos da sociedade de seu tempo\footnote{Esta questão não é de forma alguma simples, mas não entraremos em mais detalhes.}. Atualmente há indícios de que estejamos próximos da teoria física fundamental, a qual provavelmente adviria da teoria de cordas (ou teoria M). Espera-se que resultados do \emph{Large Hadron Collider} (LHC), cujas atividades começarão no próximo ano, auxiliem na escolha do caminho a se seguir para a consolidação da teoria de cordas.

A física atingiu um grau de refinamento teórico que a possibilita dar grandes saltos na compreensão da Natureza através da inspeção de sua estrutura formal e da criação de novas estruturas cuja correspondência com a Natureza só é entendida posteriormente. Nesse espírito, P. A. M. Dirac certa vez enunciou 

\begin{quote}
\emph{It is more important to have beauty in one's equations than to have them fit experiment (...) because the discrepancy may be due to minor features that are not properly taken into account and that will get cleared up with further developments of the theory.}
\end{quote}

Embora a proposta original de C. N. Yang e  R. L. Mills \cite{YM} tenha se demonstrado incorreta, se seu trabalho tivesse sido descartado hoje não teríamos o modelo padrão. A interpretação física original realmente era problemática, mas sua estrutura formal, a despeito das sérias dificuldades iniciais, continha a chave para o entendimento das forças nucleares. Segundo o próprio C. N. Yang:

\begin{quote}
\emph{We did not know how to make the theory fit experiment. It was our judment, however, that the beauty of the idea alone merited attention.}
\end{quote}

Durante a formulação da mecânica quântica, os físicos da época, com destaque para W. Heisenberg, perceberam que as grandezas clássicas expressas por funções reais deveriam ser abandonadas em favor de novas grandezas dadas por operadores cujo comutador não é necessariamente nulo. A partir de então, tornou-se claro que o espaço de fase de um sistema físico é verdadeiramente não-comutativo. Todavia, somente várias décadas mais tarde, a geometria desses espaços  veio a receber uma formulação geométrica precisa \cite{connes}. Esses avanços conceituais motivaram a busca por novas aplicações da geometria não-comutativa na física \cite{connes-lott, wittennc, connesnc} e da revisão da antiga proposta de espaço quântico de H. S. Snyder \cite{snyder}. Em particular foi descoberto que, sob certo limite de baixa energia, a teoria de cordas prevê um espaço físico não-comutativo \cite{wittennc, connesnc, sw}. 

Muito do significado dessas teorias de espaço-tempo não-comutativo não é entendido, mas seus aspectos formais tornam-se a cada dia mais ricos. Uma nova revolução pode estar bem próxima. Não é claro até que ponto tais teorias devam ser vistas como fundamentais ou como efetivas, talvez a utilidade delas seja somente para dados fenômenos específicos, como o efeito Hall quântico ou como forma de modelar um sistema no mais baixo nível de Landau; mas podem se revelar até como uma proposta de teoria fundamental alternativa à teoria de cordas. As teorias $U_*(N)$ (teorias $U(N)$ não-comutativas) possuem propriedades não encontradas em outros modelos físicos conhecidos; em particular, transformações de coordenadas são um subgrupo de suas transformações de calibre, o que as torna fortes candidatos para modelar gravitação quântica. Ademais, em um espaço não-comutativo, há uma relação de incerteza no próprio espaço físico que o torna \asp embassado" (\asp fuzzy"), pontos são dissolvidos em pequenos planos \asp enevoados", tal como ocorre no já tradicional espaço de fase quântico. Característica que é bem vinda para a compatibilização da gravitação com a teoria quântica, dado que por um lado a teoria quântica dita que é necessário uma quantidade de energia cada vez maior para examinar partes cada vez menores do espaço, enquanto a relatividade geral impõe que energia curva o espaço-tempo; o que leva a um colapso da geometria  usual do espaço-tempo em regiões da ordem do comprimento de Planck.

Mais antigo que o conceito de não-comutatividade em física é o de dualidade, cuja origem pode ser traçada desde as observações de O. Heaviside \cite{heaviside} no final do século XIX. Assim como não-comutatividade, o conceito de dualidade foi revigorado nos últimos anos. Em particular devido às conexões que as dualidades fornecem às teorias de cordas, unificando todas as formulações, o que deu origem à chamada segunda revolução (para introduções veja as Refs. \cite{introdual}).

\vspace{.4in}
Três conceitos formais físicos\footnote{São físicos no sentido de estarem associados a interpretações físicas. A abordagem desta tese é física e não matemática.} desempenham um papel de destaque nesta tese: não-comutatividade, dualidade e geometria do espaço de fase (geometria simplética). Esta tese foi elabordada sem pressupor que o leitor já tenha familiaridade com essas áreas e sua conexão com a física. Os resultados por nós obtidos podem ser divididos entre os referentes à criação do método simplético de calibre \cite{symemb, symembjf} (Seção \ref{secformsympcal}), o sobre a extensão da dualidade dos modelos Maxwell-Chern-Simons e autodual para o espaço-tempo não-comutativo \cite{nossoNCMCS} (Seção \ref{sec6ad}) e, por fim, sobre a dualidade eletromagnética sem massa topológica em 3D e 4D \cite{issues} (Seção \ref{sec6clv}). 

No próximo capítulo introduzimos ou revisamos sucintamente alguns conceitos básicos que serão úteis para toda a tese, como geometria diferencial (a notação de formas diferenciais em especial), eletromagnetismo em $D$ dimensões, grupos $SU(N)$, massa topológica e uma apresentação das dualidades eletromagnéticas com ênfase nas técnicas que serão usadas posteriormente. No Cap. 3 apresentamos uma introdução à geometria simplética no espaço de fase e ao chamado formalismo simplético, em seguida apresentamos o método simplético de calibre. O Cap. 4 é exclusivamente dedicado a uma introdução sobre não-comutatividade no espaço-tempo. Procuramos apresentar os tópicos necessários ao capítulo seguinte e mais alguns que nos parecem convenientes para dar uma melhor visão geral à questão. Quanto a esse último ponto, muito mais poderíamos acrescentar, muitos assuntos interessantes tiveram de ficar de fora, como transformações de Lorentz, mas acreditamos que o material apresentado, junto das referências indicadas, já seja o suficiente para uma compreensão razoável do quadro geral. No Cap.5 apresentamos os dois resultados independentes sobre dualidade eletromagnética no espaço-tempo não-comutativo. Por fim, o último capítulo é dedicado às nossas conclusões e perspectivas futuras.

\capitulo{Teorias $U(N)$ em $D$ dimens\~oes e modelos duais}

\vspace{.4in}
\section{Notação e tópicos sobre geometria diferencial}
\label{topicosgeo}

O principal objetivo desta seção é estabelecer a notação, para isso revisaremos de forma breve algumas propriedades de geometria diferencial. De particular interesse é a notação de formas diferenciais. É natural se questionar se essa notação realmente é relevante para a física ou para a tese em questão. Por se tratar de uma notação, os mesmos conceitos podem ser sempre expressos de através de outros meios, mas a importância de uma boa notação nunca deve ser subestimada, pois permite analisar o mesmo problema através de outros aspectos e pode possibilitar uma solução muito mais rápida, se atendo aos conceitos fundamentais em vez da extensos cálculos ou expressões. A notação de formas diferenciais é bem mais \asp limpa" $\;$ que a notação tensorial (que nada mais é do que a expressão componente por componente das formas) e trata dos objetos físicos ou matemáticos diretamente a despeito de particularidades do sistema de coordenadas. Não importa qual a transformação de coordenadas utilizada, uma $k$-forma $A$ é sempre expressa por $A$, enquanto seu tensor correspondente $A_{\mu_1 \mu_2 ... \mu_k}$ precisa se transformar de acordo. Operações freqüentemente aqui utilizadas, como dualidade Hodge e diferenciação externa, são simplesmente dadas por $\str A$ e $dA$ enquanto na notação de componentes assumem o aspecto $\ep^{\mu_1 \mu_2 .... \mu_k \nu_1 \nu_2 ... \nu_{D-k}} A_{\mu_1 \mu_2 ... \mu_k}$ e $\prt_{[\nu} A_{\mu_1 \mu_2 ... \mu_k]}$. Enfim, dada a simplicidade da notação mais moderna de formas diferenciais, a pergunta não deveria ser se é necessário recorrer a essa notação, mas sim se todas as vezes que a notação de componentes foi utilizada isso realmente foi necessário.

Esta apresentação não visa ser matematicamente rigorosa. Algum contato anterior com conceitos básicos de geometria diferencial é pressuposto. Vários resultados serão mencionados sem demonstração. Introduções mais detalhadas sobre o tema podem ser vistos, por exemplo, nas Refs. \cite{arnold, nakahara, schutz, azul}.


Vetores no espaço Euclideano $\euc^n$ se encontram nesse próprio espaço e não é importante especificar a que ponto $p \in \euc^n$ certo vetor desse espaço está associado, pois a geometria do espaço Euclideano é idêntica em todos seus pontos. Seja $M$ uma variedade $D$-dimensional, associa-se a cada ponto $p \in M$ um espaço vetorial $T_pM$ chamado de espaço tangente a $M$ em $p$. Assim, um vetor associado a $M$ pertence ao espaço tangente a $M$ em certo ponto $p \in M$. O espaço $T_p M$ tem a mesma dimensão de $M$, ou seja, $D$. O espaço dado pela união de todos os espaços tangentes a $M$ é o fibrado tangente, cuja dimensão é $2D$, e escreve-se\footnote{Usaremos o símbolo $:=$ para indicar uma igualdade por definição. $A:=B$ lê-se $A$ é definido como sendo igual a $B$. Sempre que ele for usado uma definição estará sendo feita, mas a recíproca é falsa.} 
\be
	TM := \bigcup_{p \in M} T_pM.
\ee

Como $T_pM$ é um espaço vetorial, existe um espaço vetorial dual a esse, que denotaremos por $T^\star_p M$,  cujos elementos associam um número real a cada vetor de $T_pM$, ou seja, 
\be
	\omega \in T_p^\star M \; \Leftrightarrow \; \omega: T_pM \rightarrow \real.
\ee
O espaço $T^\star_p M$ é chamado de espaço cotangente a $M$ em $p$. Reciprocamente, $V \in T_p M \; \Rightarrow \; V: T_p^\star M \rightarrow \real$. Analogamente ao fibrado tangente, $T^\star M := \bigcup_{p \in M} T^\star_pM$ é o fibrado cotangente a $M$.

Sejam $\{e_\mu \}$ base de $T_pM$ e $\{ f^\mu \}$ base de $T_p^\star M$, com $\mu = 1,2,...,D$. A atuação de $V \in T_p M$ sobre $\omega \in T^\star_p M$ define um produto interno $( \; , \; )_p: T_p^\star M \times T_p M \rightarrow \real$ cuja atuação sobre a bases desses espaços é dada por\footnote{A convenção da regra da soma sobre índices repitidos é assumida nessa equação e em todas as posteriores.}
\be
	V[\omega] = (\omega, V)_p = (\omega_\mu f^\mu, V^\nu e_\nu)_p = \omega_\mu V^\nu ( f^\mu,  e_\nu)_p.
\ee
Sabemos que $( f^\mu,  e_\nu)_p \in \real$, mas seu valor em princípio é arbitrário. A fixação desse valor determinará o valor real da atuação de qualquer vetor sobre um covetor (ou o inverso). Fixaremos\footnote{Em que $\delta^\mu_\nu = 0$ se $\mu \not= \nu$ e $\delta^\mu_\nu = 1$ caso contrário.} $( f^\mu,  e_\nu)_p = \delta^\mu_ \nu$, portanto
\be
	\label{frev4p}
	V[\omega] = \omega_\mu V^\mu.
\ee
Dado que o lado esquerdo da igualdade é invariante por transformações de coordenadas, o lado direito também é. Em particular isso mostra que as coponentes de um covetor devem se transformar de acordo com o inverso da transformação das componentes de um vetor.

Até o momento não foi explicado como o espaço tangente é construído a partir de uma variedade diferenciável, só argumentamos em favor de sua existência e apresentamos algumas definições naturais e conseqüências imediatas. Usando derivadas ao longo de curvas em $M$ que passam por $p \in M$, contrói-se o espaço $T_pM$. As componentes de um vetor $V$ tangente a $M$ em $p$ ao longo da curva $c(t) \subset M$, com $c(0) = p$,  são dadas por $V^\mu = \frac {dx^\mu(c)}{dt}|_{t=0}$. 

A taxa de variação de uma função $f: M \rightarrow \real$ em $p$ ao longo da curva $c$ é dada por
\be
	\left. \frac{d f(c(t))}{dt}\right|_{t=0} = \left. \frac{d (f \circ \varphi^{-1}\circ \varphi \circ c)(t)}{dt}\right|_{t=0}, 
\ee
em que $\varphi: U_p \subset M \rightarrow \real^D$ é a aplicação que define o mapa entre certa vizinhança de\footnote{Sua existência, assim como a de sua inversa, é pré-requisito para $M$ ser variedade.} $p$ em $M$ e o espaço $\real^D$. Há pouco escrevemos $x^\mu(c)$, mais precisamente $x^\mu(c)$ se refere a uma das $D$ componentes reais da aplicação $\varphi(c(t))$. De forma análoga, expressaremos $f \circ \varphi^{-1}:\real^D \rightarrow \real$ como $f(x^\mu)$. A última equação pode então ser escrita como
\be
	\label{ghyhh}
	\left. \frac{d f(c(t))}{dt}\right|_{t=0} =  \left. \frac {dx^\mu(c(t))}{dt}\right|_{t=0} \frac {\prt f (x)}{\prt x^\mu} = V^\mu \frac {\prt }{\prt x^\mu} f (x).
\ee
Ou seja, taxas de variação de funções definidas em $M$ são \asp orientadas" pelos vetores tangentes.

Uma forma de interpretar a equação anterior é considerar que o vetor $V$ é um operador cuja base é dada por $\prt/ \prt x^\mu$, e essa é a forma padrão da base de vetores tangentes a $M$. Comparando as Eqs. (\ref{ghyhh}) e (\ref{frev4p}), vemos que podemos considerar $\prt_\mu f$ como componentes de um covetor. Há um escalar imediato que pode-se criar com essas componentes, esse é dado por $df = \prt_\mu f dx^\mu$. Assim identifica-se $\{ dx^\mu \} $ como base de $T^\star_p M$ e $df(p)$ como elemento de $T^\star_p M$. Covetores são comumente chamados de 1-formas. Se $\omega^k$ é $k$-forma em $M$ então $\omega^k: T_pM \times T_pM \times ... \times T_pM \rightarrow \real$ é função $k$-linear totalmente anti-simétrica. Em particular, para $k=2$, vem
\ba
	\omega^2 &=& \frac 12 \omega^2_{\mu \nu} \; (dx^\mu \otimes dx^\nu - dx^\nu \otimes dx^\mu) = \frac 12 \omega^2_{\mu \nu}  dx^\mu \wedge dx^\nu, \\[.2in]
	\omega^2[V, S] &=& - \omega^2[S, V] = \frac 12 \omega_{\mu \nu}^2 V^\mu S^\nu,
\ea
em que $\omega^2_{\mu \nu} \in \real$, $\omega^2_{\mu \nu} = - \omega^2_{\nu \mu}$ e $S,V \in T_p M$.

Toda $k$-forma pode ser escrita como segue, e sempre usaremos o mesmo fator constante, 
\be
	\omega^k = \frac 1{k!} \omega^k_{\mu_1 \mu_2 ... \mu_k} dx^{\mu_1} \wedge dx^{\mu_2} \wedge ... \wedge dx^{\mu_k}.
\ee
O tensor $\omega^k_{\mu_1 \mu_2 ... \mu_k}$ é sempre totalmente anti-simétrico. 

O produto externo de uma $k$-forma por uma $r$-forma é uma $(k+r)$-forma: $\omega^k \wedge \omega^r = \omega^{k+r}= (-1)^{kr} \omega^r \wedge \omega^k$.

Uma função real $f:M \rightarrow \real$ é uma 0-forma e $df$ é uma 1-forma. ($k+1$)-formas podem ser obtidas a partir da aplicação da derivada exterior sobre uma $k$-forma segundo a regra
\be
	\label{sdfh}
	d\omega^k = \frac 1{k!} \prt_\nu \omega^k_{\mu_1... \mu_k} dx^\nu \wedge dx^{\mu_1} \wedge... \wedge dx^{\mu_k}.
\ee
Dessa regra imediatamente conclui-se que 
\be
	d d \omega^k = 0
\ee
para qualquer $k$-forma (diferenciável).

Uma $k$-forma que satisfaça $d\omega^k=0$ é chamada de fechada. Em particular, se existir uma ($k-1$)-forma tal que $d \omega^{k-1} = \omega^k$,  $\omega^k$ é forma exata. Em espaços Euclideanos, pseudo-Euclideanos ou homeomorfos a um desses dois, toda forma fechada é necessariamente exata (isso é uma conseqüência do Lema de Poincaré). 

Como conseqüência da regra (\ref{sdfh}), vem
\be
	d(\omega^k \wedge \omega^r) = d \omega^k \wedge \omega^r + (-1)^k \omega^k \wedge d\omega^r.
\ee

Comumente consideraremos variedades munidas de uma métrica $g_p: T_pM \times T_pM \rightarrow \real$. Métrica é uma forma simétria, bilinear e não-degenerada que será expressa por 
\be
	g_p = g_{\mu \nu} dx^\mu \otimes dx^\nu.
\ee
Equivalentemente, ocasionalmente ela é expressa através do quadrado do elemento de linha $ds^2$,
\be
	ds^2 = g\(dx^\mu \frac \prt {\prt x^\mu},  dx^\nu \frac \prt {\prt x^\nu} \) = g_{\mu \nu} dx^\mu dx^\nu.
\ee

Usaremos a usual convenção $(g_{\mu \nu})^{-1} =: (g^{\mu \nu})$.

A métrica estabelece um isomorfismo entre $TM$ e $T^\star M$. Seja $V \in T_pM$, logo $g_p(V, \cdot): T_pM \rightarrow \real$ e portanto $g_p(V, \cdot) \in T^\star_pM$. Sendo $V^\mu$ as componentes do vetor $V$, adota-se a regra de escrever as componentes de seu covetor associado pela métrica por $V_\mu$. Deve estar claro que $V^\mu$ e $V_\mu$ são componentes de diferentes objetos, em particular, $V = V^\mu e_\mu \not= V_\mu f^\mu = g_p(V, \cdot)$. Contudo, pode-se definir uma convenção cuja única utilidade será a de facilitar a expressão gráfica de algumas equações: $\omega= \omega_\mu f^\mu = \omega^\mu f_\mu$, em que $f_\mu := g_{\mu \nu} f^\mu$, lembrando que $f^\mu = dx^\mu$.

Convencionaremos que 
\be
	dx^1 \wedge ... \wedge dx^D =: d^Dx.
\ee
Portanto
\be
	dx^{\mu_1} \wedge ... \wedge dx^{\mu_D} = \vep^{\mu_1 ... \mu_D} d^Dx,
\ee	
em que $\vep^{1 \; 2 ... D} = 1$, $\vep$ é totalmente anti-simétrico e não se transforma por mudança de coordenadas.

Em espaços métricos, há uma forma natural de introduzir um elemento de volume invariante por transformações de coordenadas (exceto por paridade), esse é dado por
\be
	\Omega = \sqrt{|\det (g_{\mu \nu})|} \; d^Dx.
\ee

Por meio da métrica, pode-se definir um isomorfismo natural entre o espaço das $k$-formas com o das $(D-k)$-formas que é dado pela operação de dualidade de Hodge, cujo operador denotaremos por $\str$ e sua atuação é expressa em uma $k$-forma $A$ por 
\ba
	\str A &=& \frac 1{k!} \; A_{\mu_1 ... \mu_k} \str(dx^{\mu_1} \wedge ... \wedge dx^{\mu_k}) \nonumber \\[.2in]
&=& \frac 1{k!} \frac 1 {(D-k)!} \; A_{\mu_1 ... \mu_k} \; \omega^{\mu_1 ... \mu_k}_{\mbox{\hspace{.3 in} } \nu_1 ... \nu_{D-k}} \; dx^{\nu_1} \wedge ... \wedge dx^{\nu_{D-k}} \\[.2in]
	&=& \frac 1{k!} \frac 1 {(D-k)!} \; A_{\mu_1 ... \mu_k} \; \omega^{\mu_1 ... \mu_k \nu_1 ... \nu_{D-k}} \; dx_{\nu_1} \wedge ... \wedge dx_{\nu_{D-k}} \nonumber	\\[.2in]
	&=& \frac 1{k!} \frac 1 {(D-k)!} \; A^{\mu_1 ... \mu_k} \; \omega_{\mu_1 ... \mu_k \nu_1 ... \nu_{D-k}} \; dx^{\nu_1} \wedge ... \wedge dx^{\nu_{D-k}} \nonumber.	
\ea
Todas as operações acima de levantar ou abaixar índices são apenas conseqüência do emprego da métrica. $\omega$ é totalmente anti-simétrico e deve se transformar de tal forma a preservar a invariância do lado esquerdo da igualdade por transformações de coordenadas.

Considere o produto
\ba
	A \wedge \str A &=& \( \frac 1{k!} \)^2 \frac 1 {(D-k)!} A_{\mu_1... \mu_k} A^{\nu_1... \nu_k} \omega_{\nu_1 ... \nu_k \lambda_1 ... \lambda_{D-k}} \; dx^{\mu_1} ... dx^{\mu_k}\wedge dx^{\lambda_1} ...  dx^{\lambda_{D-k}} \nonumber \\[.2in]
	&=& \( \frac 1{k!} \)^2 \frac 1 {(D-k)!} A_{\mu_1... \mu_k} A^{\nu_1... \nu_k} \omega_{\nu_1 ... \nu_k \lambda_1 ... \lambda_{D-k}} \; \vep^{\mu_1... \mu_k \lambda_1 ... \lambda_{D-k}} d^Dx  \nonumber \\[.2in]
	&=& \( \frac 1{k!} \)^2 A_{\mu_1... \mu_k} A^{\nu_1... \nu_k} f \; \delta_{\nu_1 ... \nu_k}^{\mu_1 ... \mu_k} \; d^Dx \nonumber \\[.2in]
	&=& \frac 1{k!} \; A_{\mu_1... \mu_k} A^{\mu_1... \mu_k} \; f \; d^D x.
\ea
Acima foi usado que $\omega$, por ser totalmente anti-simétrico e de posto máximo, deve ser proporcional a $\vep$. Voltaremos à questão da contração de índices entre $\vep$'s em breve. Da última equação, a fim de que $A \wedge \str A$ seja invariante por transformações de coordenadas, vê-se que
\be
	f=\sqrt{|\det (g_{\mu \nu})|}.
\ee
Portanto
\be
	\omega_{\mu_1... \mu_D} = \ep_{\mu_1...\mu_D} \; \sqrt{|\det (g_{\mu \nu})|},
\ee
em que $\ep_{1 \; 2 ... D} = 1$, $\ep$ é totalmente anti-simétrico e não se transforma por mudança de coordenadas. Para evitar confusão na operação de levantar e abaixar índices é útil estabelecer a distinção entre $\vep$ e $\ep$. Em particular temos
\be
	\ep^{\mu_1... \mu_D} := g^{\mu_1 \nu_1} \; ...\;  g^{\mu_D \nu_D}\;  \ep_{\nu_1 ... \nu_D} = \vep^{\mu_1 ... \mu_D} \; g^{-1},
\ee
com $g := \det (g_{\mu \nu})$. A regra de contração dos índices dos símbolos anti-simétricos é dada por
\be
	\ep_{\mu_1... \mu_k \nu_1 ... \nu_{D-k}} \; \vep^{\lambda_1 ... \lambda_k \nu_1 ... \nu_{D-k}} = (D-k)! \; \delta_{\mu_1... \mu_k}^{\lambda_1 ... \lambda_k}.
\ee
Seguindo a notação usual, $\delta_{\mu_1... \mu_k}^{\lambda_1 ... \lambda_k}$ é o determinante da matriz $k \times k$ de $\delta$'s de elemento geral $\delta_{\mu_i}^{\lambda_j}$.

Empregando artifícios já apresentados, obtemos a seguinte útil relação:
\be
	\str \str A = (-1)^{k(D-k) + a} A,
\ee
em que $(-1)^a$ é o sinal de $g$, ou seja, $(-1)^a = |g|/g$.


Essa notação e esses resultados são suficientes para passarmos para a próxima seção e começarmos a tratar de problemas mais diretamente ligados com a física.

\vspace{.4in}
\secao{Equa\c{c}\~oes de Maxwell em $D$ dimens\~oes com métrica constante arbitrária}
\label{tu1}

Esta seção tem dois objetivos importantes para o restante da tese: estabelecer as equações de Maxwell no espaço-tempo tridimensional e servir de primeiro exemplo de emprego da notação introduzida na seção anterior. Um espaço $\real^n$ com dada métrica constante pode sempre, por meio de transformações de coordenadas, ser escrito como um espaço Euclideano ou pseudo-Euclideano. Toda a apesentação desta seção seria muito simplificadda se assumíssemos logo de início uma métrica usual para o espaço-tempo, como $\diag (+ - - - )$ em 4D ou $\diag (+ - -)$ em 3D. Todavia, a substituição logo de início da métrica pelo seu valor constante esconde diversas propriedades da manipulação de formas diferenciais e das equações que descrevem o eletromagnetismo. Ademais, meios dielétricos podem ser modelados por métricas diferentes da do vácuo (ainda que constantes) \cite{azul} e a apresentação desta seção pode ser vista como o início de uma preparação pedagógica ao eletromagnetismo em espaços curvos (métricas não constantes). Devido ao valor da métrica não ser substiuído logo de início, poderemos avaliar o comportamento das equações de Maxwell frente à inversão de sinal da métrica, relacionando dois tipos de métrica comumente usados em física $\diag (- + + ...)$ e $\diag(+ - - ...)$,  veremos também que há certa sutileza importante na definição do campo magnético em 3D que depende da adoção do primeiro ou do segundo tipo de métrica.

Seja $M$ uma variedade de dimens\~ao $D$ munida de uma métrica $g_x: T_xM \times T_xM \rightarrow \real$ constante ($g_x = g_{x'}$). Considere a seguinte ação para a $k$-forma $A \in T^\star M$ com simetria de calibre do tipo\footnote{Isto \'e, sendo $A' = A + d \lambda$, $S[A] = S[A']$.} $U(1)$ e válida para qualquer dimensão D: 
\be
	\label{sja}
	S_J[A] = a_D \int \( \frac 12 \fdu \wedge F + e \; A \wedge \jdu \).
\ee
Acima, $F:= dA$, $a_D$ \'e uma constante que pode depender da dimens\~ao de $M$, $e$ \'e a constante do acoplamento de $J$ com $A$ e $J$ \'e uma $k$-forma que satisfaz $d \str J =0$. O acoplamento entre $A$ e $J$ acima pode também ser escrito como
\be
	\tilde e \; A \wedge \tilde J,
\ee
em que $\tilde J$ é uma $(D-k)$-forma que satisfaz $d \tilde J = 0$. Para dada métrica $g$, podemos considerar a introdução de $\tilde J$ nada mais que uma nova notação, pois $\tilde J$ e $\str J$ são $(D-k)$-formas fechadas quaisquer, assim podemos em princípio simplesmente usar a substituição $ \tilde e \; \tilde J \leftrightarrow e \; \str J$. Contudo, não estamos nesta seção exclusivamente interessados em resolver esse problema para dada métrica.

Diremos que a transformação $g \rightarrow -g$ define o que chamaremos de transformação de assinatura de $M$ (ou do espaço-tempo). É imediato conferir que 
\be
	\tilde J \; \stackrel a \longrightarrow \; \tilde J \;\;\;\;\;\;\; \mbox{ e } \;\;\;\;\;\;\; \str J \; \stackrel a \longrightarrow \; (-1)^{k} \; \str J,
\ee
em que o símbolo $a$ acima designa a transformação de assinatura. O acoplamento $A \wedge \str J$, dependendo do valor de $k$, muda de sinal, enquanto $A \wedge \tilde J$ permanece constante sempre.

No momento vamos aderir à interpretação da corrente como um ente físico tal como o potencial $A$, precisando da métrica para acoplar com $A$. Não há problema em se persistir no emprego de $A \wedge \str J$, mas precisaremos da regra
\be
	\label{trane}
	e \; \stackrel a \longrightarrow \; (-1)^{k} \; e.
\ee

\vspace{.4in}

Lembrando que ${^\star} \delta F = \delta \fdu$, $ {^\star} \delta F \wedge F = \fdu \wedge \delta F$, $\delta F = d \delta A$, $d ( \fdu \wedge \delta A) = d \fdu \wedge \delta A + (-1)^{D-(k+1)} \fdu \wedge d \delta A$ e $\delta A \wedge \jdu = (-1)^{k(D-k)} \jdu \wedge \delta A$, a equa\c{c}\~ao de movimento de $S_J$ (\ref{sja}) \'e
\be
	d \fdu + (-1)^{(k+1)(D-k)} \; e \; \jdu = 0.
\ee

Somente o caso k=1 será relevante para o restante desta seção, assim, para esse valor,\footnote{Ambos os lados dessa equação são invariantes por tranformação de assinatura, mas isso ocorre somente porque $e$ se transforma segundo (\ref{trane}).}
\be
	d \fdu = -  e \; \jdu.
	\label{maxmov}
\ee

A fim de que seja poss\'{\i}vel com este formalismo modelar fen\^omenos do mundo \asp real", \'e natural a introdu\c{c}\~ao de uma diferen\c{c}a geom\'etrica para uma das coordenadas em rela\c{c}\~ao \`as demais, a qual seria associada com o tempo. Em uma m\'etrica diagonal, a diferencia\c{c}\~ao geom\'etria correntemente usada \'e a da invers\~ao de sinal da componente temporal frente \`as espaciais. No momento, a quest\~ao do sinal n\~ao ser\'a importante, exigiremos apenas uma condi\c{c}\~ao de ortogonalidade. Embora praticamente todos os resultados que seguem só precisem dessa condição de ortogonalidade, vamos sempre supor que a componente $g_{00}$ da métrica tenha sinal diferente das demais. Vamos usar\footnote{Subentende-se que $\{ dx^{\mu} \}$, com $\mu = 0,1,...,D-1$, seja, para dado $x \in M$, base de  $T_x^\star M$.} $dt := dx^0$ para denotar o elemento da base que \'e ortogonal a todos os demais.

De forma geral, para qualquer m\'etrica, vale a iqualdade
\be
	dt \wedge \ddu x^i = g^{0i} \; f   \; d^Dx,
\ee
em que $i=1,2,...,d$, $\; d := D-1$, $\; f := \sqrt{| g |} \;$ e $\; g := \det (g_{\mu \nu})$. Vamos impor a condi\c{c}\~ao de ortogonalidade 
\be
	dt \wedge \ddu x^i = 0,
	\label{ort}
\ee
ou, equivalentemente, $g^{0i} = 0$ (como conseq\"u\^encia, $g_{0i}=0$, pois $g^{\mu \nu} g_{\nu \lambda} = \delta^\mu_\lambda$).

A fim de separarmos a parte temporal de $F$ de sua parte espacial, sejam $E$ uma 1-forma e $B$ uma 2-forma, ambas pertencentes ao espa\c{c}o de base $\{ dx^i \}$, isto \'e, $E = E_i \; dx^i$ e $B=\frac 12 B_{ij} \; dx^i \wedge dx^j$, e sejam essas tais que 
\be
	E \wedge dt + B = F.
	\label{ebf}
\ee

Lembremos que a 1-forma $A$ pertence ao espa\c{c}o cotangente \`a variedade D-dimensional $M$. Diremos que $E(x)$ pertence ao espa\c{c}o cotangente a $\hat M$ em $x \in \hat M$, que \'e uma variedade d-dimensional de m\'etrica $\hat g$, com $g_{ij} = \hat g_{ij}$. Assim sendo, $B(x) \in  T^\star_x \hat M \times T^\star_x \hat M$.

Como $F$ \'e uma 2-forma exata, em particular ela \'e fechada ($dF=0$). Em espa\c{c}os topologicamente triviais (i.e., homeomorfos ao espaço Euclideano ou ao de Minkowski), vale a rec\'{\i}proca. Portanto, em vez de definir $E$ e $B$ como fun\c{c}\~oes de $A$, como sugere a Eq. (\ref{ebf}), essas formas podem ser igualmente definidas por
\be
	dE \wedge dt + dB = 0.
	\label{ebbianchi}
\ee

Obter as equações de movimento em termos das componentes de $E$ e $B$ é simples, basta avaliar diretamente a Eq. (\ref{ebbianchi}) e o dual Hodge de (\ref{maxmov}). As equações abaixo valem para qualquer métrica constante e qualquer dimensão:
\ba
	\label{MDa}
	&& \prt_{[i} E_{j]} + \prt_0 B_{ij} = 0, \\
	\label{MDb}
	&& \prt_{[i} B_{jk]} = 0, \\
	\label{MDc}
	&& \prt^j B_{ji} - \prt^0 E_i = e \; J_i, \\
	\label{MDd}
	&& \prt_i E^i = e J_0,
\ea
em que os índices contravariantes acima são obtidos a partir do emprego da métrica, em particular $E^i$ depende da métrica. Os colchetes acima indicam anti-simetrização dos índices sem fator de normalização. As duas primeiras equações, advindas da identidade de Bianchi, são topológicas, ao contrário do que ocorre com as demais.

A fim de associar as últimas equações com alguma física, é essencial determinar a relação desse formalismo com as grandezas mensuráveis experimentalmente. Queremos associar $E$ e $B$ aos campos elétrico e magnético respectivamente. No espaço-tempo quadridimensional, denotaremos por 
\be
	\vec E = E^i \; e_i\; \; \; \; \mbox{ e } \; \; \; \;  \vec B = B^i \; e_i 
\ee
os campos físicos, isto é, os que são \asp diretamente" $\;$ medidos, sendo $\{ e_i \}$ base do espaço vetorial $T_xM$. Há arbitrariedades na associação entre os pares ($\vec E$, $\vec B$) e ($E$, $B$). A fim de obtermos as equações de Maxwell com as convenções usuais, fixaremos a relação desses pares como sendo: 
\be
	E^i = g^{ij} E_j \; \; \; \; \mbox{ e } \; \; \; \; B^i = \frac 12 \hat \omega^{ijk} \; B_{jk},
\ee	
em que $\hat \omega^{ijk} = \ep^{ijk} \;  \sqrt{| \hat g |}$. Em princípio poderíamos usar, por exemplo, $B^i = \frac 12  \vep^{ijk} \; B_{jk}$, assim o vetor $\vec B$ associado se comportaria tal qual o campo eletromagnético usual, porém somente em certas métricas, como $\diag ( - + + +)$; no caso da métrica $\diag ( + - - -)$,  $\vec B$ deixa de ser o que usualmente chamamos de campo magnético, passando a ser o negativo desse (conforme pode ser verificado nas quações de Maxwell correspondentes).

As equações para os vetores $\vec E$ e $\vec B$ são\footnote{Estas equações requerem a condição $g_{0i}=0$; mas exceto por essa e por ser constante, a métrica é arbitrária.}:

\ba
\label{4Dd354a}
&& \nabla \times \vec E + \dot {\vec B} = 0, \\
\label{4Dd354b}
&& \nabla \cdot \vec B = 0, \\
\label{4Dd354c}
&& (-1)^{\hat a} \; \nabla \times \vec B + g^{00} \; \dot {\vec E}  =  - e \; \vec J, \\
\label{4Dd354d}
&& \nabla \cdot \vec E = e \; J_0 = e \; g_{00} \; J^0,
\ea
com $\dot {\vec E} := \prt_0 \vec E$,  
\be
(-1)^{\hat a} = \frac{{\hat f}^2} {\hat g} = \frac{| \hat g|} {\hat g},
\ee

\be
	\nabla \cdot \vec E := \prt_i \; E^i \; \; \; \; \; \; \mbox{ e } \; \; \; \; \; \; (\nabla \times \vec E)^i := \hat \omega^{ijk} \; \prt_j \; E_k.
\ee

Sendo $g = \diag (- + + +)$, nota-se que $\hat \omega^{ijk} = \vep^{ijk}$ e $(-1)^{\hat a} = 1$. Utilizando a convenção $\rho = J^0$, sendo $\rho$ a densidade de carga, as equações de Maxwell em unidades Gaussianas com $c=1$ são obtidas se $e = - 4 \pi$. Conseqüentemente, para $g = \diag (+ - - -)$ temos  $e = 4 \pi$.

\vspace{.4in}

Em qualquer dimensão é imediato associar um vetor espacial como campo elétrico advindo da 1-forma $E$, mas a associação da 2-forma $B$ com um campo físico de natureza escalar ou vetorial é menos direta e depende da dimensão do espaço tratado. Seguindo a prescrição de representar o campo magnético através de certa operação de dualidade Hodge no espaço, vemos que em 3D o campo magnético deve ser representado por um escalar. Isso também é natural sob um ponto de vista físico. Suponhamos que em um laboratório esteja-se interessado em fenômenos eletromagnéticos que ocorram em uma superfície bidimensional. Passam por essa campos elétricos e magnéticos do nosso espaço tridimensional. A única componente do campo magnético que nos interessa é a perpedicular à superfície estudada ($B_z$), pois as componentes paralelas só produzem forças sobre as cargas da superfície na direção perpendicular a essa (por hipótese, estamos estudando um sistema que não evolui para fora da superfície original). Analogamente, a componente do campo elétrico perpendicular à superfície pode ser desprezada. Assim, o eletomagnetismo bidimensional no espaço deve ser descrito por um vetor elétrico $\vec E$ e um escalar $B_z$; mais precisamente é um pseudo-escalar (mudanças de paridade em 3D só invertem uma componente dos vetores espaciais).

Em termos das componentes de $E$ e do escalar $B_z :=  \frac k2 \hat \omega^{ij} B_{ij}$, sendo $k$ uma constante que será determinada em breve, as Eqs. (\ref{MDa}- \ref{MDd}) são escritas como
\ba
	\label{3Dcd354a}
	&& \hat \omega^{ij} \prt_i E_j + k^{-1} \; \dot B_z = 0,  \\
	\label{3Dcd354c}
	&& (-1)^{\hat a} \; k^{-1} \; \hat \omega^{ij} \prt_j B_z + g^{00} \dot E^i = - e J^i, \\
	\label{3Dcd354d} 
	&& \prt_i E^i = e J_0. 
\ea
	
A Eq. (\ref{MDb}) é identicamente nula no espaço bidimensional.  O termo $(-1)^{\hat a}$ em 3D é sempre igual à unidade desde que $t$ esteja associado ao tempo, em 4D esse termo troca de acordo com a mudança de assinatura; nas próximas equações em 3D ele não será escrito. As três equações acima podem também ser facilmente obtidas de (\ref{4Dd354a} - \ref{4Dd354d}) se for considerado que $E_z,B_y, B_x, J_z, \prt_z = 0$, $(-1)^{\hat a} = 1$ e $k=1$. Se no caso quadridimensional tivermos $(-1)^{\hat a} = - 1$, a redução dimensional não é compatível com as três equações acima para $k=1$. A constante $k$ se faz necessária pois as teorias eletromagnéticas em 3D e 4D são independentes entre si, mas desejamos que as mesmas convenções que usamos em 4D induzam convenções em 3D. Embora seja tentador crer que o campo magnético fisicamente mensurável seja sempre dado por uma espécie de dualidade Hodge no espaço da 2-forma $B$, ao adotar-se essa convenção entra-se em contradição, de forma geral, com as convenções usuais do espaço-tempo quadridimensional. A terceira componente, ou componente $z$, de $\vec B$ é, segundo a definição original, dada por
\be
	B^3 = \frac 12 \hat \omega^{3 ij} B_{ij} = \frac {\hat f_{3d}}{\hat g_{3d}} \; B_{12},
\ee
enquanto no espaço bidimensional temos
\be
	B_z = \frac k2 \hat \omega^{ij} B_{ij} = k \frac {\hat f_{2d}}{\hat g_{2d}} \; B_{12}.
\ee
Portanto, a igualdade entre $B_z$ e $B_3$ só pode ser obtida se
\be
  k = \frac {\hat g_{2d}} {\hat f_{2d}} \; \frac {\hat f_{3d}}{\hat g_{3d}}.
\ee

Se a métrica no espaço tridimensional for $\diag ( - + + +)$, a métrica induzida no espaço bidimensional será $\diag ( - + + )$, portanto $k=1$ neste caso. Por outro lado, começando com $\diag ( + - - -)$, a métrica induzida é $\diag ( + - -)$, logo $k= -1$. 


Por motivos análogos, é necessário introduzir o mesmo fator $k$ na definição de $\nabla \times \vec E$, portanto
\be
	\nabla \cdot \vec E := \prt_i \; E^i, \; \; \; \; (\nabla \times B_z)^i :=  \hat \omega^{ij} \; \prt_j \; B_z \; \; \; \; \mbox{ e }  \; \; \; \; \nabla \times \vec E := k \; \hat \omega^{ij} \; \prt_i \; E_j,
\ee

\ba
\label{3Dd354a}
&& \nabla \times \vec E + \dot  B_z = 0, \\
\label{3Dd354c}
&&  k^{-1} \; \nabla \times B_z + g^{00} \; \dot {\vec E}  =  - e \; \vec J, \\
\label{3Dd354d}
&& \nabla \cdot \vec E = e \; J_0 = e \; g_{00} \; J^0.
\ea

No Apêndice \ref{apformasm} mostramos como, seguindo as idéias desta seção, introduzir formas diferenciais no espaço $\hat M$. Depois dessas diversas manipulações com tensores e formas diferenciais, na próxima seção trataremos das teorias de calibre não-Abelianas de forma mais objetiva.





\vspace{1in}
\secao{Teorias de calibre $SU(N)$}
\label{tu2}

O objetivo dessa seção é introduzir alguns conceitos básicos sobre teorias de calibre associadas a grupos unitários não-Abelianos. Conforme será apresentado nos próximos capítulos, as teorias de calibre não-comutativas do tipo U(1) possuem fortes semelhaças com as não-Abelianas. Há inúmeras referências sobre o assunto que o abordam de forma mais profunda ou detalhada, esta apresentação baseia-se principalmente nas Refs. \cite{YM, weinberg2, nakahara}. Sucintas introduções históricas podem ser encontradas nas Refs. \cite{weinberg2, revjackiwym}.

Em 1928, Dirac propôs sua conhecida Lagrangeana para descrever uma partícula quântica-relativística de spin 1/2 e massa $m$:
\be
	\label{dirac}
	\cl_D = \bar \psi (i \gamma^\mu \prt_\mu - m) \psi.
\ee
Essa Lagrangeana possui uma simetria global trivial, a saber: $\psi \rightarrow S \psi$, com $S = \exp(i \vep)$ e $\vep \in \real$, ou seja, $S \in U(1)$. O operador $S$ pode pertencer a outros grupos unitários $U(N >1)$ e ainda assim manter a invariância global de  $\cl_D$. Como $U(N) = U(1) \times SU(N)$, no que segue trataremos apenas dos grupos SU(N) e U(1).

A fim de implementar simetrias locais na Lagrangeana de matéria $\cl_M$, isto é, simetrias com $ \prt_\mu S \not=0$, uma conexão é necessária para se definir uma derivada covariante $D$ tal que $D \psi \stackrel {SU(N)} \longrightarrow S \; D \psi$. Essa conexão naturalmente não deve ser a conexão métrica, usualmente representada por $\Gamma^\mu_{\nu \lambda}$, pois este problema independe da geometria do espaço-tempo. Considere
\be
	D = d - i g A, 
\ee
em que $g$ é uma constante (adimensional em 4D) chamada de constante de acoplamento de calibre e $A$ é uma 1-forma\footnote{Mais precisamente, veremos que $A$ é uma 1-forma que pertence ao espaço da álgebra de SU(N).} que atuará como conexão e se transforma segundo a regra
\be
	\label{tcalibre}
	A \rightarrow A' = S A S^\dagger + \frac ig S \; dS^\dagger,
\ee
com $S \in SU(N)$, ou seja, $\det S = 1$ e $S^\dagger \; S = S \; S^\dagger = \id$.

Assim, substituindo $\prt$ por $D$ no setor de matéria da ação, este torna-se invariante por transformações locais $SU(N)$.

\vspace{.4in}

Um elemento qualquer $S \in SU(N)$ pode ser univocamente representado por uma exponencial $ S = \exp (i \vep (x))$, em que $\vep$ é dito pertencer à álgebra de $SU(N)$ [i.e., $\vep \in su(N)$]. No caso $U(1)$, $\vep \in \real$. Para $S \in SU(N)$, a fim de que $S S^\dagger = \id$ e $\det S = 1$, $\vep$ deve ser Hermiteano e de traço nulo. 

Para o caso de transformações infinitesimais, isto é, $\vep$ suficientemente pequeno, temos
\ba
	\label{tcalibreinf}
	A \rightarrow A' &\approx& (1 + i \vep) A (1 - i \vep) + \frac 1g d\vep \nonumber \\
	&\approx& A + \frac 1g d\vep + i [\vep, A].
\ea

Sendo $\{ t^a \}$ base de $su(N)$, os $t^a$'s são chamados de geradores do grupo $SU(N)$. Dados $S, T \in SU(N)$, sabemos que o produto $ST$ também pertence a $SU(N)$ [admitindo que $SU(N)$ seja grupo, isso é imediato pela definição de grupo; paralelamente, pode-se verificar essa propriedade facilmente pela definição dos elementos de $SU(N)$ já apresentada]. Conseqüentemente, como $\{ t^a \}$ é base de $su(N)$ existe $f^{abc} \in \real$, denominado constante de estrutura do grupo, tal que
\be
	[ t^a, t^b] = i f^{abc} t^c.
\ee
(A regra da soma se aplica, embora todos os índices estejam no mesmo nível.)

Vamos assumir que os geradores e as constantes de estruturas são normalizadas de tal forma que a relação anterior seja preservada e 
\be
	\tr (t^a \; t^b) = n \delta^{ab},
\ee
em que $n \in \real$. Como um traço sempre aparece nas ações não-Abelianas, a escolha de $n$ influi na constante global da acão. A prescrição usual é trocar a constante global Abeliana por ela própria divida por $n$. Para maior semelhança com o caso Abeliano, e menor confusão com constantes, vamos seguir a convenção de \cite{YM, weinberg2} e adotar $n=1$, ou, equivalentemente, vamos usar $g^2$ no lugar de $n g^2$.

\vspace{.4in}
Retomando agora a análise da transformação de $A$ (\ref{tcalibre}), vemos que $A$ deve pertencer ao mesmo espaço que o $\vep$ introduzido há pouco, ou seja, $A \in su(N)$; ademais, qualquer transformação de calibre de $A$ nunca o leva para fora de $su(N)$, como pode ser diretamente verificado. Portanto, sempre podemos expandir a 1-forma $A$ usando a base de $su(N)$, ou seja,
\be
	A = A_\mu dx^\mu = A_\mu^a \; t^a \; dx^\mu = A^a t^a,
\ee
conseqüentemente 
\be
	A \wedge A = A^a \wedge A^b t^a t^b = \frac 12 A^a \wedge A^b [t^a, t^b] = \frac i2 A^a \wedge A^b f^{abc}t^c.
\ee

No caso $U(1)$, o termo da ação responsável pela dinâmica da parte de calibre é $DA \wedge \str DA = dA \wedge \str dA$, termo que é trivialmente invariante por transformações de calibre (\ref{tcalibre}), pois $F=dA$ é invariante. De forma geral, o termo responsável pela dinâmica do campo de calibre é
\be
	\label{symc}
	S_{YM} = -\frac 1{2g^2} \int \tr \; F \wedge \str F,
\ee
com 
\be
	F = dA - i A \wedge A,  \; \;  \mbox{ou seja,}  \; \; F_{\mu \nu} = \prt_\mu A_\nu - \prt_\nu A_\mu - i[A_\mu, A_\nu].
\ee

Como, para dado $A$, a ação é um número, a presença do traço é natural. Por outro lado, sua presença é imprescindível para garantir a invariância da ação de Yang-Mills $S_{YM}$, pois perante transformações de calibre $F$ não é invariante, a saber: 
\be
	F \; \longrightarrow \; S \; F \; S^\dagger,
\ee
mas $\tr \; F$ e $\tr \; F \wedge \str F$ são invariantes. 

Nas últimas equações, uma redefinição de $A$ foi feita: $A \rightarrow \frac 1g A \; \; \Rightarrow D = d - i A$. Normalmente usaremos esta notação.

A ação localmente invariante por $SU(N)$ encontrada é portanto
\be
	S = S_{YM}[A] + S_{M} [\psi, \bar \psi, A],
\ee
em que $A$ ocorre em $S_{YM}$ somente interiormente a $F$ e na ação de matéria $S_{M}$ somente interiormente à derivada covariante $D$.

Em teorias $U(1)$, $F$ satisfaz a identidade $dF=0$. Em $SU(N)$ essa identidade é substituída pela seguinte (também topológica): 
\be
	D' F = 0,
\ee
com $D' = d - i [A, \; ]$ e $[A, B] = A \wedge B - B \wedge A$. As equações de movimento são
\be
	D' \str F = - \tilde J, \; \; \; \mbox{ou seja,} \; \; \; D'_\mu F^{\mu \nu a} = - J^{\nu a},
\ee
sendo 
\be
	J^{\nu a} := -i \frac {\delta \cl_M}{\delta A_\nu^a}.
\ee
É interessante notar que, sendo $\cl_M$ invariante de calibre, a corrente $J^{\nu a}$ só é invariante no caso $U(1)$. Para $SU(N)$, $\tilde J$ é covariante. A corrente $\tilde J$ satisfaz a lei de conservação
\be
	D' \tilde J = 0, \; \; \; \mbox{ou seja,} \; \; \; D'_\nu J^{\nu a} = 0.
\ee

Dentre várias outras observarções a serem feitas é importante notar que teorias de Yang-Mills, mesmo na ausência de matéria, são teorias interagentes. A revisão de teorias não-Abelianas aqui exposta terá alguma utilidade direta nas próximase seções e será especialmente útil para o entendimento das teorias não-comutativas.


\vspace{1in}
\secao{Massa topológica e o termo de Chern-Simons}
\label{tu3}

A primeira referência na física ao termo de Chern-Simons (CS), embora sem utilizar essa nomenclatura, ocorreu no final da década de 70 e se deve a Siegel \cite{siegelcs}. Desde o início da década de 80, especialmente devido aos trabalhos de Deser, Jackiw, Templeton e Schonfeld \cite{jtcs, schonfeld, djtcs}, as teorias que fazem uso desse termo têm despertado  atenção de áreas diversas como $QED_3$, teorias de cordas, gravitação e matéria condensada \cite{aplicacoescs, nccsdbranes, jtcs, djtcs, revjackiw3d}. O termo de CS tem uma ligação direta com a classe característica secundária de Chern-Simons \cite{matcs, nakahara}, que já era conhecida no meio matemático.

Conforme introduzido em \cite{jtcs, schonfeld, djtcs}, espaço-tempos de dimensão ímpar permitem o aparecimento de uma estrutura topológica invariante de calibre capaz de conferir massa ao bóson correspondente. Um fóton com massa topológica no espaço-tempo tridimensional é descrito pelo modelo chamado de Maxwell-Chern-Simons (MCS), esse possui uma série de diferenças em relação ao eletromagnetismo sem massa e ao modelo de Proca: os spins das partículas associadas a esses modelos são, respectivamente, dois singletos de spin 1, spin 0 e um dubleto de spin 1; todos os três são invariantes por conjugação de carga (C), os dois últimos são invariantes também por transformações de paridade (P) e de inversão temporal (T), enquanto o modelo MCS viola tanto P quanto T, sendo invariante por PT; o tensor energia-momento de MCS é diferente do de Proca e igual ao do eletromagnetismo puro em três dimensões; dentre outras particularidades que serão apresentadas nesta seção.


Sendo $A \in u(1)$ uma 1-forma em um espaço-tempo tridimensional, definimos o chamado termo de Chern-Simons (CS) pela seguinte ação: 
\be
	S_{CSA} = \frac m {2 g^2} \int A \wedge dA,
\ee
em que $g$ é a constante de acoplamento de calibre e $m$ é uma constante com dimensão de massa.

Sendo $A$ elemento da álgebra $su(N)$, escreve-se 
\be
	\label{CSnA}
	S_{CSnA} = \frac m {2g^2} \int \tr \( A \wedge dA - \frac {2i}3 A \wedge A \wedge A \).
\ee

Dependendo das convenções usadas, as constantes que aparecem acima podem ser diferentes. Independentemente das convenções, é essencial que 
\be
\frac {\delta S_{CS}} {\delta A}  \propto F.
\ee

Verifica-se diretamente que a Lagrangeana do termo de CS não é de forma geral invariante por transformações de calibre. A menos de termos de superfície, para transformações infinitesimais de calibre, as ações $S_{SCA}$ e $S_{SCnA}$ são invariantes. Para transformações de calibre finitas, há uma distinção clara entre o caso Abeliano e o não-Abeliano. A ação do primeiro mantém-se como invariante a menos de termos de superfície, a ação do segundo, porém, a menos de termos de superfície, transforma-se por uma constante global \cite{djtcs} 
\be
	S_{CSnA} \; \longrightarrow \; \frac {8m \; \pi^2 \; \omega}{g^2} S_{CSnA},
\ee
em que $\omega \in \nat$. Impondo que a função partição de $S_{CSnA}$ não se altera por transformações de calibre, a constante $\frac {8m \; \pi^2 \; \omega}{g^2}$ tem de ser um múltiplo inteiro de $2 \pi$, ou seja,
\be
	4 \pi \frac m {g^2} = n, \; \; \; \; \mbox{com } n = 0, \pm 1, \pm 2...
\ee

Esta relação de quantização, naturalmente associada a uma quantização de $m$, estabelece uma distinção dramática entre as ações $S_{CSA}$ e $S_{CSnA}$, pois, para dado $g^2$, $m$ na primeira pode ser qualquer. Essa propriedade é chamada de quantização de nível \cite{djtcs, revjackiw3d}.

As Lagrangeanas $L_{CSA}$ e $L_{CSnA}$ se conservam por C, mas trocam de sinal global por P ou T, sendo invariantes por PT e CPT \cite{djtcs}.

O termo de CS é chamado de topológico devido a independer da métrica, essa não aparece nem entre as contrações dos índices e nem como fator do elemento de volume, como o caso Abeliano ilustra abaixo\footnote{Deve-se lembrar que  $\vep^{\mu \nu \lambda}$ não depende da métrica, ao contrário de $\ep^{\mu \nu \lambda} = (\det g_{\mu \nu})^{-1} \vep^{\mu \nu \lambda}$.}:
\ba
	S_{CSA} && = \frac m {2g^2} \int A_\mu \prt_\nu A_\lambda \; dx^\mu \wedge dx^\nu \wedge dx^\lambda \nonumber \\[.2in]
	&& = \frac m {2g^2} \int A_\mu \prt_\nu A_\lambda \; \vep^{\mu \nu \lambda} d^3x.
\ea
Portanto o tensor energia-momento do termo de CS é nulo. A adição desse termo a qualquer modelo não altera o tensor energia-momento do modelo original.

\vspace{.4in}
O termo de CS, seja Abeliano ou não-Abeliano, possui zero grau de liberdade. Isso pode ser minuciosamente verificado através do método de Dirac\footnote{Ao fazer a contagem de graus de liberdade, deve-se estar atento à interdependência linear de seus vínculos.}\cite{mdirac} ou pelo simplético \cite{mfj, mbw}. Para o caso Abeliano, essa propriedade pode ser antevista dada sua trivial equação de movimento: $F = dA = 0$. 

Para conferir dinâmica ao termo de CS é natural adicioná-lo ao termo de Maxwell, assim formando a teoria Maxwell-Chern-Simons (MCS), que com fonte é dada por
\be
	\label{acaomcso}
	S_{MCS}[A] = - \frac 1 {2g^2} \int  \( \str F \wedge F \pm m A \wedge F \) + \int A \wedge \str J,  
\ee
com $F=dA$ e $d \str J = 0$. Assumiremos que a métrica é $(g_{\mu \nu}) = \diag (+ \; - \; -)$ e $m > 0 $. O sinal do termo de CS pode ser positivo ou negativo, um se transforma no outro por mudança de paridade. A equação de movimento é dada por
\be
	d \str F \mp  m F = - g^2 \; \str J. 
	\label{mcseqmf}
\ee

A equação acima é consistente com a conservação da corrente $d \str J = 0$. Assim como no eletromagnetismo usual, e contrariamente ao modelo de Proca, o potencial $A$ só ocorre nas equações de movimento internamente a $F$. Nota-se que $m$ não só tem dimensão de massa como realmente está associada a um pólo no propagador \cite{djtcs}. A última equação pode ser escrita como 
\be
	(d \str d \str +  m^2) F = - g^2 \; dJ \mp g^2 \; m\; \str J. 
\ee
Essa equação é obtida ao se aplicar o \asp divergente" $\; d \str$ em (\ref{mcseqmf}) e substituir  $m d \str F$ pela sua própria expressão dada por (\ref{mcseqmf}). Não foi antes comentado, mas $d \str d \str F$ é proporcional ao d'Alambertiano de $F_{\mu \nu}$. 

Dividindo a 2-forma $F$ em duas formas no espaço $E$ e $B$, como feito em (\ref{ebf}), as equações advindas da identidade de Bianchi permanecem as mesmas (\ref{MDa}, \ref{MDb}), enquanto a Eq. (\ref{mcseqmf}) se torna
\ba
	&&  \prt^jB_{ji} - \prt_0 E^i  \mp  m \; E^j \; \ep_{ji} = g^2 \; J_i,  \\[.2in]  
	&&  \prt^i E_i \pm  m \; B_{ij} \; \ep^{ij} =  g^2 \; J_0. 
\ea

Vamos introduzir a seguinte notação escalar-vetorial\footnote{O sinal que aparece na frente de $B_z$ e $\nabla \times \vec E$ é condizente com as observações da Seção \ref{tu1} em relação à métrica adotada, ou seja, $k= -1$.}:

\ba
	B_z &:=& - \frac 12 B_{ij} \ep^{ij} = - B_{12}, \\
	\nabla \times \vec E &:=& - \ep^{ij} \prt_i E_j, \\
	(\nabla \times B_z)^i &:=& \ep^{ij} \prt_j B_z, \\
	(\vec{ \tilde E})^i &:=& \ep^{ij} E_j.
\ea

Conseqüentemente, nesta notação, as equações de movimento do modelo de MCS são
\ba
\label{MCSa}
&& \nabla \times \vec E + \dot  B_z = 0, \\[.2in]
\label{MCSc}
&&  \nabla \times B_z -  \dot {\vec E} \pm m \; \vec{\tilde E} =  g^2 \; \vec J, \\[.2in] 
\label{MCSd}
&& \nabla \cdot \vec E \pm m \; B_z = g^2 \; J_0. 
\ea

A violação de paridade é clara nas equações acima, pois $B_z$ e $\vec{\tilde E}$ são pseudo-escalar e pseudo-vetor. Na próxima seção veremos que, por meio da dualidade do modelo de MCS com o autodual, a união dos dois modos propagantes do modelo de MCS é descrita pelo modelo de Proca em 3D. A versão não-Abeliana do termo de CS será retomada no Cap. \ref{cap6}, ao tratamos do termo de CS não-comutativo.

\vspace{.4in}

\section{Dualidades eletromagnéticas}
\label{tu4}

Dualidade é por si só um tema muito vasto com diversas aplicações em física. Há vários tipos de dualidades empregadas na física e muitas vezes o conceito é utilizado de forma vaga, algumas introduções sobre o assunto podem ser vistas na Refs. \cite{introdual, intromaster}. De forma geral, dualidade se refere a descrições distintas de um mesmo fenômeno ou sistema físico. Há diversas expectativas quanto ao papel das dualidades na física, incluindo o entendimento de confinamento, unificação das teorias de cordas, verificação de monopólos magnéticos etc. Em teorias de cordas costuma-se distingüir três tipos de dualidades: $S$, $T$ e $U$. Essencialmente, dualidades do tipo $S$ (\asp strong/weak dualities") são as que relacionam modelos por meio da inversão da constante de acoplamento, dualidades do tipo $T$ (\asp target-space dualities") relacionam modelos por meio da inversão do raio de compactificação e dualidades do tipo $U$ são uma mistura dos dois casos anteriores. As dualidades eletromagnéticas são manifestações de dualidades do tipo $S$, são autodualidades das cordas do tipo $IIB$ e podem ser descritas no contexto de teorias de campos. As dualidades eletromagnéticas, contudo, foram percebidas originalmente bem antes do surgimento das teorias de cordas, já no século XIX por Heaviside \cite{heaviside}, e além de sua relevância intrínseca (que será comentada em breve) são úteis também como laboratórios para esclarecer o funcionamento de outras dualidades. 

Dualidade eletromagnética é também um tema amplo e, embora antigo, continua a se desenvolver nos dias de hoje. Dentre os avanços das últimas décadas encontram-se a conjectura de Montonen-Olive (${\cal N} = 4$) \cite{montonenolive}, dualidades com massa topológica \cite{master}, dualidade de Seiberg-Witten (${\cal N}=2$) \cite{swd}, dualidade de Seiberg (${\cal N}=1$) \cite{seibergd}, dualidades Abelianas não-lineares (DBI inclusive) \cite{nonlineard}, dualidades em espaço-tempos não-comutativos \cite{ganor, gmms} e mais recentemente a conexão com o programa de Langlands descoberta por Kapustin e Witten \cite{wittenl}. Revisões \asp pedagógicas" sobre os três primeiros tipos de dualidade se encontram nas Refs. \cite{introdual, intromaster}. Somente alguns aspectos básicos das dualidades eletromagnéticas serão necessárias para os objetivos desta tese, nesta apresentação não iremos tratar de nenhum caso supersimétrico e nem da conexão com Langlands.

\vspace{.2in}
Essa complexa estrutura atual de dualidades eletromagnéticas teve início com uma simples observação das equações de Maxwell em 4D e sem fontes\footnote{Veja as Eqs.(\ref{4Dd354a}-\ref{4Dd354d}) com $g=\diag (+ - - - )$.},
\ba
\label{4Dd354an}
&& \nabla \times \vec E + \dot {\vec B} = 0, \\
\label{4Dd354bn}
&& \nabla \cdot \vec B = 0, \\
\label{4Dd354cn}
&& \nabla \times \vec B - \; \dot {\vec E}  = 0, \\
\label{4Dd354dn}
&& \nabla \cdot \vec E = 0.
\ea
É imediato notar que essas equações são invariantes pelas transformações $\vec E \rightarrow \vec B$ e $\vec B \rightarrow -\vec E$ [equivalentemente escreve-se $(\vec E, \vec B) \rightarrow (\vec B, -\vec E)$]. Caso esteja-se analisando uma onda eletromagnética no vácuo, essa transformação é equivalente a uma rotação de $\pi/2$. Essa correspondência possui também uma forma contínua, dada por
\be
	\pmatrix { \vec E' \cr \vec B'}  = \pmatrix{ \cos \alpha & \sen \alpha \cr - \sen \alpha & \cos \alpha} \pmatrix{ \vec E \cr \vec B},
\ee
em que o caso discreto anterior corresponde a $\alpha = \pi/2$. Pode-se ainda escrever vetores complexos que se mantém invariantes por essas transformações, a menos de uma fase, e expressar as quatro equações de Maxwell através de duas apenas. Ademais, esse quadro ainda pode ser estendido para o caso de meios dielétricos e de teorias eletromagnéticas não-lineares, como a de Dirac-Born-Infeld. Nesse quadro geral, as equações de movimento são
\ba
	\nabla \cdot ( \vec D + i \vec B) &=& 0, \\
	\nabla \times (\vec E + i \vec H) &=& i\frac \prt {\prt t} (\vec D + i \vec B),
\ea
em que $D^i = G^{i0}$, $H^i = \frac 12 \ep^{ijk} G_{jk}$ e $G^{\mu \nu} = -2 \frac {\prt \cl}{\prt F_{\mu \nu}}$ \cite{nonlineard}. E essas equações são invariantes por 
\ba
	( \vec D + i \vec B) &\rightarrow & e^{i \alpha} ( \vec D + i \vec B), \\
	(\vec E + i \vec H) &\rightarrow & e^{i \alpha} (\vec E + i \vec H),
\ea
com $\alpha$ constante. Os detalhes dessas passagens podem ser vistos nas Refs. \cite{nonlineard, nonlineard2}. Embora consideravelmente gerais e elegantes, essas correspondências não são imediatamente preservadas ao inserirmos fontes, não só por \asp questões formais", mas também é evidente sob um ponto de vista físico. Em particular, frente a essas correspondências ondas eletromagnéticas são rodadas por uma ângulo $\alpha$. No vácuo sem fontes, qualquer rotação da onda eletromagnética é indiferente, mas havendo um polarizador essa simetria é quebrada, a menos que ela passe a envolver o polarizador também. Outra forma de analisar essa questão, talvez mais óbvia, é considerar um sistema composto de carga elétrica junto de seu campo elétrico; assim, a fim de que a transformação $(\vec E, \vec B) \rightarrow (\vec B, -\vec E)$ produza um sistema físico coerente, monopólos elétricos devem tornar-se monopólos magnéticos. Essa propriedade, em parte, levou a propostas de vários tipos de monopólos magnéticos.

\vspace{.2in}
Iremos analisar agora como essa simetria se apresenta na ação. Em termos da 2-forma eletromagnética, a simetria $(\vec E, \vec B) \rightarrow (\vec B, -\vec E)$ é simplesmente descrita por $ F \rightarrow \str F$. Trocar $F$ por $\str F$ apenas altera o sinal global da ação do eletromagnetismo, pois $\str \str F=-F$ no espaço-tempo tratado. Existe uma forma mais precisa e geral de analisar essa invariância física. Seja
\be
	S[A] = \frac 1{2g^2} \int F \wedge \str F
\ee
a ação do eletromagnetismo em 4D com $F:=dA$, $g=\diag(+ - - - )$ e $g^2$ a constante de acoplamento. Considere a seguinte ação
\be
	\label{primestra}
	S_M[A_D,F] = \int \frac 1 {2g^2} F \wedge \str F + dA_D \wedge F.
\ee
Acima, $F$ é uma 2-forma qualquer e $A_D$ é uma 1-forma qualquer, cujas dinâmicas são determinadas pelo princípio de mínima ação (assume-se que a integral atua em ambas as parcelas acima). A variação dessa ação com respeito a $A_D$ produz
\be
	\label{eqma1}
	dF=0.
\ee
Como a topologia do espaço-tempo tratado é trivial, existe uma 1-forma $A$ tal que $F=dA$. Substituindo esse resultado em $S_M$ vem
\be
	S_M \lra S[A] = \frac 1{2g^2} \int d A \wedge \str dA.
\ee

Considerando agora a variação de $S_M$ com respeito a $F$, temos
\be
	\label{eqma2}
	dA_D = - \frac 1 {g^2} \str F.
\ee
Essa relação nos permite eliminar $F$ de $S_M[A_D,F]$, conseqüentemente
\be
	S_M \lra \tilde S [A_D] = \frac {g^2}2 \int d A_D \wedge \str dA_D.
\ee
Ou seja,
\be
	S \lra \tilde S.
\ee
O símbolo $\lra$ foi introduzido para indicar equivalência entre funcionais quando suas variações são tomadas como nulas. A equivalência obtida através da técnica acima é mais forte do que essa correspondência, mas não entraremos em detalhes no momento. O mapa que fornece a correspondência entre os campos de $S$ e $\tilde S$ é dado pela combinação das Eqs.(\ref{eqma1}, \ref{eqma2}),
\be
	dA_D = - \frac 1 {g^2} \str dA.
\ee
Nota-se que ao aplicar $d$ na equação acima obtém-se a equação de movimento de $S$, enquanto a aplicação de $d \str$ leva à equação de movimento de $\tilde S$.

Esse mapa, a menos do fator $- \frac 1 {g^2}$, descreve exatamente a substituição de $F$ por $\str F$ anteriomente mencionada. Embora classicamente esse fator não tenha importância, essa mesma dedução pode ser feita em funções partição (as variações acima correspondem a integrais Gaussianas) e vê-se que a constante de acoplamento é realmente invertida. Enfim, partindo da invariância $(\vec E, \vec B) \rightarrow (\vec B, - \vec E)$ obtivemos um primeiro indício de existência de certa autodualidade que relaciona regiões de acoplamento forte com acoplamento fraco. 

Com um pouco de experiência, percebe-se que essa técnica também pode ser usada para o caso com fontes. Considere a seguinte ação:
\be
	S_{M_\Lambda} [F,A_D]= \int  \frac 1 {2g^2} F \wedge \str F+  F \wedge(\Lambda + dA_D).
\ee

A variação com respeito a $A_D$ implica que $dF=0$ e portanto existe uma 1-forma $A$ tal que
\be
	F = dA.
\ee
Substituindo esse resultado em $S_{M_\Lambda}$, vem
\be
	S_{M_\Lambda} \lra S_J[A]= \int \frac 1 {2g^2} dA \wedge \str dA + A \wedge \str J ,
\ee
em que usamos $\str J := d \Lambda$, o que é consistente com $d \str J = 0$. A variação de $S_{M_\Lambda}$ com respeito a $F$ fornece a teoria dual ao eletromagnetismo com fonte $J$. A equação de movimento provinda dessa variação lê-se
\be
	\label{mapafonte}
	\frac 1{g^2} \str F + \Lambda + dA_D = 0.
\ee
Seja $F_{D\Lambda} := dA_D + \Lambda$, a ação dual é portanto
\be
	S_{M_\Lambda} \lra \tilde S_\Lambda [A_D] = \frac {g^2}2\int F_{D\Lambda} \wedge \str F_{D\Lambda}.
\ee
A teoria acima está acoplada a uma fonte, porém não através de um acoplamento mínimo. A identidade de Bianchi e a equação de movimento de $S_J$ são
\be
	dF = 0 \; \; \; \; \; \; \; \; \; \; \; \; \; \; d \str F = - g^2 \; \str J,
\ee
enquanto que no quadro dual a identidade de Bianchi e a equação de movimento são respectivamente
\be
	dF_{D\Lambda} =  \str J \; \; \; \; \; \; \; \; \; \; \; \; \; \; d \str F_{D\Lambda} = 0.
\ee
	
Um detalhe importante quanto a esse tipo de dualidade com fontes é a presença de uma simetria envolvendo $\Lambda$, chamado de \emph{Chern-kernel} \cite{chernkernel}. Enquanto $S_J$ é invariante por $\Lambda \rightarrow \Lambda + d \lambda$, para qualquer 1-forma $\lambda$, a ação $\tilde S_\Lambda$ é invariante pela combinação $\Lambda \rightarrow \Lambda + d \lambda$ e $A_D \rightarrow A_D - \lambda$. Pode-se optar pela fixação dessa simetria usada na Ref. \cite{marcelomcs}, ou seja, $d \str \Lambda = 0$. Essa escolha leva à equação de movimento $d \str d A_D = 0$, que lembra a expressão usual do eletromagnetismo, exceto pela fonte não se encontrar na equação de movimento, mas sim na identidade $dF_{D\Lambda} =  \str J$ acima apresentada.

Quanto ao significado da dualidade formal acima, alguns comentários precisam ser feitos. Em primeiro lugar nota-se que o mapa $F \rightarrow \str F$ não se manteve inalterado, foi generalizado para $F \rightarrow \str \Lambda + \str F$ (\ref{mapafonte}) (ignorando as constantes globais), ou seja, depende explicitamente do \emph{Chern-kernel}. Em certo sentido, pode-se dizer que a teoria dual $\tilde S$ contém monopólos magnéticos, isso ocorre se identificarmos o campo magnético no quadro dual como $\tilde B^i = \frac 12 \ep^{ijk} F_{D \Lambda_{jk}}$, assim devido a $dF_{D \Lambda} = \str J$ temos $\nabla \cdot \vec{\tilde B} = J^0$. Por outro lado, pode-se corretamente argumentar que a questão de monopólos magnéticos físicos não está sendo abordada, apenas se está lidando com monopólos elétricos de uma forma não convencional. Foi obtida uma equivalência formal entre Lagrangianas, monopólos magnéticos simplesmente não foram introduzidos; todavia, no quadro dual, esses mesmos monopólos elétricos são descritos tal como se fossem monopólos magnéticos ($dF \not=0$). Uma forma de introduzir monopólos magnéticos físicos é, na ação $\tilde S_\Lambda$, inserir uma nova corrente que se acople minimamente com $A_D$, tal como feito na Ref. \cite{clovisdiag}. 

\vspace{.4in}

A técnica acima utilizada, chamada de Lagrangiana (ou ação) mestra, foi introduzida em \cite{master} para verificar a dualidade entre os modelos MCS e autodual. O modelo autodual foi proposto na Ref. \cite{autodual} como uma espécie de \asp raiz quadrada" (segundo os próprios autores) do modelo de Proca em 3D. Seguindo a Ref. \cite{autodual}, sendo a ação do modelo de Proca dada por\footnote{Como nenhuma constante de acoplamento com dimensão foi inserida, nesta parte $A$ tem dimensão de massa$^{\frac 12}$.} 
\be
	S_{\mbox{\tiny Proca}}[A]= -\frac 1 2 \int F \wedge \str F - m^2 A \wedge \str A,
\ee
sua equação de movimento é
\be
	\label{elki2}
	d \str dA - m^2 \str A =0,
\ee

Considere a seguinte equação:
\be
	\label{elki3}
	A = \pm \frac 1m \str dA.
\ee
Aplicando a última equação sobre ela mesma, a Eq. (\ref{elki2}) é obtida (lembrar que $\str \str = 1$ em 3D). Substituindo $A$ por $f$, a ação que tem (\ref{elki3}) como equação de movimento é
\be
	\label{acaoadf}
	S_{AD}[f] = \frac 12 \int \mp \frac 1m f \wedge df + f \wedge \str f.
\ee

A ação acima descreve o chamado modelo autodual. Esse modelo propaga 1 grau de liberdade para cada sinal do termo de CS $f \wedge df$, enquanto o modelo de Proca em 3D possui 2 graus de liberdade. Embora a motivação original desse modelo tenha sido puramente formal, posteriormente, por meio de um mecanismo de bosonização para $m$ suficientemente grande, foi demonstrada sua correspondência com o modelo de Thirring \cite{bosSchaposnik}. Nessa referência os autores utilizam a dualidade entre as teorias MCS e autodual (obtendo a correspondência do modelo de Thirring com o MCS) que iremos agora apresentar e foi introduzida em \cite{master}. Por brevidade, demonstraremos a dualidade apenas usando a técnica da ação mestra; mais detalhes sobre essa correspodência podem ser encontrados nas Refs. \cite{master, quanmcsad}. No Cap.\ref{cap6} a dualidade entre as extensões não-comutativas dos modelos MCS e autodual será demonstrada através de outra técnica, a projeção dual. 

Considere a seguinte transformação de Legendre na ação de MCS (\ref{acaomcso}):
\be
	S_M[A,f] = \int \frac {g^2}2 f \wedge \str f + dA \wedge f - \frac m{2g^2} A \wedge dA.
\ee
Trata-se essencialmente da mesma ação mestra que aparece na Eq. (\ref{primestra}) com $F$ substituído por $f$ e com a adição do termo de CS. A variação com respeito a $f$ leva $S_M$ de volta a (\ref{acaomcso}). A equação de movimento advinda da variação com respeito a $f$ é
\be
	\label{mapamcsad}
	dA + g^2 \str f =0.
\ee
Eliminando $f$ de $S_M$ retorna-se à ação original de MCS. Por outro lado, a variação de $A$ tem como equação de movimento
\be
	df - \frac m { g^2} dA=0,
\ee
a qual, ao ser empregada para elimiar $A$ de $S_M$, leva justamente à ação do modelo autodual, a saber
\be
	S_{AD} = \frac {g^2}2 \int \frac 1m df \wedge f + f \wedge \str f.
\ee
Substituindo $m$ por $-m$ obtem-se a dualidade do outro modo propagante. Verifica-se que tanto o modelo MCS quanto o autodual violam $P$ e $T$, mas preservam $PT$, $C$ e $CPT$.

A dualidade entre os modelos MCS e autodual é uma das mais conhecidas, ela tem servido de paradigma para o estudo de várias outras dualidades. Para uma detalhada introdução à técnica da Lagrangiana mestra veja \cite{intromaster}. Embora haja correspondência física entre os modelos MCS e autodual, suas estruturas formais são consideravelmente distintas, em particular MCS é uma teoria de calibre de segunda ordem nas velocidades, enquanto o modelo autodual é de primeira ordem nas velocidades e não possui simetria de calibe. Ademais, o mapa dado pela Eq. (\ref{mapamcsad}) não descreve uma fixação de calibre (veja a Subseção \ref{sssimetriacalibre}), mas sim um colapso de todas as órbitas de calibre de MCS, cada uma para um único ponto da superfície de vínculo do modelo autodual. Devido às suas diferenças formais, ocasionalmente pode ser preferível usar um modelo em relação ao outro. Na Ref. \cite{bosSchaposnik} dá-se preferência à formulação de MCS para a avaliação de certo $loop$ fermiônico advindo do modelo de Thirring, enquanto a Ref. \cite{diamantini} utiliza a formulação autodual para efetuar cálculos na rede referentes ao efeito Hall. 

\pagebreak
Outra diferença importante quanto aos modelos MCS e autodual se deve às suas extensões não-Abelianas e/ou não-comutativas. Na Ref. \cite{master} a questão da dualidade não-Abeliana é considerada, mas a abordagem utilizada se demonstrou inconclusiva. Nas Refs. \cite{namcsduality} os autores mostram que as extensões não-Abelianas desses modelos não são duais entre si. As extensões não-comutativas serão tratadas na Seção \ref{sec6ad}.

\chapter{Formalismos Simpléticos}
\label{capfs}

Na primeira parte deste capítulo apresentaremos uma revisão do formalismo simplético \cite{mfj, mbw, montani}, trata-se de um formalismo alternativo ao de Dirac de dedução da dinâmica Hamiltoniana de sistemas vinculados \cite{mdirac} (redução Hamiltoniana). Na seção seguinte será apresentado o chamado formalismo simplético de calibre, originalmente introduzido, como uma aplicação particular,  na Ref. \cite{symembskyrme}. Esse outro formalismo lida com a questão da \asp imersão em calibre" $\;$ seguindo os princípios do método simplético, em vez dos de Dirac. A segunda seção apresenta esse formalismo e aplicações em acordo especialmente com a Ref. \cite{symemb}.

\vspace{.4in}
\section{Formalismo simplético e parênteses generalizados}

\subsection{Noções de geometria simplética}

Embora seja possível avaliar toda a física clássica de um sistema utilizando Hamiltonianas e parênteses de Dirac (ou generalizados) sem mencionar estruturas simpléticas ou simplectomorfismos, noções de geometria simplética possibilitam uma nova e mais geral forma de analisar problemas Hamiltonianos. Esta seção segue em especial os princípios utilizados na Ref. \cite{arnold} (Cap. 9), outras referências úteis incluem \cite{teitelboim, varsymp}.

Considere uma variedade $N$ de dimensão par munida de certa 2-forma $f$ não-degenerada, ou seja, 
\be
	\label{degdef}
	\forall \zeta \not= 0, \exists \eta \; | \;  f( \zeta, \eta) \not= 0 \; \; (\zeta, \eta \in T_xN).
\ee
Equivalentemente, a matriz $(f_{\alpha \beta})$ que a representa possui determinante não-nulo ($f = \frac 12 f_{\alpha \beta} d\xi^\alpha \wedge d\xi^\beta$). 

Uma diferença crucial entre $f$ e uma possível métrica $g$ no mesmo espaço é o fato do último ser simétrico, enquanto o primeiro é anti-simétrico. Por outro lado, $g$ e $f$ são tranformações bilineares e ambos têm igual direito a definir regras de  levantar ou abaixar índices, assim como suas próprias regras de ortogonalidade. Para evitar confusão, a regra de ser a métrica a única responsável por essa alteração nos índices será mantida\footnote{$f^{\alpha \beta}$ será identificado como elementos da inversa de $(f_{\alpha \beta})$, mas a contração $\xi^\alpha f_{\alpha \beta}$ não  será expressa por $\xi_\beta$.}. Há dois motivos por estarmos considerando uma variedade de dimensão par: i) desejamos  associar $N$ ao espaço de fase, cuja dimensão é o dobro da do espaço de configuração; ii) embora possa-se definir métricas em variedades de dimensão arbitrária, 2-formas não-degeneradas só ocorrem em variedades de dimensão par. A demonstração é imediata, pois, para qualquer matriz anti-simétrica $f$, 
\be
	\det (f_{\alpha \beta}) = \det [(f_{\alpha \beta})^T] = \det (f_{\beta \alpha}) = (-1)^{\mbox{\tiny dim} N} \det (f_{\alpha \beta}),
\ee
Logo $\det f = 0$ se a dimensão de $N$ for ímpar.

Sendo a 2-forma não-degenerada $f$ constante para todos os pontos da variedade $N$, por meio de transformações de coordenadas, a matriz $(f_{\alpha \beta})$ pode sempre ser escrita na chamada forma canônica,
\be
	\label{fcano}
	(f_{\alpha \beta}) = \pmatrix{ 0_{n \times n} & -I_{n \times n} \cr
	I_{n \times n} & 0_{n \times n}},
\ee
com $\mbox{dim} N = 2n$.

\vspace{.4in}
Consideremos que $N$ é uma \asp fatia" $2n$-dimensional da variedade $2n+1$-dimensional $N_e$ com $t$ constante, em que $t$ é a variável adicional necessária para descrever $N_e$. Seja $f_e$ uma 2-forma de posto máximo em $N_e$, ou seja, de posto $2n$. Conseqüentemente, a matriz de $f_e$, em cada ponto $\xi \in N_e$, possui um único autovetor $\nu(\xi)$ de autovalor nulo linearmente independente (autovetor esse chamado de modo-zero). Ou seja, $f_e$ determina uma direção privilegiada em $N_e$ em cada um de seus pontos, dada pelo campo vetorial $\nu(\xi)$. Essa direção privilegiada iremos associar à evolução temporal do sistema no espaço de fase. Para determinar como essa evolução se dará,  precisa-se definir certa notável 1-forma em $N_e$,
\be
	\label{priL}
	L = p_i \; dq^i - H(q,p)dt,
\ee
em que $i=1,2,...,n$, $\{q,p,t\}$ são coordenadas locais de $N_e$ e $H(q,p)$ é dada função (0-forma) definida em $N_e$ (constante para todo $t$). A diferenciação externa de $L$ leva à seguinte 2-forma:
\be	
	dL = dp_i \wedge dq^i - \frac {\prt H}{\prt q^i} dq^i \wedge dt - \frac {\prt H}{\prt p_i} dp_i \wedge dt.
\ee

Em particular, podemos identificar a 2-forma acima com $f_e$ ($dL$ tem posto $2n$, como será visto). Usando o ordenamento $\{q^i \}, \{p_i \}, t$ para as coordenadas de $N_e$ e $dL = f_e = \frac 12 f_{e_{\alpha \beta}} d\xi^\alpha \wedge d \xi^\beta$, vem
\be
	(f_{e_{\alpha \beta}}) = \pmatrix {0 & -I & -\frac  {\prt H}{\prt q^j} \cr
	I & 0 & -\frac  {\prt H}{\prt p_j} \cr
	\frac  {\prt H}{\prt q^i} & \frac  {\prt H}{\prt p_i} & 0},
\ee
em que $I$ é a identidade $n \times n$, o índice $i$ simboliza vetores linha, enquanto $j$ vetores coluna. Comparando a matriz acima com a Eq. (\ref{fcano}) é imediato ver que $dL$ é uma extensão natural de $f$ para $N_e$. A matriz $f_e$ acima possui o seguinte modo-zero\footnote{Esse modo-zero poderia ser também representado na forma coluna, mas futuramente trataremos de modos-zero que são operadores, o que torna mais conveniente sua atuação pela esquerda da matriz $f_e$, ao invés  da direita.}:
\be
	\label{nugeos}
	(\nu^\alpha) = \pmatrix{ \frac  {\prt H}{\prt p_i} & - \frac  {\prt H}{\prt q^i} & 1}.
\ee
Qualquer outro vetor não-nulo cuja atuação em $f_e$ produza um resultado nulo é necessariamente proporcional a $\nu$.

Se a evolução temporal no espaço de fase de dado sistema é dada localmente em cada ponto de $N$ pelo campo vetorial $\nu \in TN$, temos $\dot \xi = \nu(\xi)$, ou seja, obtém-se as seguintes equações de familiar aspecto:
\be
	\dot q^i = \frac  {\prt H}{\prt p_i} \; \; \; \; \; \; \; \; \; \; \dot p_i = - \frac  {\prt H}{\prt q^i}.
\ee

Há várias observações a serem feitas. As equações acima devem ser vistas como as equações de movimento de um sistema físico, em que a função Hamiltoniana $H$ é escolhida de forma a reproduzir as simetrias e equações de movimento do sistema físico estudado, determinando conseqüentemente a Lagrangiana, isto é, 1-forma $L$ (veremos em breve o caso inverso e mais comum: a determinação de $H$ a partir de $L$). Resultados equivalentes apareceriam se tivéssemos usado algum princípio variacional, mas para o objetivo desta seção essa outra dedução das equações de movimento nos parece mais esclarecedora. Nas equações acima consideramos que $q$ e $p$ podem ser vistos como funções de $t$. Evoluções arbitrárias de um sistema em $N_e$ não possibilitam de forma geral escrever $q$ e $p$ como funções de $t$, mas a evolução dada por $\nu$ segue sempre a direção de $t$ crescente e sempre com a mesma velocidade (\ref{nugeos}), o que permite a associação de $t$ com o tempo. Escolhemos uma forma muito particular para a 2-forma $f_e$, falta saber até que ponto essa escolha não é excessivamente restritiva, isto é, se não há sistemas físicos que são descritos por outras 2-formas $f_e$; como veremos agora, essa escolha inicial realmente foi restritiva demais.

Sendo $f_e$ uma 2-forma exata de posto máximo em $N_e$ ($2n$), considere uma curva (unidimensional) fechada $\gamma_1$ em $N_e$. A evolução temporal de $\gamma_1$ gera, a menos de alguns casos singulares, um tubo bidimensional de superfície $\sigma$ no espaço de fase estendido $N_e$ (como sempre, estamos supondo que essa evolução temporal seja dada pelo modo-zero $\nu$ de $f_e$). Seja $\gamma_2$ uma outra curva fechada que envolva o tubo gerado por $\gamma_1$. Em particular, $\gamma_1$ e $\gamma_2$ podem ser curvas de $N$ em tempos $t$ distintos, como mostram as figuras abaixo:

\vspace{.1in}

\begin{figure*}[htbp]
	\centering
		\includegraphics[width=0.30\textwidth]{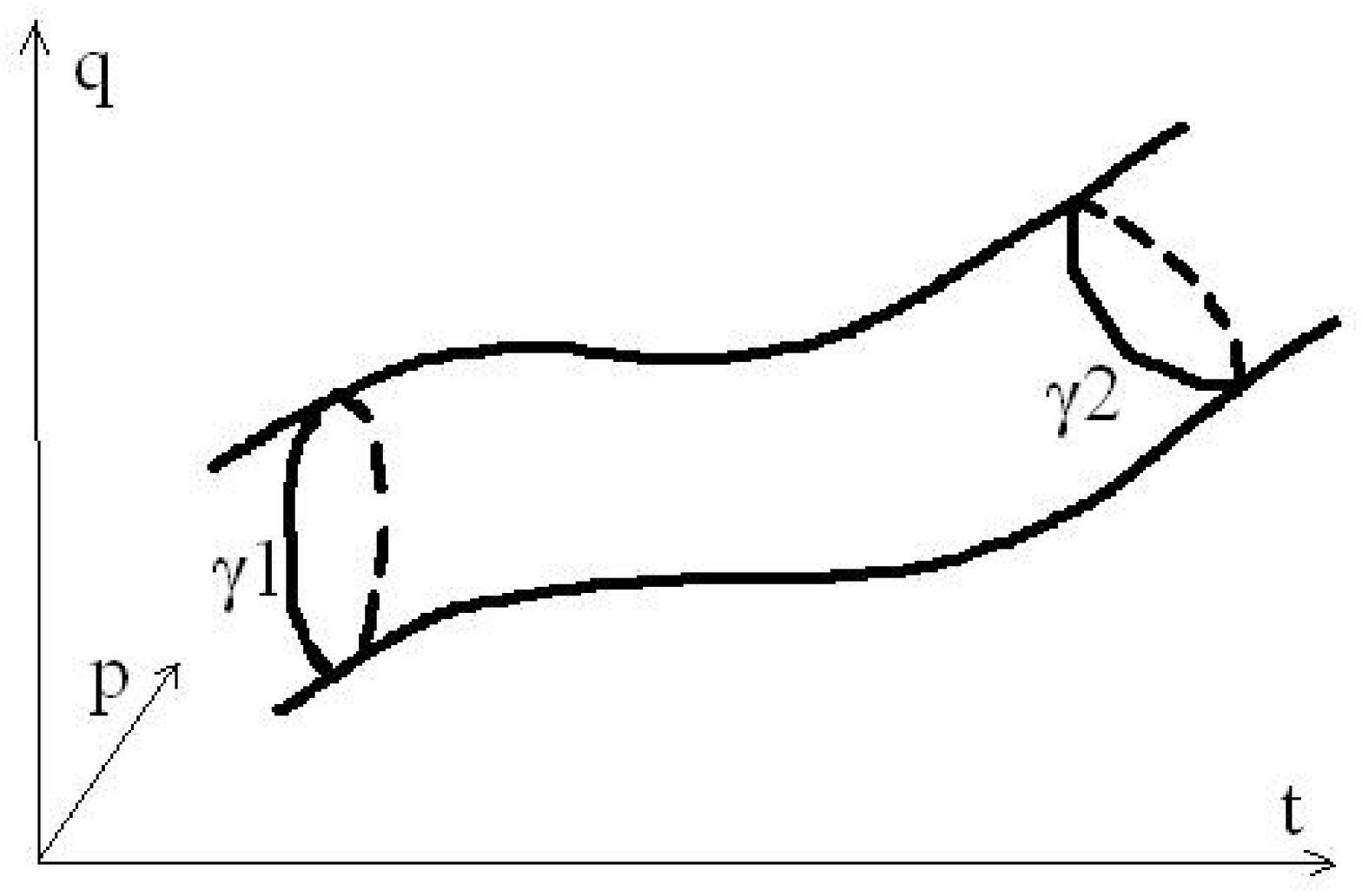}
		\hspace{1.1in}		\includegraphics[width=0.30\textwidth]{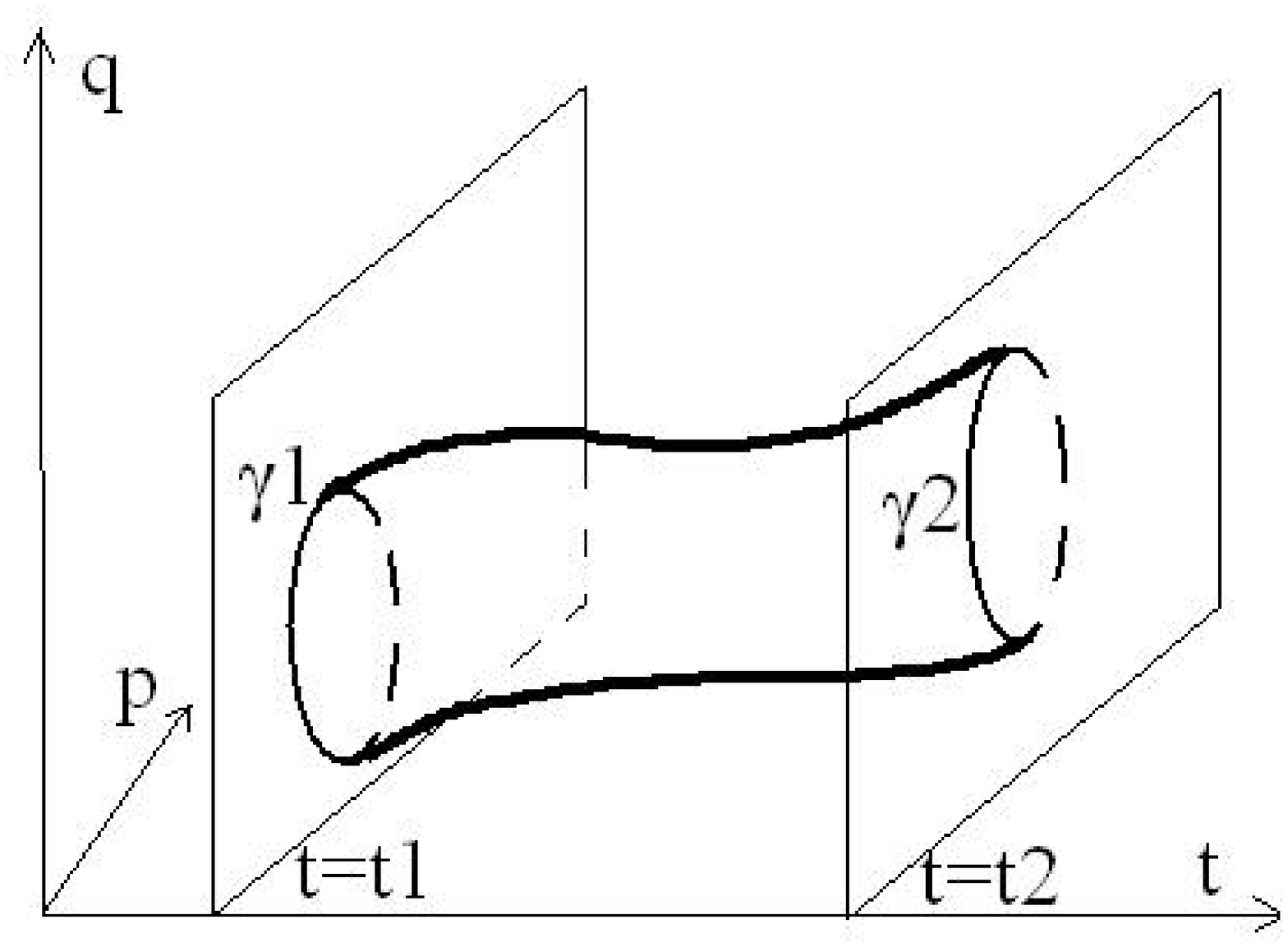}
\end{figure*}
\vspace{-.2in}
\begin{center}
	Fig.1 \hspace{2in} Fig.2
\end{center}	

\vspace{.2in}

Para ambos os casos, sendo $f_e$ exata, existe uma 1-forma $L$, não necessariamente dada pela Eq.(\ref{priL}), tal que $f_e = dL$, e portanto
\be
	\label{es39}
	\oint_{\gamma_1} L - \oint_{\gamma_2} L = \oint_{\prt \sigma} L = \int_\sigma f_e =0.
\ee
A última igualdade se deve a $f_e$ ser nulo para quaisquer vetores da superfície bidimensional $\sigma$, isto é, $f_e(\zeta, \eta) = 0 \; \; \forall \zeta, \eta \in T\sigma$, pois $\nu \in T\sigma$. Para ser mais específico, $\forall \eta \in T\sigma$ temos $f(\nu, \eta) = 0$ (\ref{degdef}); e devido a $\sigma$ ser bidimensional (assim como $T_x\sigma$), temos $f_e(\zeta, \eta) = 0 \; \; \forall \zeta, \eta \in T\sigma$. Da Eq. (\ref{es39}), para curvas $\gamma_1$ e $\gamma_2$ tais que $dt=0$ (Fig. 2), segue imediatamente a conservação da integral da \asp parte cinética"  da 1-forma $L$,
\be
	\label{presintL}
	\oint_{\gamma_1} a_\alpha (\xi) \; d\xi^\alpha = \oint_{\gamma_2} a_\alpha(\xi) \; d\xi^\alpha,
\ee
em que $\alpha = 1,2,...,2n$ e usamos a seguinte expressão geral para $L$ (desconsiderando apenas qualquer dependência explícita em $t$):
\be
	\label{Lger}
	L(\xi) = a_\alpha(\xi) \; d\xi^\alpha - H(\xi) \; dt.
\ee

Seja $\sigma_1 \in N_e$ a superfície cujo contorno é $\gamma_1$ em $t=t_1$. Em $t=t_2$, a superfície que $\gamma_2$ envolve é denotada por $\sigma_2$. Como
\be
	\oint_{\gamma_{1,2}} a_\alpha (\xi) \; d\xi^\alpha = \int_{\sigma_{1,2}} d a_\alpha (\xi) \wedge d\xi^\alpha,
\ee
usando a Eq. (\ref{presintL}) vem
\be
	\label{consarea}
	\int_{\sigma_1} d a_\alpha(\xi) \wedge d \xi^\alpha = \int_{\sigma_2} d a_\alpha(\xi) \wedge d \xi^\alpha.
\ee

A última equação tem como caso particular (caso canônico)
\be
	\label{consareac}
	\int_{\sigma_1} dp_i \wedge dq^i = \int_{\sigma_2} dp_i \wedge dq^i,
\ee
com $i=1,2,...,n$. Dado que o produto externo de $n$ dos elementos de área acima é proporcional ao elemento de volume $2n$-dimensional 
\be
	dp_{i_1}  \wedge ... \wedge dp_{i_n} \wedge dq^{i_1} \wedge ...\wedge dq^{i_n},
\ee
das Eqs. (\ref{consarea}, \ref{consareac}) deduz-se o teorema de Liouville, ou seja, a evolução de estados no espaço de fase é tal que o volume é sempre preservado. Em (\ref{consarea}), $n$ produtos externos do termo $da_\alpha \wedge d\xi^\alpha = \prt_\beta a_\alpha d\xi^\beta \wedge d\xi^\alpha$ atuam como elemento de volume, desde que essa 2-forma seja não-degenerada (mais detalhes serão apresentados em breve).

Nota-se que a condição de $f_e$ ser forma exata foi importante para a dedução do teorema de Liouville, contudo essa condição pode ser um pouco abrandada sem invalidar o teorema de Liouville, pois para o obter é necessário apenas que  $f_e$ seja fechada \cite{arnold}. Essa maior generalidade não será de importância para nossos objetivos; como veremos, mesmo para sistemas vinculados, por fim a variedade estudada é topologicamente trivial. Enfim, continuaremos a considerar que $f_e$ é 2-forma exata, sem perda de generalidade para nossos propósitos. Como veremos agora, a generalização de $L$ dada pela Eq. (\ref{Lger}) será especialmente útil, o caso dado por (\ref{priL}) não é suficientemente geral para abarcar grande parte dos problemas de interesse físico.

A fim de que a enésima potência de $da_\alpha \wedge d\xi^\alpha$ seja proporcional ao elemento de volume do espaço de fase $2n$-dimensional, o qual desejamos associar a $N$, a 2-forma $da(\xi)$ não pode ser degenerada. Como essa 2-forma é defina em $N$ e é igual a $f_e$ em dado $t$, pois 
\be
	f_e = dL = da_\alpha \wedge d\xi^\alpha - dH \wedge dt,
\ee
identificamos 
\be
	f = da = \frac 12 \( \frac {\prt a_\alpha}{\prt \xi^\beta} - \frac {\prt a_\beta}{\prt \xi^\alpha} \) d \xi^\alpha \wedge d \xi^\beta.
\ee
Assim, $f \in N$ é 2-forma não-degenerada e exata, enquanto $f_e$ é 2-forma exata de posto máximo na variedade $2n+1$-dimensional $N_e$.

Sendo $f$ não-degenerado, $f_e$ possui apenas um modo-zero e a dinâmica no espaço de fase encontra-se univocamente determinada. A Lagrangiana do eletromagnetismo é um bom exemplo de que nem sempre o termo cinético de $L$ tem a forma $p_i \wedge dq^i$ [mesmo quando expressa em primeira ordem nas velocidades (\ref{Leletro1ordem})]. Disso conclui-se que a generalização de $L$, como dada por (\ref{Lger}), é necessária. Formulamos agora as seguintes definições:

\vspace{.2in}
\noindent
\textbf{Definição \ref{capfs}.1.} Uma estrutura simplética em uma variedade $N$ é uma 2-forma $f_\xi: T_\xi N \times T_\xi N \rightarrow \real$ fechada e não-degenerada, ou seja, $df=0$ e
\be
	\label{degdef2}
	\forall \zeta \not= 0, \; \exists \eta \; \;  | \; \;  f( \zeta, \eta) \not= 0 \; \; \; \; (\zeta, \eta \in T_xN).
\ee
Uma estrutura simplética pode também ser chamada de 2-forma simplética ou simplesmente forma simplética. Sendo $f$ forma simplética e exata, a 1-forma $a_\xi:T_\xi N \rightarrow \real$ que satisfaz $f=da$ é ocasionalmente chamada de 1-forma canônica.

\vspace{.2in}
\noindent
\textbf{Definição \ref{capfs}.2.} Por variedade simplética chama-se o par $(N,f)$, em que $N$ é uma variedade e $f$ é uma estrutura simplética.

\vspace{.2in}
\noindent
\textbf{Definição \ref{capfs}.3.} O par $(N,h)$, em que $N$ é variedade e $h$ é 2-forma fechada degenerada, é uma variedade pré-simplética. $h$ é forma pré-simplética.

\vspace{.2in}

Conforme será visto nas próximas seções, da parte cinética de diversas Lagrangianas, quando expressas em primeira ordem nas velocidades (\ref{Lger}), extrai-se de imediato uma 2-forma pré-simplética $h$, com $h=da$ e degenerado. Em princípio isso levaria a ambigüidades ou inconsistências dinâmicas, pois a extensão de $h$ para o espaço de fase estendido $N_e$ pode gerar uma 2-forma cujo posto não é máximo em $N_e$, ou seja, teríamos mais de uma modo-zero responsável pela evolução temporal do sistema no espaço de fase. Como será visto, isto é um sinal de que há vincúlos e/ou simetria de calibre na Lagrangiana tratada. Em geral, não se assume que a Lagrangiana original seja problemática, mas sim que essa dificuldade não pode ser  superada em uma variedade trivial como $\real^{2n}$, ou seja, supõe-se que interdependências entre as coordenadas que aparecem na Lagrangiana devem ser consideradas. Passa-se portanto para outra variedade trivial de dimensão maior $\real^{2n+m}$ na qual a variedade não-trivial do espaço de fase físico estaria imersa (superfície de vínculos). Assim procedendo, encontra-se por fim uma verdadeira estrutura simplética $f$ associada à variedade $\real^{2n+m}$, cuja extensão $f_e$ para o espaço $\real^{2n + m +1}$ possui um único modo-zero\footnote{Nota-se que $m$ tem de ser par, isso será avaliado posteriormente}.

\vspace{.4in}
\subsection{Parênteses generalizados}
Seja $\real^{2n}$ um espaço de fase dado pelas coordenadas $q^1,...,q^n,p_1,...,p_n$ cujos parênteses de Poisson satisfazem 
$$
	\{q^i, q^j\} = \{p_i,p_j\} = 0,
$$
\be
	\{q^i, p_j\} = - \{p_j, q^i\} = \delta^i_j,
\ee 
$$
	\dot q^i = \{q^i,H\}, \;\;\; \dot p_i = \{p_i,H\},
$$
com $i,j = 1,2,...,n$ e $H = H(q,p)$ \'{e} a fun\c{c}\~{a}o Hamiltoniana.

	Essas propriedades podem ser apresentadas de forma mais compacta se a seguinte nota\c{c}\~{a}o for introduzida:
\be
	\xi^i := q^i, \;\;\;\; \xi^{n+i} := p_i,
	\label{vsc}
\ee
\be
	(\sigma^{\alpha \beta}) := \pmatrix{ 0 & \delta^i_j \cr
-\delta^i_j & 0}.
	\label{ms}
\ee
Com efeito,
\bq 
	&&\{\xi^\alpha, \xi^\beta\} = \sigma^{\alpha \beta}, \nonumber \\
	\label{sgfj}
	&&\dot \xi^\alpha = \sigma^{\alpha \beta} \frac{\partial H}{\partial \xi^\beta}, \hspace{0.5in}\alpha, \beta = 1,2,...,2n.
\eq

	Com esta nota\c{c}\~{a}o, os par\^{e}nteses de Poisson entre as fun\c{c}\~{o}es $A(\xi)$ e $B(\xi)$ assumem a forma
\be
	\{A,B\} = \frac{\partial A}{\partial \xi^\alpha} \sigma^{\alpha \beta} \frac{\partial B}{\partial \xi^\beta}.
	\label{ppes}
\ee

	Nota-se que a matriz $\sigma$ acima é igual à inversa da matriz simplética (\ref{fcano}) na forma canônica. Como a estrutura simplética, independentemente da base utilizada, é a responsável por determinar a evolução do sistema no espaço de fase, define-se a seguinte generalização dos parênteses de Poisson:

\be
	\{A,B\}^* := \frac{\partial A}{\partial \xi^\alpha}f^{\alpha \beta}\frac{\partial B}{\partial \xi^\beta},
	\label{ppes2}
\ee
em que $f^{\alpha \beta}$ são elementos da inversa da matriz $(f_{\alpha \beta})$. Estruturas simpléticas induzem  estruturas de Poisson \cite{arnold, varsymp}. Esses parênteses generalizados são os que contém uma direta relação com a dinâmica de um sistema físico no espaço de fase. 
 
\vspace{.4in}
Dado que um vínculo é escrito como uma função do espaço de fase que é igualada a zero ($\Omega_m=0$), a quantização dessas funções leva a operadores nulos e o comutador de operadores nulos é sempre nulo. Por outro lado, os parênteses de Poisson de um vínculo com outra grandeza não são necessariamente nulos. Essa questão motivou Dirac a definir novos parênteses clássicos, os quais, em particular, produziriam sempre um resultado nulo quando envolvessem qualquer vínculo. Dirac propôs que, em teorias com vínculos, os parênteses de Poisson devem ser substituídos por \cite{mdirac, teitelboim}
\be
	\{A,B\}_D := \{A,B\} - \{A,\Omega_m\} C^{mn} \{\Omega_n, B\},
	\label{pdirac}
\ee
chamados de parênteses de Dirac. Acima, os $\Omega_m$'s s\~{a}o os v\'{\i}nculos da teoria e $(C^{mn})$ \'{e} definida como a inversa da matriz $(\{\Omega_m, \Omega_n\})$, chamada de matriz de Dirac. Segundo Dirac, seriam esses parênteses acima os que deveriam ser substituídos por comutadores ao proceder com a quantização de sistemas vinculados, em vez dos de Poisson, isto é,
\be
	\{A,B\}_D \longrightarrow -i[ \hat A, \hat B].
\ee

O formalismo de Dirac, embora correto, possui alguns inconvenientes, além de ser trabalhoso. Esse formalismo insere uma distinção entre vínculos primários e secundários que é arbitrária; isto é, diferentes aplicações do método ditingüem um mesmo vínculo ora como primário ora como secundário \cite{mdirac, teitelboim}. Os vínculos segundo esse formalismo são também classificados entre primeira classe e segunda classe. Os de primeira classe possuem duas propriedades distintas: além de imporem certa relação de dependência entre as coordenadas do espaço de fase, esses ainda são geradores de simetria de calibre, segundo a conjectura de Dirac \cite{mdirac, teitelboim}. 

Décadas depois, Faddeev e Jackiw indicaram que a quantização de sistemas vinculados pode também  ser feita através de métodos matemáticos mais modernos \cite{mfj}. Em vez dos parênteses de Dirac, pode-se usar os parênteses induzidos pela inversa da estrutura simplética (\ref{ppes2}). Essa abordagem, que veio a ser conhecida como método simplético, foi posteriormente estendida por Barcelos Neto e Wotzasek \cite{mbw} e a correspondência com os parênteses de Dirac foi demonstrada por Montani \cite{montani}. De forma geral, em vista da equivalência, chamaremos tanto os parênteses do método simplético quanto os de Dirac de parênteses generalizados.

\vspace{.4in}
\subsection{Formalismo de Faddeev-Jackiw}
Para encontrar a matriz simpl\'{e}tica de um sistema vinculado e conseqüentemente seus parênteses generalizados, Faddeev e Jackiw \cite{mfj} indicaram um m\'{e}todo que se baseia na estreita rela\c{c}\~{a}o dessa com a Lagrangiana de primeira ordem nas velocidades. A fim de esclarecer esta rela\c{c}\~{a}o, continuemos a tratar de um sistema com $2n$ vari\'{a}veis can\^{o}nicas independentes. A  Lagrangiana pode ser escrita como
\begin{equation}
	L = p_i\dot q^i - H, \hspace{0.5in} i=1,2,...,n.
	\label{L123}
\end{equation}

	Ao introduzir as vari\'{a}veis simpl\'{e}ticas $\xi^i := q^i$, $\xi^{i+n} := p_i$, a 1-forma $Ldt$ \'{e} escrita como
\begin{eqnarray}
	Ldt & = & \frac{1}{2} \xi^\alpha f_{\alpha\beta} d\xi^\beta + \frac{1}{2}d(p_iq^i) - Hdt  \nonumber \\
	& = &  \frac{1}{2} \xi^\alpha f_{\alpha\beta} d\xi^\beta - Hdt, \hspace{0.5in}\alpha,\beta = 1,2,...,2n,
\end{eqnarray}
em que
\begin{equation}
	(f_{\alpha \beta}) = \left (\begin{array}{rcl} 
	0 & -\delta^j_i \\
	\delta^j_i & 0
\end{array}\right ).
\end{equation}
Essa é matriz simplética em sua forma canônica (\ref{fcano}).

	Neste exemplo, a chamada 1-forma can\^{o}nica \'{e} $a = \frac{1}{2} \xi^\alpha f_{\alpha\beta} d\xi^\beta$ (linear em $\xi^\alpha$) e a 2-forma simpl\'{e}tica $f = \frac{1}{2} f_{\alpha\beta} d\xi^\alpha \wedge d\xi^\beta = da$ \'{e} constante. Pela defini\c{c}\~{a}o de estrutura simpl\'{e}tica, sabemos n\~{a}o ser este o caso mais geral: a diferencial externa de $f$ deve se anular, mas $f$ n\~{a}o precisa ser constante. Consideremos um caso mais geral de Lagrangiana de primeira ordem\footnote{Nesta equa\c{c}\~{a}o $\xi^\alpha$ deve ser visto como  coordenada de um sistema qualquer, as equa\c{c}\~{o}es de (\ref{vsc}) n\~{a}o s\~{a}o necessariamente v\'{a}lidas.}:
\begin{equation}  
	L = a_\alpha(\xi) \dot \xi^\alpha - V(\xi), \hspace{0.5in} \alpha = 1,2,...,N.
	\label{Llinear}
\end{equation}

	As equa\c{c}\~{o}es de Euler-Lagrange para a Lagrangiana (\ref{Llinear}) s\~{a}o\footnote{As coordenadas simpl\'{e}ticas s\~{a}o, a princ\'{\i}pio, tratadas como independentes, o que justifica o emprego das equa\c{c}\~{o}es de Euler-Lagrange.}:
\begin{equation}
	\frac{\partial a_\beta}{\partial \xi^\alpha} \dot{\xi}^\beta - \frac{\partial V}{\partial \xi^\alpha} =  \frac {\partial a_\alpha}{\partial \xi^\beta} \dot{\xi}^\beta, 
\end{equation}
logo
\begin{eqnarray}
	\partial_\alpha V & = & (\partial_\alpha a_\beta - \partial_\beta a_\alpha) \dot{\xi}^\beta \nonumber \\
	 & = & h_{\alpha \beta} \dot{\xi}^\beta,
	\label{gradpot}
\end{eqnarray}
onde foi usado $\partial_\alpha := \frac{\partial}{\partial \xi^\alpha}$ e 
\begin{equation}
	h_{\alpha \beta} := \partial_\alpha a_\beta - \partial_\beta a_\alpha.
\end{equation}

	A defini\c{c}\~{a}o de $h_{\alpha \beta}$ coloca este tensor em proximidade com o tensor simpl\'{e}tico. Seja $h$ uma 2-forma diferencial dada pela derivada externa da 1-forma can\^{o}nica $a= a_\alpha(\xi) d\xi^\alpha$, logo
\bq
	h & = & da_\beta \wedge d\xi^\beta \nonumber \\
	& = & (\partial_\alpha a_\beta) d\xi^\alpha \wedge d\xi^\beta \nonumber \\
	& = & \frac{1}{2} (\partial_\alpha a_\beta - \partial_\beta a_\alpha)d\xi^\alpha \wedge d\xi^\beta  \\
	& = & \frac{1}{2} h_{\alpha \beta}d\xi^\alpha \wedge d\xi^\beta. \nonumber 
\eq
A 2-forma $h$ \'{e} fechada, pois $d(da)=0$, ou seja, para ser poss\'{\i}vel identificar $h$ com $f$ (a 2-forma simpl\'{e}tica), \'{e} necess\'{a}rio apenas que o determinante de $(h_{\alpha \beta})$ n\~{a}o se anule. Infelizmente isto nem sempre \'{e} verdade, por isto a matriz $(h_{\alpha \beta})$ ser\'{a} chamada de matriz pr\'{e}-simpl\'{e}tica.
	
	A equa\c{c}\~{a}o de movimento (\ref{gradpot}) e a equa\c{c}\~{a}o (\ref{sgfj}) sugerem uma rela\c{c}\~{a}o entre o potencial $V$ e a Hamiltoniana. A Hamiltoniana de $L$ \'{e}\footnote{Utilizando a nomenclatura de Dirac, essa Hamiltoniana \'{e} a can\^{o}nica. Veremos que n\~{a}o ser\'{a} necess\'{a}rio definir novas Hamiltonianas an\'{a}logas \`{a} total ou \`{a} estendida. Todo o procedimento simpl\'{e}tico n\~{a}o \'e afetado diretamente, ou melhor, de forma expl\'{\i}cita, pelo comportamento das fun\c{c}\~{o}es fora da superf\'{\i}cie de v\'{\i}nculo.}
\be
	H(\Pi, \xi) = \Pi_\alpha \dot \xi^\alpha - L(\xi, \dot \xi),
\ee
em que $\Pi_\alpha = \frac{\partial L}{\partial \dot \xi^\alpha}$. Como $\frac{\partial L}{\partial \dot \xi^\alpha}= a_\alpha(\xi)$, essa teoria possui, segundo o m\'{e}todo de Dirac, $N$ v\'{\i}nculos prim\'{a}rios: $\Omega_\alpha (\Pi, \xi) = \Pi_\alpha - a_\alpha(\xi)$. Nosso objetivo n\~{a}o \'{e} proceder com o formalismo de Dirac, n\~{a}o vamos considerar tais rela\c{c}\~{o}es de v\'{\i}nculos. Fa\c{c}amos simplesmente a substitui\c{c}\~{a}o dos momentos $\Pi_\alpha$ pelas fun\c{c}\~{o}es $a_\alpha(\xi)$ (o que parece ser, ao menos intuitivamente, mais sensato). Ao faz\^{e}-lo, $H(\Pi, \xi)$ passa a ser $H(\xi)$ e temos
\be
	H(\xi) = a_\alpha(\xi)\dot \xi^\alpha - L(\xi,\dot\xi) = V(\xi).
\ee

	Sabemos que o operador Hamiltoniano \'{e} respons\'{a}vel pela evolu\c{c}\~{a}o temporal dos operadores qu\^{a}nticos, portanto, em vista da rela\c{c}\~{a}o que deve existir entre o comutador e os par\^{e}nteses generalizados, a seguinte rela\c{c}\~{a}o \'{e} esperada:
\be
	\dot \xi^\alpha = \{ \xi^\alpha, H\}^*.
	\label{pxid}
\ee
Substituindo $H$ por $V$ comprova-se sua coer\^{e}ncia, pois, usando a defini\c{c}\~{a}o dos par\^{e}nteses generalizados seguida da equa\c{c}\~{a}o de movimento, temos
	
\be
	\dot \xi^\alpha = \{\xi^\alpha,V\}^* = f^{\alpha \beta}\partial_\beta V = \dot \xi^\alpha.
	\label{xid}
\ee

\vspace{10pt}
\noindent
{\bf Conclus\~{a}o:} Para obter os par\^{e}nteses generalizados, \'{e} suficiente escrever a Lagrangiana da teoria em primeira ordem nas velocidades; donde infere-se as componentes da 1-forma can\^{o}nica e determina-se o tensor pr\'{e}-simpl\'{e}tico. Se a matriz (pr\'{e}-)simpl\'{e}tica n\~{a}o for degenerada, esta ser\'{a} a matriz simpl\'{e}tica, e sua inversa, por meio de (\ref{ppes2}), determinar\'{a} os par\^{e}nteses generalizados.
\vspace{10pt}

No caso de ser encontrada uma matriz $h$ degenerada, a Ref. \cite{mfj} sugere o emprego de redefinições de coordenadas (via teorema de Darboux) ou do próprio método de Dirac caso essa abordagem torne-se excessivamente complicada. Uma forma sistemática de lidar com teorias pré-simpléticas, isto é, que possuem \asp vínculos verdadeiros" segundo o formalismo simplético, foi proposta na Ref. \cite{mbw} e será na próxima subseção exposta.

\vspace{.4in}
\subsection{Vínculos e o formalismo de Barcelos Neto-Wotzasek-Montani}

At\'{e} o presente momento, o termo ``v\'{\i}nculo" s\'{o} foi usado no sentido empregado pelo formalismo de Dirac. A se\c{c}\~{a}o anterior indica que tais v\'{\i}nculos n\~{a}o desempenham, a princ\'{\i}pio, um papel importante no m\'{e}todo simpl\'{e}tico. A equa\c{c}\~{a}o de movimento (\ref{gradpot}), em especial, est\'{a} de acordo com a \'{u}ltima afirmativa; afinal, observa-se que a determina\c{c}\~{a}o das velocidades como fun\c{c}\~{o}es das coordenadas depende exclusivamente da exist\^{e}ncia da inversa da matriz pr\'{e}-simpl\'{e}tica.

	Seja $P< N$ o posto da matriz $(h_{\alpha \beta})_{N \times N}$,  logo existem $N - P$ vetores n\~{a}o nulos e linearmente independentes, chamados modos-zero, que satisfazem\footnote{Os modos-zero ser\~{a}o sempre vistos como vetores linha. Tratando-se de campos, esta considera\c{c}\~{a}o \'{e} importante, pois os modos-zero podem ser operadores.}
\be
	\nu_m^\alpha h_{\alpha \beta} = 0,
	\label{modzer}
\ee
onde $m=1,2...N-P$. Portanto, de acordo com (\ref{gradpot}), temos
\be
	\nu^\alpha_m \partial_\alpha V = 0.
	\label{vs}
\ee

	Entretanto, o potencial pode n\~{a}o satisfazer estas rela\c{c}\~{o}es. A fim de eliminar a contradi\c{c}\~{a}o, iremos impor (\ref{vs}). Desta imposi\c{c}\~ao adv\'em rela\c{c}\~oes de depend\^encia entre as coordenadas simpl\'eticas, que constituem os v\'{\i}nculos do formalismo simpl\'etico.

	\'{E} claro que, utilizando a imposi\c{c}\~ao anterior, estamos resolvendo a inconsist\^{e}ncia, por\'{e}m sua origem ainda precisa ser apontada. Conforme j\'a exposto, (\ref{gradpot}) foi deduzida desconsiderando qualquer rela\c{c}\~{a}o de depend\^{e}ncia entre as coordenadas simpl\'{e}ticas; agora vemos que essas rela\c{c}\~{o}es s\~{a}o encontradas a posteriori. N\~{a}o h\'{a}, todavia, garantia de que todas as rela\c{c}\~{o}es inicialmente omitidas est\~{a}o contidas nessas $N-P$ equa\c{c}\~{o}es (e de fato n\~{a}o est\~{a}o necessariamente).

	Chama-se de v\'{\i}nculo verdadeiro o termo $\nu_m^\alpha \partial_\alpha V$ que n\~{a}o \'{e} nulo a priori. 

	Apesar dos v\'{\i}nculos de Dirac n\~{a}o terem import\^ancia direta para a formula\c{c}\~{a}o simpl\'{e}tica, h\'{a}, sim, uma rela\c{c}\~{a}o entre esses e a inversibilidade da matriz pr\'{e}-simpl\'{e}tica. Seja a Lagrangiana
\be
	L = a_\alpha(\xi) \dot \xi^\alpha - V(\xi), \hspace{0.5in} \alpha = 1,2,...,N.
	\label{lsimp}
\ee
Esta possui os seguintes v\'{\i}nculos de Dirac: $\Omega_\alpha = \Pi_\alpha - a_\alpha (\xi)$. Os par\^{e}nteses de Poisson destes, no espa\c{c}o das coordenadas e dos momentos simpl\'{e}ticos, s\~{a}o
\begin{eqnarray}
	\{\Omega_\alpha, \Omega_\beta\}^{\xi,\Pi} & = & -\frac{\partial a_\alpha}{\partial \xi^\gamma}\frac{\partial \Pi_\beta}{\partial \Pi_\gamma} + \frac{\partial \Pi_\alpha}{\partial \Pi_\gamma}\frac{\partial a_\beta}{\partial \xi^\gamma}, \nonumber \\
	& = & -\partial_\beta a_\alpha + \partial_\alpha a_\beta, \nonumber \\
	& = & h_{\alpha \beta}.
\end{eqnarray}

	A matriz $(\{\Omega_\alpha, \Omega_\beta\}^{\xi,\Pi})$ possui inversa se, e somente se, os v\'{\i}nculos forem de segunda classe\footnote{De acordo com a nomenclatura de Dirac, uma fun\c{c}\~{a}o do espa\c{c}o de fase \'{e} de primeira classe se os par\^{e}nteses de Poisson dessa com os v\'{\i}nculos da teoria forem nulos na superf\'{\i}cie de v\'{\i}nculos. Caso contr\'{a}rio, isto \'{e}, se a fun\c{c}\~{a}o do espa\c{c}o de fase n\~{a}o for de primeira classe, ela \'{e} dita ser de segunda classe.}\cite{mdirac,teitelboim}, logo o mesmo pode ser conclu\'{\i}do a respeito da matriz $(h_{\alpha \beta})$.

\vspace{.4in}
		Sobre a correspond\^{e}ncia entre v\'{\i}nculos verdadeiros e a interdepend\^{e}ncia das coordenadas simpl\'{e}ticas, acrescentamos aqui uma pequena nota. Sabemos que a exist\^{e}ncia de v\'{\i}nculos verdadeiros implica a exist\^{e}ncia de rela\c{c}\~{o}es de depend\^{e}ncia entre as coordenadas simpl\'{e}ticas ($\Phi_m(\xi) = 0$), averigüemos se a recíproca é verdadeira. Segundo a t\'{e}cnica dos multiplicadores de Lagrange, havendo $M$ v\'{\i}nculos entre as coordenadas, as equa\c{c}\~{o}es de movimento s\~{a}o $\Phi_m = 0$ e
\be
	h_{\alpha \beta} \dot \xi^\beta = \partial_\alpha V + \lambda^m \partial_\alpha \Phi_m, \hspace{0.5in} m=1,2,...,M,
\ee
onde os $\lambda^m$'s s\~{a}o multiplicadores de Lagrange e os $M$ v\'{\i}nculos s\~{a}o dados pelas fun\c{c}\~{o}es $\Phi_m(\xi)$, isto \'{e}, $\Phi_m (\xi) = 0$; conseq\"{u}entemente, $\dot \Phi_m = \partial_\alpha \Phi_m \dot \xi^\alpha = 0$.

	Suponhamos, por absurdo, que $(h_{\alpha \beta})$ possua inversa, isto \'{e}, seja a matriz simpl\'{e}tica $(f_{\alpha \beta})$, logo
\be
	\partial_\beta \Phi_m \dot \xi^\beta = \partial_\beta \Phi_m f^{\beta \alpha}(\partial_\alpha V + \lambda^m \partial_\alpha \Phi_m).
\ee

	Sabemos que o lado esquerdo dessa igualdade \'{e} nulo, portanto, devido ao termo que figura entre par\^{e}nteses ser qualquer, conclui-se que $(f^{\beta \alpha})$ possui $M$ modos-zero: $(\partial_\beta \Phi)_m$. Isto contraria a hip\'{o}tese de $(f_{\alpha \beta})$ ser n\~{a}o-degenerada, logo h\'{a} rela\c{c}\~{o}es de depend\^{e}ncia entre as coordenadas simpl\'{e}ticas se, e somente se, houver v\'{\i}nculos verdadeiros.

\vspace{.4in}

	Apresentemos agora o algoritmo proposto em \cite{mbw}. O objetivo deste m\'{e}todo \'{e} partir de uma Lagrangiana $L^\0$ com v\'{\i}nculos verdadeiros e, ap\'{o}s $n$ itera\c{c}\~{o}es, obter uma Lagrangiana $L^\n$ da qual infere-se a matriz simplética responsável por sua dinâmica.

	Segundo esse formalismo, adiciona-se a derivada temporal dos v\'{\i}nculos \`{a} Lagrangiana por meio de multiplicadores de Lagrange, incorporando-os \`{a} parte cin\'{e}tica:
\be
	L^\1 \equiv L^\0 - \lambda^{\0 m} \dot \Phi^\0_m, \hspace{0.5in} m=1,2,...,M.
\ee
Esse não é o procedimento usual de lidar com vínculos através de multiplicadores de Lagrange, pois normalmente adiciona-se os vínculos somente, e não suas derivadas. O método de Dirac impõe que a evolução temporal dos vínculos deve também se anular, mas o faz por outros caminhos.

	A menos de uma derivada temporal total,
\bq
	L^\1 & = & L^\0 + \dot \lambda^{\0 m}\Phi^\0_m \nonumber \\
	& = & a^\1_\alpha {\dot \xi}^{\1 \alpha} - V^\1, \hspace{0.5in} \alpha = 1,2,...,N+M,
\eq
onde\footnote{Com isto quero dizer que as $N$ primeiras componentes do vetor $\xi^\1$ s\~{a}o as do vetor $\xi^\0$ e suas $M$ \'{u}ltimas s\~{a}o as de $\lambda$. Rigorosamente, um outro \'{\i}ndice diferente de $\alpha$ deveria ter sido escolhido para se associar ao vetor $\xi^\1$, mas nota\c{c}\~{a}o utilizada \'{e} mais pr\'{a}tica.}
\bq
	(\xi^{\1 \alpha}) & = & \pmatrix{\xi^{\0 \alpha} & \lambda^m}, \nonumber \\
	(a_\alpha^\1)^T & = & \pmatrix{a^\0_\alpha & \Phi_m}, \\
	V^\1 (\xi^\0) & = & V^\0(\xi^\0)|_{\Phi_m = 0}, \nonumber
\eq
isto \'{e}, no novo potencial os v\'{\i}nculos verdadeiros s\~{a}o removidos (caso eles se encontrem expl\'{\i}citos no potencial original). Esta remo\c{c}\~{a}o \'{e} feita para facilitar os c\'{a}lculos; tendo em vista exclusivamente a Lagrangiana final, ela é indiferente. Para determinar o inverso da matriz simplética, contudo, esse procedimento pode ser importante em alguns casos \cite{rothe}.

	Com isto, partimos de $L^\0$ e obtivemos $L^\1$, cuja matriz pr\'{e}-simpl\'{e}tica \'{e}
\begin{equation}
(h^{(1)}_{\alpha \beta}) = \pmatrix{
h^{(0)}_{\alpha \beta}  & \partial_\alpha \Phi^{(0)}_m\cr
-\partial_\beta \Phi^{(0)}_n & 0 \cr},
\end{equation}
com $m,n = 1,2,...,M$, os \'{\i}ndices $\alpha$ e $\beta$ do lado direito da igualdade assumem $N$ valores e
\be
	h^\1_{\alpha \beta} \equiv \frac{\partial a^\1_\beta}{\partial \xi^{\1 \alpha}} - \frac{\partial a^\1_\alpha}{\partial \xi^{\1 \beta}}.
\ee
Os v\'{\i}nculos adicionados na parte cin\'{e}tica da Lagrangiana modificam a 1-forma can\^{o}nica e, conseq\"{u}entemente, a matriz pr\'{e}-simpl\'{e}tica\footnote{Inserir os v\'{\i}nculos na parte potencial tamb\'{e}m modifica a matriz pr\'{e}-simpl\'{e}tica, por\'{e}m apenas a acrescenta $M$ linhas e $M$ colunas nulas.}. 

	Se $(h_{\alpha \beta}^\1)$ for n\~{a}o degenerada, o problema ter\'{a} sido resolvido: a matriz simpl\'{e}tica da teoria foi encontrada. 

	Pode ocorrer de $L^\1$ ainda possuir v\'{\i}nculos a serem descobertos, nesse caso repete-se os mesmos procedimentos, obtendo $L^\2$. Pode também ocorrer de alguns dos modos-zero de $(h_{\alpha \beta})$ não gerarem novos vínculos, isto é, de $\nu_m^{\0 \alpha} \partial_\alpha V^\0$ ser nulo \emph{a priori}. Isto est\'{a} associado a simetrias de calibre da teoria. Para prosseguir com o m\'{e}todo de obten\c{c}\~{a}o da matriz simpl\'{e}tica, deve-se inserir condições que fixem o calibre, condições essas que são inseridas tais como vínculos, e assim parte-se para a nova iteração. A partir de agora vamos considerar que dentre os $M$ v\'{\i}nculos $\Phi^\0_m$ tamb\'{e}m se encontram os fixadores de calibre.

	Se ap\'{o}s a inser\c{c}\~{a}o de todos os $M$ v\'{\i}nculos $(h_{\alpha \beta}^\1)$ persistir degenerada, faz-se a segunda itera\c{c}\~{a}o: os modos-zero de $(h_{\alpha \beta}^\1)$ levam a novos v\'{\i}nculos ou a simetrias de calibre (que devem ser fixadas); os v\'{\i}nculos $\Phi^\1_m$ (incluindo os fixadores de calibre) s\~{a}o adicionados \`{a} parte cin\'{e}tica de $L^\1$ e eliminados de $V^\1$; em seguida obt\'{e}m-se $L^\2$, cuja matriz pr\'{e}-simpl\'{e}tica \'{e} $(h_{\alpha \beta}^\2)$... E assim por diante, at\'{e} encontrar a matriz simpl\'{e}tica. 

	Em cada itera\c{c}\~{a}o, uma quantidade n\~{a}o nula de v\'{\i}nculos \'{e} obtida. A Lagrangiana $L^\0$ possui um n\'{u}mero finito de coordenadas\footnote{Tratando-se de campos, $L^\0$ possui um n\'{u}mero finito de coordenadas em cada ponto do espa\c{c}o (um n\'{u}mero infinito para todo o espa\c{c}o), mas isso n\~{a}o altera esta argumenta\c{c}\~{a}o, pois para qualquer ponto do espa\c{c}o ela \'{e} v\'{a}lida.}, portanto possui no m\'{a}ximo tantos v\'{\i}nculos linearmente independentes quantas forem essas. Sendo assim, espera-se que os par\^{e}nteses generalizados ser\~{a}o encontrados ap\'{o}s um n\'{u}mero finito de itera\c{c}\~{o}es.

\vspace{.4in}
\subsection{Simetrias de calibre}
\label{sssimetriacalibre}


	Consideremos agora certa varia\c{c}\~{a}o infinitesimal das coordenadas,
\be
	\xi'^\alpha = \xi^\alpha + \delta_\vep \xi^\alpha.
\ee
O \'{\i}ndice $\vep$ serve para indicar que a varia\c{c}\~{a}o acima n\~{a}o \'{e} arbitr\'{a}ria, deve ser tal que produza o vetor $\xi'$ a partir de $\xi$. 

	A expans\~{a}o em s\'{e}rie de Taylor de $S\{\xi'\}$ subtraida de $S\{\xi\}$ \'{e}
\be
	S\{\xi'\} - S\{\xi\} = \int_{t_1}^{t_2} \left [ \frac{\partial L}{\partial \xi^\alpha}(\xi, \dot \xi) \delta_\vep \xi^\alpha + \frac{\partial L}{\partial \dot \xi^\alpha}(\xi, \dot \xi) \delta_\vep \dot \xi^\alpha \right ] dt + O((\delta_\vep \xi)^2).
\ee

	Se $\delta_\vep \xi$ for pequeno o suficiente para a contribui\c{c}\~{a}o de $O((\delta_\vep \xi)^2)$ ser insignificante, e se $\delta_\vep \xi(t_1) = \delta_\vep \xi(t_2) = 0$, temos
\bq
	S\{\xi'\} - S\{\xi\} \approx \delta_\vep S & \equiv & \int_{t_1}^{t_2} \left [ \frac{\partial L}{\partial \xi^\alpha}(\xi, \dot \xi) \delta_\vep \xi^\alpha + \frac{\partial L}{\partial \dot \xi^\alpha}(\xi, \dot \xi) \delta_\vep \dot \xi^\alpha \right ] dt \nonumber \\
	\label{s01}
	& = & \int_{t_1}^{t_2} \left [ \frac{\partial L}{\partial \xi^\alpha}(\xi, \dot \xi)  - \frac{d}{dt}\frac{\partial L}{\partial \dot \xi^\alpha}(\xi, \dot \xi) \right ] \delta_\vep \xi^\alpha dt.
\eq

	Independentemente dos valores dos instantes final e inicial, a integral acima deve se anular. Sendo $L$ a Lagrangiana de primeira ordem (\ref{Lger}) e impondo $\delta_\vep S = 0$ (identicamente), ou seja, exigindo que $\xi$ e $\xi'$ estejam relacionados por transformação de calibre, conclui-se que
\be
	0 = \left ( \frac{\partial a_\beta}{\partial \xi^\alpha} \dot \xi^\beta - \frac{\partial V}{\partial \xi^\alpha} - \frac{da_\alpha}{dt} \right ) \delta_\vep \xi^\alpha = \left ( h_{\alpha \beta} \dot \xi^\beta - \partial_\alpha V \right ) \delta_\vep \xi^\alpha.
	\label{hjuy034}
\ee
Para esta equa\c{c}\~{a}o ser satisfeita h\'{a} duas possibilidades: a express\~{a}o acima entre par\^{e}nteses se anula, ou  
\be
	\delta_\vep \xi^\alpha \frac{\partial V}{\partial \xi^\alpha} = 0
	\label{varv}
\ee
e
\be
	\delta_\vep \xi^\alpha h_{\alpha \beta} \dot \xi^\beta = 0.
	\label{varh}
\ee

A primeira possibilidade tem o aspecto de uma equação de movimento e é impossível de ser atingida identicamente. Explicando de outra forma: as velocidades e coordenadas simpl\'eticas s\~ao vari\'aveis independentes, portanto \'e imposs\'{\i}vel aplicar a derivada $\partial/\partial \xi^\alpha$ em uma fun\c{c}\~ao $V(\xi)$ e obter-se $h_{\alpha \beta}(\xi) \dot \xi^\beta$ como resposta. 

O segundo caso é uma caracteriza\c{c}\~{a}o de  transforma\c{c}\~{a}o de calibre para a Lagrangiana tratada.
	
	Em alguns sistemas, conforme antes mencionado, a contra\c{c}\~{a}o de um modo-zero de $(h_{\alpha \beta})$ com o grandiente do potencial n\~{a}o \'{e} um v\'{\i}nculo, ou seja, o modo-zero \'{e} ortogonal ao gradiente do potencial. Quando isto ocorre, as equa\c{c}\~{o}es (\ref{varv}) e (\ref{varh}) s\~{a}o v\'{a}lidas. Agora \'{e} claro que a anula\c{c}\~{a}o \emph{a priori} de $\nu^\alpha \partial_\alpha V$ est\'{a} associada a uma simetria infinitesimal de calibre, cuja varia\c{c}\~{a}o das coordenadas \'{e} dada pelo modo-zero (desde que os valores de suas componentes sejam pequenos). Devido a esta \'{u}ltima condi\c{c}\~{a}o e a um m\'{u}ltiplo de um modo-zero ser tamb\'{e}m um modo-zero, costuma-se efetuar a seguinte identifica\c{c}\~{a}o:
\be
	\delta_\varepsilon \xi^\alpha = \varepsilon \nu^\alpha,
	\label{nmvmraee}
\ee
em que $\varepsilon = \varepsilon(t)$ [ou $\varepsilon = \varepsilon (\vec{x},t)$ para campos] possui a finalidade de tornar as componentes do modo-zero suficientemente pequenas. Devido \`{a} equa\c{c}\~{a}o acima, diz-se que os modos-zero s\~{a}o os geradores das simetrias de calibre do m\'{e}todo simpl\'{e}tico.

\vspace{.4in}

\noindent
\textbf{Fixação de calibre}

	N\~{a}o existe matriz simpl\'{e}tica em teorias com simetrias calibre, logo, a fim de obt\^{e}-la, estas devem ser eliminadas tendo o cuidado de n\~{a}o alterar a f\'{\i}sica da teoria. Este processo \'{e} chamado de fixa\c{c}\~{a}o de calibre. A seguir o enunciaremos de forma mais precisa\footnote{Esta apresentação sobre fixação de calibre segue a Ref. \cite{teitelboim} com a substituição de termos próprios ao método de Dirac pelos termos do método simplético.}.

	O conjunto de fun\c{c}\~{o}es independentes
\be
	\Lambda_c(\xi) = 0, \hspace{0.5in} c=1,2,...,C,
	\label{fix}
\ee
fixa o calibre de certa teoria se satisfizer:

\noindent
$i$) {\it Acessibilidade do calibre}: para cada vetor simpl\'{e}tico (que satisfaz as equa\c{c}\~{o}es de movimento) deve existir uma sucess\~{a}o de transforma\c{c}\~{o}es infinitesimais de calibre capaz de mapear este vetor nas vari\'{a}veis que obedecem as $C$ equa\c{c}\~{o}es de (\ref{fix}). 

	A partir de um vetor simpl\'{e}tico $\xi'$, um outro pode ser obtido atrav\'{e}s de transforma\c{c}\~{o}es infinitesimais de calibre da seguinte forma:
\be
	\xi''^\alpha = \xi'^\alpha + \varepsilon^g \nu_g^\alpha,
	\label{tic}
\ee
sendo $\{\nu_g^\alpha\}$, com $g=1,2...G$, o conjunto de todos os geradores de transforma\c{c}\~{o}es de calibre (cada $\nu_g$ \'{e} tangente \`{a} superf\'{\i}cie $V(\xi) =$ constante).

	Se $\xi$ for um vetor que cumpre as $C$ equa\c{c}\~{o}es de (\ref{fix}) e $\xi'$ for um vetor solu\c{c}\~{a}o das equa\c{c}\~{o}es de movimento, a condi\c{c}\~{a}o $i$ imp\~{o}e a exist\^{e}ncia de uma combina\c{c}\~{a}o linear dos geradores tal que
\be
	\xi^\alpha = \xi'^\alpha + a^g \nu_g^\alpha,
\ee
com
\be
	a^g = \varepsilon^g_1 + \varepsilon^g_2 + ... + \varepsilon^g_T
\ee
e $T$ \'{e} o n\'{u}mero total de transforma\c{c}\~{o}es do tipo (\ref{tic}) que foram usadas (não necessariamente finito).

	O n\'{u}mero de componentes independentes do vetor de transforma\c{c}\~{a}o de calibre $(a^g \nu_g^\alpha)$ informa quantos $\xi'^\alpha$'s podem ser independentemente modificados sem alterar a f\'{\i}sica do problema. Logo, obtivemos um limite superior para o n\'{u}mero $C$:
\be
	C \le \mbox{n\'{u}mero de componentes independentes de }(a^g \nu^\alpha_g).
	\label{cle}
\ee

\noindent
$ii$) {\it Quebra da simetria}: dentre todos os vetores simpl\'{e}ticos equivalentes por transforma\c{c}\~{o}es de calibre, somente um deve satisfazer as rela\c{c}\~{o}es de (\ref{fix}). Se $\xi$ for este vetor simpl\'{e}tico, e $\xi'$ for obtido por meio de uma transforma\c{c}\~{a}o infinitesimal de calibre, as equa\c{c}\~{o}es
\be
	\Lambda_c(\xi) - \Lambda_c(\xi') \approx \delta_\vep \Lambda_c = \frac{\partial \Lambda_c}{\partial \xi^\alpha} \varepsilon^g \nu_g^\alpha = 0
	\label{varlam}
\ee
devem implicar
\be
	\varepsilon^g \nu_g^\alpha = 0,
\ee
isto \'{e}, $\xi = \xi'$.

	Mas esta implica\c{c}\~{a}o s\'{o} pode ser verdadeira se o n\'{u}mero de equa\c{c}\~{o}es em (\ref{varlam}) for maior ou igual ao n\'{u}mero de componentes independentes do vetor $(\varepsilon^g \nu_g^\alpha)$. Unindo este resultado com o obtido em (\ref{cle}), temos, enfim:
\be
	C = \mbox{n\'{u}mero de componentes independentes de }(a^g \nu^\alpha_g).
\ee

	No formalismo de Dirac, algo muito similar \'{e} encontrado, por\'{e}m envolvendo os v\'{\i}nculos de primeira classe ao inv\'{e}s das componentes do modo-zero \cite{teitelboim, montani}.

	Cada gerador de transforma\c{c}\~{a}o de calibre pode ser associado a uma \'{o}rbita de calibre, isto \'{e}, \`{a} regi\~{a}o da superf\'{\i}cie de v\'{\i}nculo cujos pontos correspondem a diferentes vetores simpl\'{e}ticos equivalentes por certa transforma\c{c}\~{a}o de calibre; logo, geometricamente, a variedade determinada pelos fixadores de calibre (\ref{fix}) est\'{a} contida na superf\'{\i}cie de v\'{\i}nculo e intercepta uma \'{u}nica vez cada \'{o}rbita independente de calibre.

\vspace{.4in}
\subsection{Exemplo 1: Eletromagnetismo}

	Nesta subseção apresentamos um primeiro exemplo de aplicação do método simplético. Seu objetivo é exclusivamente ilustrar como aplicar o método simplético em dada Lagrangiana. 

 A densidade de Lagrangiana da teoria eletromagn\'{e}tica de Maxwell, na aus\^{e}ncia de fontes, \'{e} 
\begin{equation}
	{\cal L} = -\frac{1}{4} F^{\mu \nu}F_{\mu \nu}, \hspace{0.5in} \mu,\nu = 0,1,2,3.
\end{equation}
O tensor eletromag\'{e}tico \'{e} $F_{\mu \nu} = \partial_\mu A_\nu - \partial_\nu A_\mu$, com $A_\mu = A_\mu(\vx ,t)$, e a  m\'{e}trica \'{e} $(g_{\mu \nu}) = \diag \pmatrix{+ & - & - & -}$.

	O primeiro passo \'{e} lineariz\'{a}-la nas velocidades. Os momentos s\~{a}o
\bq
	\pi_\mu = \frac{\partial {\cal L}}{\partial \dot A^\mu} & = & - \frac{1}{2} F_{\gamma \nu} \frac{\partial F^{\gamma \nu}}{\partial \dot A^\mu} \nonumber \\
	& = & - \frac{1}{2} F_{\gamma \nu} (\delta^\gamma_0 \delta^\nu_\mu - \delta^\gamma_\mu \delta^\nu_0) \\
	& = & F_{\mu 0} = \partial_\mu A_0 - \dot A_\mu. \nonumber
\eq

	Sendo $i,j=1,2,3$, a Lagrangiana de primeira ordem \'{e}, ent\~{a}o,
\bq
	{\cal L}^\0 & = & \dot A^i\pi_i - \pi_i(\partial^i A_0 - \pi^i) - \frac{1}{2} F^{i0}F_{i0} - \frac{1}{4}F^{ij}F_{ij} \nonumber \\
	\label{Leletro1ordem}
	&=& \dot A^i \pi_i - \pi_i \partial^i A_0 + \frac{1}{2} \pi_i \pi^i - \frac{1}{4}F^{ij}F_{ij}.
\eq
Logo, identifica-se 
\bq
	V^\0 & = & \pi_i \partial^i A_0 - \frac{1}{2} \pi_i \pi^i + \frac{1}{4}F^{ij}F_{ij}, \nonumber \\
	(\xi^{\0 \alpha}) & = & \pmatrix{A^0 & A^i & \pi^i},\\
	(a^\0_\alpha)^T & = & \pmatrix{ 0 & \pi_i & 0}, \nonumber
\eq 
em que o momento conjugado a $A^0$ foi eliminado do vetor simpl\'{e}tico por n\~{a}o aparecer na Lagrangiana. Aquele que desejar manter $\pi^0$ no vetor simpl\'{e}tico deve, durante a lineariza\c{c}\~{a}o, inserir o v\'{\i}nculo $\pi_0 = 0$ por meio de um multiplicador de Lagrange ou sua derivada. 
	
	A matriz pr\'{e}-simpl\'{e}tica \'{e}\footnote{Se o vetor $\xi^\0$ fosse $\pmatrix{A^0 & A^i & \pi_i}$, no lugar de $g_{ij}$ ter\'{\i}amos $\delta^i_j$. A estrutura usada para o vetor simplético tem o mérito de não misturar índices contravariantes com covariantes.}
\be
	(h_{\alpha \beta}^\0) = \pmatrix{ 0 & 0 & 0 \cr
0 & 0 & -g_{ij} \cr
0 & g_{ji} & 0} \delta^3(\vx - \vy).
\ee
Esta possui o modo-zero
\be
	\nu^\0 = \pmatrix{1 & 0 & 0},
\ee
que produz o v\'{\i}nculo
\be
	\int \frac{\delta V^\0(\vy)}{\delta A^0(\vx)} d^3y = -\partial^i \pi_i = \Phi^\0.
\ee
O v\'{\i}nculo obtido nada mais \'{e} que a lei de Gauss no v\'{a}cuo.

	Estando o v\'{\i}nculo inserido no setor cin\'{e}tico de ${\cal L}^\0$, ele pode ser eliminado do setor potencial, causando o desaparecimento de $A_0$:
\be
	\label{l1eletro}
	{\cal L}^\1 = \pi^i \dot A_i + \dot \eta \partial_i \pi^i - V^\1,
\ee
com
\be
	V^\1 = -\frac{1}{2} \pi^i\pi_i + \frac{1}{4} F^{ij}F_{ij}.
\ee
	
	De ${\cal L}^\1$ obt\'{e}m-se
\bq
	(\xi^{\1 \alpha}) & = & \pmatrix{A^i & \pi^i & \eta}, \nonumber \\
	(a_\alpha^\1)^T & = & \pmatrix{ \pi_i & 0 & \partial_i \pi^i}, \\
	(h^\1_{\alpha \beta}) & = & \pmatrix{ 0 & -g_{ij} & 0 \cr
g_{ji} & 0 & \sy{\partial}_i \cr
0 & -\sx{\partial}_j & 0} \delta^3(\vx - \vy), \nonumber
\eq
os s\'{\i}mbolos $\sx{\partial}_j$ e $\sy{\partial}_i$ indicam respectivamente: $\partial/\partial x^j$ e $\partial/\partial y^i$.

O modo-zero de $(h^\1_{\alpha \beta})$ \'{e} 
\be
	\nu^\1 = \pmatrix{-\prt^i & 0 & 1}.
\ee

	Averig\"{u}emos se um novo v\'{\i}nculo \'{e} obtido:
\bq
	&&\int \left(-\sx{\prt}^i_x \frac{\delta V^\1(\vy)}{\delta A^i(\vx)}  +   \frac{\delta V^1(\vy)}{\delta \eta (\vx)} \right ) d^3y  = \nonumber \\
	& =&   -\int \sx{\prt}^i\frac{1}{2} F_{kl}(\vy)(\sy{\partial^k} \delta_i^l \delta^3(\vx-\vy) - \sy{\partial^l} \delta^k_i \delta^3(\vx - \vy)) d^3y  \nonumber \\
	& = & -\int \sx{\prt}^i F_{ik}(\vy)\sy{\partial^k} \delta^3(\vx - \vy) d^3y \\
	& = & 0. \nonumber
\eq

	Desse resultado, conclui-se a exist\^{e}ncia de uma transforma\c{c}\~{a}o de calibre dada por:
\bq
	\delta_\vep A^i & = & \prt^i \vep, \nonumber \\
	\delta_\vep \eta  & = & \vep.
\eq
É imediato checar que essa simetria realmente está presente em $\cl^\1$ (\ref{l1eletro}) para $\vep$ arbitrário. Dado que $\vep$ é arbitrário, em particular pode-se escolher $\prt_i \prt^i \vep = - \prt_i A^i$, conseqüentemente há campos $A'^i$ que diferem de $A^i$ por uma transformação de calibre tais que
\be
	\prt_i A'^i  = \prt_i A^i + \prt_i \prt^i \vep = 0.
\ee
Escolheremos o calibre $\prt_i A^i=0$. Nota-se que essa escolha está de acordo com os critérios apresentados na subseção anterior, pois esse calibre é acessível (como mostrado pela última equação) e a simetria é quebrada (não há outra escolha de $\vep$ condizente com $\prt_i A^i=0$; assim, ao substituirmos $A$ por $A'$, a nova Lagrangiana não possui simetria de calibre).

	A itera\c{c}\~{a}o seguinte da Lagrangiana \'{e}
\be
	{\cal L}^\2 = \pi_i\dot A^i + \dot \eta \partial_i \pi^i + \dot \gamma \partial_i A^i - V^\2,
\ee 
com
\be
	V^\2 = V^\1 = -\frac{1}{2} \pi^i \pi_i + \frac{1}{4}F^{ij}F_{ij}.
\ee
Desta, obt\'{e}m-se
\bq
	(\xi^{\2 \alpha}) & = & \pmatrix{A^i & \pi^i & \eta & \gamma}, \nonumber \\
	(a_\alpha^\2)^T & = & \pmatrix{\pi_i & 0 & \partial_i \pi^i & \partial_i A^i}, \\
	(f_{\alpha \beta}) & = & \pmatrix{ 0 & -g_{ij} & 0 & \sy{\partial}_i \cr
g_{ji} & 0 & \sy{\partial}_i & 0 \cr
0 & -\sx{\partial}_j & 0 & 0 \cr
-\sx{\partial}_j & 0 & 0 & 0} \delta^3(\vx -\vy). \nonumber
\eq

	A inversa da matriz simpl\'{e}tica \'{e}
\be
(f^{\alpha \beta}) = \pmatrix{
0  & g^{ij} - \frac{\partial^i \partial^j}{\partial^k \partial_k} & 0 & \frac{\partial^i}{\partial^k \partial_k} \cr
- g^{ij} + \frac{\partial^i \partial^j}{\partial^k \partial_k} & 0 & -\frac{\partial^i}{\partial^k \partial_k} & 0 \cr
0 & \frac{\partial^j}{\partial^k \partial_k} & 0 & \frac{1}{\partial^k \partial_k} \cr
\frac{\partial^j}{\partial^k \partial_k} & 0 & -\frac{1}{\partial^k \partial_k} & 0 \cr} \delta^3(\vec{x} - \vec{y}).
\ee
Nesta, todas as derivadas atuam em $\vx$.

	Os par\^{e}nteses generalizados entre $A^i$ e $\pi^j$ s\~{a}o
\be
	\{A^i(\vx,t), \pi^j(\vy,t)\}^* = \left ( g^{ij} - \frac{\partial^i \partial^j}{\partial^k \partial_k} \right ) \delta^3(\vec{x} - \vec{y}).
\ee

	Esses são exatamente os parênteses obtidos pelo método de Dirac. O objetivo desta subseção foi atingido. Referências sobre quantização do eletromagnetismo usando os parênteses de Dirac podem ser vistos em \cite{eletrodirac}. Grande parte dos livros textos de teoria quântica de campos utilizam o formalismo de Gupta-Bleuler para a quantização canônica do eletromagnetismo; esse tem o mérito de ser mais simples, porém é menos geral que o de Dirac ou o simplético.

	Originalmente a teoria possu\'{\i}a oito campos (4 campos $A_\mu$ e 4 momentos conjugados), dois foram eliminadas ($A_0$ e $\pi_0$), um v\'{\i}nculo foi encontrado e foi necess\'{a}rio introduzir um fixador de calibre, portanto esta teoria possui $(8-2-1-1)/2 = 2$ graus de liberdade, como era de se esperar \cite{teitelboim}.

\vspace{.4in}
\subsection{Exemplo 2: Modelo de Proca}

A densidade de Lagrangiana deste sistema \'{e}
\be
	{\cal L} = -\frac{1}{4} F^{\mu \nu} F_{\mu \nu} + \frac{1}{2} m^2 A^\mu A_\mu,
	\label{procacl}
\ee
$m$ \'{e} a massa do campo $A^\mu = A^\mu(x)$, $x$ \'{e} vetor do espa\c{c}o-tempo, a m\'{e}trica deste \'{e} $(g_{\mu \nu}) = \diag \pmatrix{+ & - & - & -}$ e o tensor eletromagn\'{e}tico \'{e} $F_{\mu \nu} = \partial_\mu A_\nu - \partial_\nu A_\mu$. 

	Este modelo designa a massa $m$ ao f\'{o}ton e possui, como ser\'{a} visto, um grau de liberdade a mais que a teoria de Maxwell. Devido \`{a} presen\c{c}a de uma massa n\~{a}o nula, a transforma\c{c}\~{a}o de calibre $A_\mu \rightarrow A_\mu + \partial_\mu \Lambda$ da teoria de Maxwell inexiste nesta.

	\'{E} necess\'{a}rio escrever a Lagrangiana sob a forma
\be
	{\cal L}^\0 = a^{\0}_\alpha \dot\xi^{\0 \alpha} - V^{\0}.
\ee
Para isto, o momento conjugado ao quadrivetor potencial ser\'{a} calculado.
\be
	\pi_\mu = \frac{\partial {\cal L}}{\partial \dot A^\mu} = F_{\mu 0},
\ee
logo
\be
	F^{\mu \nu}F_{\mu \nu} = 2 \pi^i\pi_i + F^{ij}F_{ij},
\ee
e
\be
	\dot A^i \pi_i = (-\pi^i + \partial^i A_0) \pi_i.
\ee
Portanto
\be
	{\cal L} = {\cal L}^\0 = \dot A^i \pi_i - V^\0,
\ee
onde
\be
	V^\0 = -A_0\partial_i \pi^i - \frac{1}{2} \pi_i \pi^i + \frac{1}{4}F^{ij}F_{ij} - \frac{1}{2}m^2 A^\mu A_\mu.
\ee

	Os vetores simpl\'{e}tico e potencial s\~{a}o
\bq
	(\xi^{\0 \alpha}) & =& \pmatrix{A^i & \pi^i & A_0}, \nonumber \\
	(a^\0_\alpha)^T & =& \pmatrix{\pi_i & 0 & 0}.
\eq 

	A matriz pr\'{e}-simpl\'{e}tica \'{e}
\be
(h^{(0)}_{\alpha \beta}) = \left(
\begin{array}{ccc}
0           & -g_{ij} & 0 \\
g_{ji}&         0     & 0 \\
0           &         0     & 0
\end{array}
\right)\,\delta^3({\vec x} - {\vec y}).
\ee
Esta possui um modo-zero, que leva a um v\'{\i}nculo que expressa a lei de Gauss:
\begin{equation}
\Omega = \partial_i\pi^i + m^2A_0.
\end{equation}

	Assim, obt\'{e}m-se a Lagrangena de primeira itera\c{c}\~{a}o
\begin{equation}
{\cal L}^{(1)} = \pi_{i}\dot {A^{i}} + \Omega\dot{\beta} - V^{(1)},
\end{equation}
com o seguinte potencial simpl\'{e}tico:
\be
V^{(1)} = - \frac{1}{2}{\pi_{i}}{\pi^{i}} + \frac 14 F_{ij}F^{ij} + \frac{1}{2}\,m^2\,\({A_{0}}{A_{0}} - {A_{i}}{A^{i}}\).
\ee

	Os vetores $\xi^\1$ e $a^\1$ desta Lagrangiana s\~{a}o
\bq
	(\xi^{(1)\alpha}) & = & \pmatrix{A^i & \pi^i & A_0 & \beta}, \nonumber \\
	(a_\alpha^\1)^T & = & \pmatrix{\pi_i & 0 & 0 & \Omega},
\eq
dos quais obt\'{e}m-se a matriz simpl\'{e}tica
\begin{equation}
(f_{\alpha \beta})=\left(
\begin{array}{cccc}
0           & -g_{ij}  &  0   &   0 \\
g_{ji} &         0     &   0    &    \partial^y_i \\
0         &         0     &   0    &     m^2   \\
0       &        -\partial^x_j    &  -m^2     &     0  
\end{array}
\right)\,\delta^3({\vec x} - {\vec y}).
\end{equation}

	A matriz simpl\'{e}tica foi obtida sem a necessidade de qualquer fixa\c{c}\~{a}o de calibre, de acordo com o coment\'{a}rio feito no in\'{\i}cio desta se\c{c}\~{a}o.

	Os par\^{e}nteses generalizados s\~{a}o extra\'{\i}dos da inversa da matriz simpl\'{e}tica: 
\bq
\lbrace A_i(\vec x),A_j(\vec y)\rbrace^* &=& 0,\nonumber\\
\label{ppc}
\lbrace A_i(\vec x),\pi_j(\vec y)\rbrace^* &=& g_{ij}\delta^3(\vec x - \vec y), \nonumber\\
\lbrace \pi_i(\vec x),\pi_j(\vec y)\rbrace^* &=& 0.\\
\{A_0 (\vx), A_i(\vy)\}^* &=& \frac 1{m^2} \sx{\partial_i} \delta (\vx-\vy), \nonumber\\
\{A_0 (\vx),\pi_i(\vy) \}^* &=& 0. \nonumber
\eq

	Observe que somente um v\'{\i}nculo foi encontrado e $\pi_0$ foi excluído do vetor simpl\'{e}tico. O modelo de Proca possui tr\^{e}s graus de liberdade $[(8-1-1)/2]$.
	
	Parte dos cálculos acima serão úteis para a próxima seção na qual trataremos da \asp imersão em calibre" $\;$  do modelo de Proca usando o formalismo simplético.

\vspace{.4in}
\section{Formalismo simplético de calibre}
\label{secformsympcal}

Na seção anterior comentamos sobre simetria de calibre e sua fixação no contexto do formalismo simplético. Ocorre que o procedimento oposto ao da fixação, ou seja, o da implementação de novas simetria de calibre, também é útil para diversos problemas em física. Dentre as possíveis aplicações, pode-se contornar problemas de quantização (como ordenamento dos operadores ou anomalias) \cite{wz, fs, bfft, embquan}, encontrar simetrias escondidas \cite{symembskyrme, hidsym}, possibilitar o emprego do mapa de Seiberg-Witten (caso o espaço seja NC) \cite{abf, hr}, determinar teorias duais \cite{symembdual, banerjdual} \emph{etc}.

Um método recente nesse contexto é o chamado método simplético de calibre, esse usa a \asp filosofia" $\;$ do método simplético anteriormente apresentado para inserir campos auxiliares (campos de Wess-Zumino) de forma consistente com a dinâmica da teoria original e com os desejados geradores da nova simetria de calibre. O método simplético de calibre foi empregado pela primeira vez no modelo de Skyrme  \cite{symembskyrme}, nessa aplicação certa simetria escondida foi avaliada e o espectro da teoria em sua versão de calibre foi diretamente demonstrada como sendo equivalente à da original. Posteriormente esse método foi generalizado e outras aplicações foram encontradas \cite{symemb, symembjf, symemboutros, symembproca, symembdual, symembcor}. Em \cite{symemb, symembjf} o formalismo foi apresentado e sistematizado em sua forma mais geral, na próxima subseção iremos apresentar esse formalismo. A menos de algumas sutilezas, todas as referências mencionadas seguem o mesmo algoritmo. Conforme veremos, a maior diferença entre seus algoritmos encontra-se na escolha de empregar um ou dois campos de Wess-Zumino (WZ) \cite{wz}.

\vspace{.4in}
\subsection{A versão mais simples: uma variável de Wess-Zumino}
Segundo o formalismo simpl\'{e}tico, teoria de calibre \'{e} aquela que, para algum $n$, $(h_{\alpha \beta}^{(n)})$ \'{e} degenerada e seus modos-zero n\~{a}o produzem novos v\'{\i}nculos. A fim de conceder esta propriedade a uma teoria que n\~{a}o a possui, acrescentaremos uma vari\'{a}vel auxiliar,
\be
	\xi^\alpha \; \longrightarrow \;\; \tilde\xi^{\tilde\alpha}= \pmatrix{\xi^{\alpha} & \theta}
\ee
e duas novas fun\c{c}\~{o}es: $\Psi(\xi)$ e $G(\xi,\theta)$, a primeira na parte cin\'{e}tica da Lagrangiana e a segunda na parte potencial; de tal forma que, ao tomar-se  $\theta = 0$, retorne-se à Lagrangiana original.

	Ap\'{o}s a introdu\c{c}\~{a}o de $\theta$, $\Psi$ e $G$, o primeiro passo \'{e} impor que a nova matriz pr\'{e}-simpl\'{e}tica $(\tilde h_{\tilde \alpha \tilde \beta})$ seja degenerada, assim $\Psi$ ser\'{a} determinado. O segundo passo consiste em impor que seus modos-zero n\~{a}o produzam novos v\'{\i}nculos, o que vem a determinar $G$. Com isto, obtém-se uma Lagrangiana com simetria de calibre, cuja fixa\c{c}\~{a}o com a condi\c{c}\~{a}o $\theta=0$ (calibre que, por construção, será atingível)  promove as mesmas equa\c{c}\~{o}es de movimento da Lagrangiana original. Detalharemos esse procedimento agora.

\vspace{.4in}
Considere a Lagrangiana,  
\be
	L = a_\alpha\dot\xi^{\alpha} - V.
\ee
Acrescentaremos a vari\'{a}vel auxiliar $\theta(t)$ e as fun\c{c}\~{o}es $\Psi = \Psi(\xi)$ e $G = G(\xi,\theta)$ da seguinte forma:
\be
	\label{lsymemb1}
	\tilde L= a_\alpha\dot\xi^{\alpha} + \Psi\dot\theta - V - G,
\ee
e consideraremos que $G$ possa ser expandida em uma série de potências em $\theta$ com a condição $G(\theta = 0) = 0$, ou seja,
\be
	\label{symG}
	G(\xi,\theta) = \sum^\infty_{n=1}g_n(\xi)\theta^n.
\ee

	O tensor pr\'{e}-simpl\'{e}tico associado a $\tilde L$ é dado por
\be
	\tilde h_{\tal \tbe} \equiv \frac{\partial\tilde a_{\tbe}}{\partial \tilde \xi^{\tal}} - \frac{\partial\tilde a_{\tal}}{\partial \tilde \xi^{\tbe}},
\ee
em que $(\tilde \xi^{\tal}) = \pmatrix{\xi^{\alpha} & \theta}$ e $(\tilde a_{\tal})^T = \pmatrix{a_\alpha &  \Psi}$. Matricialmente temos

\be
	(\tilde h_{\tal \tbe}) = \pmatrix{
h_{\alpha \beta} & \partial_\alpha \Psi \cr
- \partial_\beta \Psi & 0 \cr}.
\ee
Com $\partial_\alpha := \partial / \partial \xi^\alpha$.

	Imporemos agora que a matriz acima possua um modo-zero do tipo
\be
	(\tilde \nu^\tal) = \pmatrix{ \mu^\alpha & 1},
\ee
sendo $\mu$ um vetor n\~{a}o nulo e diferente dos modos-zero de $(h_{\alpha \beta})$ (se esta possuir algum).

	A \emph{escolha} de $\mu$ determinar\'{a} $\Psi$ atrav\'{e}s das equa\c{c}\~{o}es
\be
	\tilde \nu^{\tal}\tilde h_{\tal \tbe} = 0.
\ee
Esse modo-zero ser\'{a} respons\'{a}vel pela transforma\c{c}\~{a}o de calibre da nova teoria, logo a simetria que ser\'{a} concedida \'{e} escolhida no momento em que define-se o vetor $\mu$. O fato da \'{u}ltima componente de $(\tilde \nu^{\tal})$ ser a unidade assegura a exist\^{e}ncia de uma transforama\c{c}\~{a}o de calibre envolvendo $\theta$.

	Estando os modos-zero escolhidos e a fun\c{c}\~{a}o $\Psi$ determinada, d\'{a}-se in\'{\i}cio ao passo seguinte do m\'{e}todo, o cálculo de $G$. Para termos uma teoria de calibre, temos de impor que novos vínculos não surjam, ou seja,
\be
	\tilde \nu^{\tal}\partial_{\tal}\tilde V = 0,
\ee
com $\tilde V := V + G$. Assim,
\be
	\mu^{\alpha} \left( \frac{\partial V}{\partial \xi^{\alpha}} + \frac{\partial G}{\partial \xi^{\alpha}} \right) + \frac{\partial G}{\partial \theta} = 0.
\ee

	A partir da equa\c{c}\~{a}o anterior, a fun\c{c}\~{a}o $G$ pode ser encontrada. Usando que $G$ pode ser escrita como uma série de potências (\ref{symG}), temos
\be
	\mu^{\alpha} \partial_\alpha V + \frac{\partial {\cal G}_1}{\partial \theta}=0
\ee
e
\be
	\mu^{\alpha} \partial_\alpha {\cal G}_n + \frac{\partial {\cal G}_{n+1}}{\partial \theta} = 0,
\ee
em que ${\cal G}_n := g_n(\xi)\theta^n$ e $n\geq 1$.

	Com isto encerra-se o algoritmo do formalismo simpl\'{e}tico de calibre para uma variável de WZ. A matriz pr\'{e}-simpl\'{e}tica $(\tilde h_{\tal \tbe})$ n\~{a}o pode ser invertida e possui um modo-zero que n\~{a}o gera um novo v\'{\i}nculo. A teoria dada por $\tilde L$ é de calibre e invariante por  
\be
	\label{simetria1}
	\delta_\vep \xi^\alpha = \mu^\alpha \vep \;\;\;\;\;\;\;\;\;\;\;\;\; \delta_\vep \theta = \vep.
\ee

Esta versão do formalismo simplético de calibre é bem simples, mas suficientemente forte para lidar com vários problemas. Sua passagem para o contínuo não tem maiores dificuldades. Como somente uma variável de WZ foi empregada, em particular não é a priori natural comparar qualquer resultado desse formalismo com o BFFT \cite{bfft}. Mas o formalismo simplético pode ser estendido para duas variáveis de WZ

Isso pode ser feito estendendo o formalimo simplético para dois campos de WZ, como veremos na próxima subseção.

\vspace{.4in}
\subsection{Generalizando: dois campos de Wess-Zumino}

A fim de preparar a notação para as aplicações subseqüentes, esta subseção será apresentada diretamente para o caso contínuo, ao invés do mecânico. Como antes anunciado, há simetrias de calibre mais complexas que requerem o emprego de mais campos de WZ. Desde que para cada campo de WZ esteja associado um gerador de transformações de calibre independente, nenhum problema de violação de graus de liberdade é esperado. Analogamente, no BFFT, para cada campo de WZ inserido, um vínculo de segunda classe é convertido em um de primeira classe. 

Em particular, com o formalismo anterior não é possível obter qualquer Lagrangiana invariante de calibre que seja quadrática em $\dot \theta$. Para generalizar a expressão (\ref{lsymemb1}) para dois campos de WZ pode-se usar
\be
	\tilde \cl_{\theta, \gamma}^{\mbox{\tiny geral}} = \cl + \Psi(\xi, \theta, \gamma) \; \dot \theta + \Sigma (\xi, \theta, \gamma) \; \dot \gamma - G(\xi, \theta, \gamma), 
\ee
mas os cálculos tornam-se incrivelmente complicados. Para as aplicações que seguem, precisaremos no máximo da seguinte forma:
\be 
	\label{ltg}
	\tilde \cl_{\theta,\gamma} = \cl + (\Psi(\xi, \theta) + \gamma) \dot \theta - G(\xi, \theta) - \frac k2 \gamma \gamma,
\ee
ou seja, a presença de $\gamma$ permite que $\dot \theta^2$ apareça em $\tilde \cl$.

Seja $(f_{\alpha \beta})$ a matriz simplética\footnote{Ocasionalmente, por simplicidade e de acordo com uso corrente na literatura, chamaremos $f$ de matriz simplética no lugar de (pré-)simplética.} de $\cl$, a matriz simplética de $\tilde \cl_{\theta, \gamma}$ é dada por
\be
	\tilde f_{\theta, \gamma} (\vx,\vy) = \pmatrix{f_{\alpha \beta} & \frac{\delta \Psi(\vy)}{\delta \xi^\alpha(\vx)} & 0_\alpha \\[0.1in] \cr 
	- \frac{\delta \Psi(\vx)}{\delta \xi^\beta(\vy)} & \Theta_{xy} & -\dirac \\[0.1in] \cr 
	0_\beta & \dirac & 0},
\ee
em que $0_\alpha$ é uma coluna nula, $0_\beta$ uma linha nula e o símbolo $\Theta_{xy}$ é definido por
\be
	\Theta_{xy} := \frac{\delta \Psi(\vy)}{\delta \theta (\vx)} - \frac{\delta \Psi(\vx)}{\delta \theta (\vy)}.
\ee

	No formalimo com um campo de WZ, selecionaria-se agora o modo-zero que seria o gerador da simetria de calibre. No presente caso isso não é possível, devido ao aparecimento das $\delta$'s em $\tilde f$. Para dar seguimento, será preciso supor que $\cl$ possua algum vínculo, de forma análoga ao que ocorre no formalismo BFFT, e contrariamente ao observado no caso com somente um campo de WZ. Assim, o único modo-zero à disposição tem o seguinte aspecto:
\be
	\label{introzmmod}
	\tilde \nu^\alpha(\vx) = \pmatrix{ \nu^\alpha(\vx) &  0 &  b(\vx)},
\ee
em que $(\nu^\alpha)$ é modo-zero de $(f_{\alpha \beta})$ e $b$ em geral é uma função de $\tilde \xi$. A fim de que $\tilde \nu$ seja modo-zero de $\tilde f$, é suficiente e necessário que 
\be
	\label{nucond2}
	\int d^nx \; \left ( \nu^\alpha \frac {\delta \Psi(\vy)}{\delta \xi^\alpha(\vx)} + b \dirac \right ) = 0.
\ee
Por simplicidade, comumente consideraremos que $b$ é uma constante não-nula. A equação acima impõe uma primeira condição que $\Psi$ deve satisfazer.

O modo-zero $\tilde \nu$ não serve como gerador de transformações de calibre, mas servirá para modificar o vínculo da teoria original $\cl$ gerado por $\nu$, como segue
\ba
	\int d^ny \; \tilde \nu^\alpha(\vx) \frac {\delta \tilde V_{\theta,\gamma}(\vy)}{\delta \tilde \xi^\alpha(\vx)} & = & \int d^ny \; \tilde \nu^\alpha(\vx) \frac \delta {\delta \tilde \xi^\alpha(\vx)} \left( V(\vy) + G(\vy) + \frac k2 \gamma(\vy)\gamma(\vy) \right ) \nonumber \\[0.1in]
	\label{introGnu}
	& = & \Omega(\vx) + \int d^ny \; \nu^\alpha (\vx) \frac {\delta G(\vy)}{\delta  \xi^\alpha(\vx)} + k b \gamma (\vx)  \\[0.1in]
	& = & \Omega  + G_\nu + kb \gamma =: \tilde \Omega. \nonumber
\ea
Acima, $\tilde \xi^\alpha = (\xi^\alpha, \theta, \gamma)$, $\Omega$ é o vínculo da teoria original que é gerado por $\nu$ e $G_\nu$ econtra-se implicitamente definido.

Prosseguindo com as etapas usuais do método simplético, o vínculo modificado $\tilde \Omega$ é adicionado ao setor cinético de $\tilde \cl_{\theta, \gamma}$ por meio de um multiplicador de Lagrange, definindo $\tilde \cl_{\theta, \gamma}^\1$, ou seja,
\be
	\tilde \cl_{\theta, \gamma}^\1 = a_\alpha \dot \xi^\alpha + (\Psi + \gamma) \dot \theta + \tilde \Omega \dot \lambda - \tilde V_{\theta, \gamma}.
\ee

Poderia-se modificar $\tilde V_{\theta, \gamma}$ usando $\tilde \Omega = 0$, como indicado em \cite{mbw}, mas esse passo não é indiferente para este formalismo. Vamos considerar que $\tilde V_{\theta, \gamma}$ ainda é dado por $V + G + \frac 12 k \gamma \gamma$.

Sendo $\tilde \xi^{\1 \alpha} = ( \xi^\alpha, \theta, \gamma, \lambda)$, a nova matriz simplética é  
\be
	\tilde f_{\theta, \gamma}^\1 (\vx,\vy) = \pmatrix{(f_{\alpha \beta}) & \frac{\delta \Psi(\vy)}{\delta \xi^\alpha(\vx)} & 0_\alpha & \frac{\delta (\Omega + G_\nu)(\vy)}{\delta \xi^\alpha(\vx)}\\[0.1in] \cr 
	- \frac{\delta \Psi(\vx)}{\delta \xi^\beta(\vy)} & \Theta_{xy} & -\dirac & \frac{\delta G_\nu(\vy)}{\delta \theta(\vx)}\\[0.1in] \cr 
	0_\beta & \dirac & 0 & kb(\vy) \dirac  \\[0.1in] \cr
	- \frac{\delta (\Omega + G_\nu)(\vx)}{\delta \xi^\beta(\vy)} & - \frac{\delta G_\nu(\vx)}{\delta \theta(\vy)} & - kb(\vx) \dirac & 0}.
\ee

	Essa matriz possibilita a seleção de modos-zero que serão os geradores da simetria. Selecionaremos os dois seguintes geradores independentes:
\ba
	\label{intronugamma}
	\tilde \nu_\gamma^\alpha &=& \pmatrix{\nu^\alpha(\vx) & \;\;\; 0 & \;\; b & 0} = \pmatrix{\tilde \nu^\alpha(\vx) & 0}, \\
	\label{intronutheta}
	\tilde \nu_\theta^\alpha &=& \pmatrix{\mu^\alpha(\vx) &  -kb & 0 &  1}.
\ea
Acima, $\mu^\alpha$ pode ser neste ponto fixado de acordo com a simetria desejada ou pode ser carregado como uma incógnita a ser fixada somente na Lagrangiana final do método. No primeiro caso, chega-se à resposta final mais rapidamente, mas corre-se o risco de se impor uma simetria incompatível com $\tilde \cl$. 

	A condição de que $\tilde \nu_\gamma$ não deve gerar novos vínculos é imediatamente satisfeita, pois
\be
	\int d^ny \; \tilde \nu_\gamma^\alpha(\vx) \frac {\delta \tilde V_{\theta,\gamma}(\vy)}{\delta \tilde \xi^{\1 \alpha}(\vx)} = \int d^ny \; \tilde \nu^\alpha(\vx) \frac {\delta \tilde V_{\theta,\gamma}(\vy)}{\delta \tilde \xi^\alpha(\vx)} \nonumber = \tilde \Omega(\vx).
\ee

	A condição de que $\tilde \nu_\theta$ não gere novos vínculos leva à equação diferencial
\ba
	0 & = & \int d^ny \; \tilde \nu_\theta^\alpha(\vx) \frac {\delta \tilde V_{\theta,\gamma}(\vy)}{\delta \tilde \xi^{\1 \alpha}(\vx)} \nonumber \\[0.1in]
	\label{condG2}
	& = & \int d^ny \; \left ( \mu^\alpha(\vx) \frac{\delta (V + G)(\vy)}{\delta \xi^\alpha(\vx)} - kb \frac{\delta G(\vy)}{\delta \theta(\vx)} \right ).
\ea

	Por fim, $\Psi$ e a constante $k$ são determinadas através das equações
\ba
	\label{fdif1}
	\int d^nx \; \tilde \nu_\gamma^\alpha (\vx) \tilde f^{\1 \theta, \gamma}_{\alpha \beta} (\vx, \vy) &=& 0, \\[0.1in]
	\label{fdif2}
	\int d^nx \; \tilde \nu_\theta^\alpha (\vx) \tilde f^{\1 \theta, \gamma}_{\alpha \beta} (\vx, \vy) &=& 0.
\ea

\vspace{.4in}
	Encontradas as soluções para as equações diferenciais (\ref{nucond2}, \ref{condG2}, \ref{fdif1}, \ref{fdif2}), $\tilde \cl$ terá dois geradores independetes de transformações de calibre, dados por $\tilde \nu_\theta$ e $\tilde \nu_\gamma$. Usaremos $\delta_{\vep_\theta}$ para designar variações de calibre devido a $\tilde \nu_\theta$ e $\delta_{\vep_\gamma}$ para as de $\tilde \nu_\gamma$. Somando essas variações independentes, vem 
\ba
	(\delta_{\vep_\gamma} + \delta_{\vep_\theta})\xi^\alpha (\vx) & = & \int d^ny \; [\vep_{\gamma} (\vy) \nu^\alpha (\vy) + \vep_{\theta} (\vy) \mu^\alpha (\vy)] \dirac, \nonumber \\
	(\delta_{\vep_\gamma} + \delta_{\vep_\theta})\theta (\vx) & = & - \vep_{\theta} (\vx) kb,  \nonumber \\
	(\delta_{\vep_\gamma} + \delta_{\vep_\theta})\gamma (\vx) & = & \vep_{\gamma} (\vx) b, \\
	(\delta_{\vep_\gamma} + \delta_{\vep_\theta})\lambda (\vx) & = & \vep_{\theta} (\vx). \nonumber
\ea 

	Eliminando $\gamma$ por meio de sua equação de movimento $\gamma = \dot \theta /k$, da igualdade 
\be
	(\delta_{\vep_\gamma} + \delta_{\vep_\theta}) \dot \theta = k (\delta_{\vep_\gamma} + \delta_{\vep_\theta}) \gamma	
\ee	
vem $\vep_\gamma = - \dot \vep_\theta$. Seja $\vep := \vep_\theta$ e $\delta_\vep := (\delta_{\vep_\gamma} + \delta_{\vep_\theta})$, logo	
\ba
	\delta_\vep \xi^\alpha (\vx) & = & \int d^ny \; [- \dot \vep (\vy) \nu^\alpha (\vy) + \vep (\vy)\mu^\alpha (\vy)] \dirac,  \nonumber \\
	\label{introsymdotvep}
	\delta_\vep\theta (\vx) & = & - \vep (\vx) kb, \\
	\delta_\vep\dot \theta (\vx) & = & - \dot \vep (\vx) kb. \nonumber
\ea 

Esse caso deve ser contraposto ao apresentado na Eq. (\ref{simetria1}). Não é de se esperar que $\mu^\alpha$ em (\ref{simetria1}) envolva $\prt_0$, pois essa derivada em geral impede que $\mu^\alpha$ seja simultaneamente modo-zero de $f$ e produza soluções para $G$ que não envolvam derivadas temporais (pois do contrário G não é um termo do potencial). Graças à introdução de um segundo campo de WZ, foi possível de forma razoavelmente geral tratar do caso em que a Lagrangiana invariante de calibre envolve $\dot \theta^2$.

Nas próximas subseções ilustraremos esse formalismo em duas teorias bem distintas ambas sem simetria de calibre: o fluido irrotacional e o modelo de Proca. Estas duas aplicações se encontram em \cite{symemb}, outras aplicações podem ser vistas nas Refs. \cite{symemb, symembproca, symemboutros, symembdual}.

\vspace{.4in}
\subsection{Exemplo 1: fluido irrotacional}

Nesta seção o método simplético de calibre será aplicado em uma teoria sem vínculos (no sentido do método simplético) e já linear nas velocidades. Ilustraremos o caso mais simples do método que emprega apenas um campo de WZ.

	A Lagrangiana do modelo tratado em $d$ dimensões espaciais é dada por \cite{irrotfluid}
\begin{equation}
\label{00010}
{\cal L} = - \rho\dot\eta + \frac 12 \rho(\partial_a\eta)(\partial^a\eta) - \frac g\rho,
\end{equation}
em que $a=1,2,...,d$, $\;\; \rho$ é a densidade de massa, $\eta$ é o potencial de velocidade e $g$ é uma constante. A métrica é Euclideana. Essa Lagrangiana não possui nem vínculos\footnote{Por ser uma Lagrangiana de primeira ordem nas velocidades, uma aplicação imediata do algoritmo de Dirac introduz um campo extra e um vínculo de segunda classe. Esse procedimento circular não faz parte do algoritmo simplético.} e nem simetria de calibre. 

A Lagrangiana invariante de calibre $\tilde \cl$ possui o aspecto
\be
\label{00020}
{\tilde{\cal L}} = - \rho\dot\eta + \Psi \dot \theta + \frac 12 \rho \; \partial_a\eta\partial^a\eta - \frac g\rho - G,
\ee
com $\Psi= \Psi(\rho, \eta, \theta)$ e $G = G(\rho, \eta, \theta)$. 

Sendo $\tilde \xi^{ \alpha} = (\rho, \eta, \theta)$ o vetor simplético de coordenadas, o vetor simplético dos momentos e a matriz correspondente são $\tilde a_\alpha = (0, -\rho, \Psi)$ e
\be
\label{00030}
{\tilde f }= \pmatrix{0 & - \delta(\vec x - \vec y) & \frac{\delta\Psi(\vec y)}{\delta\rho(\vec x)} \\[0.1in] \cr
\delta(\vec x - \vec y) & 0 & \frac{\delta\Psi(\vec y)}{\delta\eta(\vec x)}  \\[0.1in] \cr
- \frac{\delta\Psi(\vec x)}{\delta\rho(\vec y)} & -  \frac{\delta\Psi(\vec x)}{\delta\eta(\vec y)} & \Theta_{xy}  }.
\ee
Acima, $\vx$ e $\vy$ são vetories $d$-dimensionais, as deltas também são d-dimensionais e $\Theta_{xy} := \delta \Psi (\vy)/ \delta \theta (\vx) - \delta \Psi (\vx)/ \delta \theta (\vy)$  (no caso mecânico $\Theta_{xy}$ é sempre nulo).

Selecionando o modo-zero mais geral, 
\be
	\tilde \nu = \pmatrix {a & b & 1},
\ee
as seguintes condições são impostas a $\Psi$:
\ba
	\frac {\delta \Psi (\vx)}{\delta \rho(\vy)} & = &  b \dirac \nonumber \\
	\label{fluidpsieqs}
	\frac {\delta \Psi (\vx)}{\delta \eta(\vy)} & = & - a \dirac\\
	\Theta_{xy} & = & 0 \nonumber.
\ea

Por simplicidade, assumiremos que $a$ e $b$ são constantes.

Das Eqs. (\ref{fluidpsieqs}), encontra-se $\Psi$ como sendo
\be
	\Psi = b \rho - a \eta + f(\theta),
\ee
em que $f(\theta)$ é uma função arbitrária de $\theta$ somente. Essa função, como pode ser facilmente verificado, apenas contribui com termos de superfície para $\tilde \cl$, portanto ela não será mais escrita.

O último passo para concluir o método simplético de calibre é obter a função $G$. Essa função é determinada ao se exigir que $\tilde \nu$ não gere novos vínculos, ou seja,
\be
\label{condGfluid}
\int d^d y\,\,\tilde \nu^\alpha(\vec x) \frac{\delta \tilde V(\vec y)}{\delta \tilde \xi^\alpha(\vec x)} = 0,
\ee
sendo $\tilde V$ a parte potencial de $\tilde {\cal L}$, 
\be
	\tilde V = - \frac 12 \rho \; \partial_a\eta\partial^a\eta + \frac g\rho + G.
\ee
Portanto,
\ba
\label{00090}
&&\int d^d y\,\,\left\{ a \left ( - \frac 12\partial_a\eta \; \partial^a\eta \; \dirac - \frac g {\rho^2} \dirac + \frac {\delta G (\vy)}{\delta \rho (\vx)} \right ) + \right. \nonumber \\[.2in]
&& \left.  + b \left ( - \rho \; \partial_a \eta \; \partial^a \dirac + \frac {\delta G (\vy)}{\delta \eta (\vx)} \right ) + \frac {\delta G (\vy)}{\delta \theta (\vx)} \right\}=0.
\ea
Acima, toda dependência implícita do vetor espacial se refere ao vertor $\vy$.

Expandindo $G$ em potências de  $\theta$, $G = \sum \cg_n$, com  $\cg_n \propto \theta^n$ e $n \ge 1$ [devido a $G(\theta=0)=0$], temos
\ba
\label{00100}
	\cg_1 &=& a \left ( \frac 12 \partial_a\eta \; \partial^a\eta \; \theta + \frac g {\rho^2} \theta \right ) + b \rho \partial_a \eta \; \partial^a \theta, \nonumber\\[0.1in]
	\cg_2 &=&  - a \left ( -a \frac g {\rho^3} \theta^2 + b \partial^a \eta \; \partial_a \theta \; \theta \right ) - \frac {b^2 }2 \rho \partial_a \theta \; \partial^a \theta, \nonumber\\[0.1in]
	\cg_3 & = & a \left ( a^2 \frac g {\rho^4} \theta^3 + \frac {b^2} 2\theta \partial^a \theta \; \partial_a \theta \right ), \\[0.1in]
	\cg_n & = & a^n \frac {g}{\rho^{n+1}} \theta^n \;\;\;\;\;\;\; \forall \;\; n \ge 4. \nonumber 
\ea

Sendo $\rho > a \theta$ a série $\sum \cg_n$ converge e a seguinte Lagrangiana é encontrada: 
\ba
	\tilde \cl &=& - \rho \dot \eta + (b \rho - a \eta) \dot \theta - \frac g {\rho - a \theta} + \nonumber \\[.2in]
	&& +  (\rho - a\theta) \left ( \frac 12 \partial_a \eta \; \partial^a \eta - 	b \partial^a \eta \; \partial_a \theta + \frac {b^2}2  \partial^a \theta \; \partial_a \theta \right ) .
\ea 

Essa Lagrangiana é invariante pelas transformações dadas pelo modo-zero $\tilde \nu$, ou seja,
\ba
\label{00150}
\delta_\vep \rho &=& a \vep,\nonumber\\
\delta_\vep \eta &=& b \vep,\\
\delta_\vep \theta&=& \vep. \nonumber
\ea

	Pode-se verificar diretamente que realmente $\delta_\vep \tilde \cl = 0$.

\vspace{.4in}


\subsection{Exemplo 2: modelo de Proca}

O termo de massa inserido pelo modelo de Proca quebra a simetria $U(1)$ do eletromagnetismo usual. Nesta seção vamos empregar o método simplético com dois campos de WZ para recuperar essa simetria. Sendo mais específico, queremos implementar uma simetria em que $\delta_\vep A^\mu = \partial^\mu \vep$, $\delta_\vep \tilde \cl = 0$ e a dinâmica dada por $\tilde \cl$ seja a mesma do modelo de Proca; ou seja, por meio de certo fixador de calibre, pode-se retornar à Lagrangiana do modelo de Proca. Como $\delta_\vep A_0 = \dot \vep$, isso sugere que precisamos da versão do formalismo com dois campos de WZ (\ref{introsymdotvep}).

Antes de começar o método simplético de calibre é necessário introduzir campos auxiliares para obter uma Lagrangiana em primeira ordem nas velocidades. A Lagrangiana do modelo de Proca
\be
	{\cal L}(A_\mu,\partial_\nu A_\mu)  =  -\frac 14 F^{\mu \nu}F_{\mu \nu} + \frac {m^2}2 A^\mu A_\mu, 
	\label{proca1}
\ee
pode ser escrita como
\be
	{\cal L}(A_\mu,\partial_\nu A_i, \pi_i, \partial_j \pi_i) =  \pi^i \dot A_i + \frac 12 \pi^i \pi_i - \pi^i\partial_i A_0 - \frac 14 F^{ij}F_{ij} + \frac {m^2}2 A^\mu A_\mu,
	\label{proca2}
\ee
em que $\mu=0,1,2,3$, $i=1,2,3$, $g = \diag \pmatrix{+ & - & - & -}$ e $F_{\mu \nu} := \partial_\mu A_\nu - \partial_\nu A_\mu$. 

Agora pode-se dar início ao formalismo. O primeiro passo é introduzir os campos de WZ $\theta$ e $\gamma$ (\ref{ltg}),
\be
	\tilde {\cal L} = \pi_i \dot A^i + (\Psi + \gamma) \dot \theta - \tilde V,
\ee 
com
\be
	\tilde V =  - \frac 12 \pi^i \pi_i + \pi^i \partial_i A_0 + \frac 14 F^{ij}F_{ij} - \frac {m^2}2 A^\mu A_\mu + G  + \frac k2 \gamma \gamma .
	\label{vproca}
\ee
Precisaremos agora determinar $k$, $\Psi$ e $G$.

Seja $\tilde \xi^\alpha = (A^0, A^i, \pi^i, \theta, \gamma)$, logo os momentos associados são  $\tilde a_\alpha = (0, \pi_j, 0, \Psi + \gamma, 0)$ e 
\be
	\tilde f = \pmatrix{ 0 & 0 & 0 & \frac{\delta \Psi(\vy)}{\delta A^0(\vx)} & 0 \cr
	0 & 0 & -g_{ji} \dirac & \frac{\delta \Psi(\vy)}{\delta A^i(\vx)} & 0 \cr
	0 & g_{ij} \dirac & 0 & \frac{\delta \Psi(\vy)}{\delta \pi^i(\vx)} & 0 \cr
	- \frac{\delta \Psi(\vx)}{\delta A^0(\vy)} & - \frac{\delta \Psi(\vx)}{\delta A^j(\vy)} & - \frac{\delta \Psi(\vx)}{\delta \pi^j(\vy)} & \Theta_{xy} & - \dirac \cr
	0 & 0 & 0 & \dirac & 0}
\ee
é a matriz simplética, cujas componentes acima são determinadas, como usual, a partir de
\be
	\tilde f_{\alpha \beta} (\vx,\vy) \equiv \frac{\delta \tilde a_\beta(\vy)}{\delta \tilde \xi^{\alpha}(\vx)} - \frac{\delta \tilde a_\alpha(\vx)}{\delta \tilde \xi^{\beta}(\vy)}.
\ee
Alguns zeros que aparecem acima são verdadeiramente vetores linhas, colunas ou matrizes cujos  elementos são todos nulos.

O modelo de Proca possui um vínculo que é gerado a partir do modo-zero $\pmatrix{ 1 & 0_{1 \times 3} & 0_{1 \times 3}}$. Como antes apresentado, encontraremos um vínculo modificado a partir do seguinte vetor:
\be
	\tilde \nu  = \pmatrix {1 & 0_{1 \times 3} & 0_{1 \times 3} & 0 & b}, 
\ee
em que consideraremos que $b$ é constante. 

Exigindo que $\tilde \nu$ seja modo-zero de  $\tilde f$, encontra-se 
\be
	\frac {\delta \Psi(\vy)}{\delta A^0(\vx)} = -b \dirac.
	\label{psi_a0proca}
\ee

O vínculo associado a $\tilde \nu$ é dado por
\be
	\tilde \Omega (\vx) = \int d^3y \; \tilde \nu^\alpha(\vx) \frac{\delta \tilde V(\vy)}{\delta \tilde \xi^\alpha (\vx)} = -\partial_i \pi^i - m^2A_0 + \int d^3y \; \frac{\delta G(\vy)}{\delta A_0 (\vx)}  + bk \gamma.
	\label{modconstproca}
\ee
De forma mais compacta escreveremos $\tilde \Omega = \Omega + G_0 + bk\gamma$. $G_0$ é o $G_\nu$ usado na apresentação do formalismo geral.

Seguindo o procedimento padrão de lidar com vínculos no formalismo simplético \cite{mbw}, adicionamos $\dot \lambda \tilde \Omega$ a $\tilde \cl$. Portanto,
\be
	\tilde \cl^{(1)} = \pi^i \dot A_i + (\Psi + \gamma) \dot \theta + \dot \lambda \tilde \Omega - \tilde V.
\ee
Costuma ser conveniente eliminar o vínculo do setor potencial, devido a ele já ter sido imposto no setor cinético \cite{mbw}, mas esse procedimento não será útil, portanto $\tilde V$ permanece inalterado.

Sendo $\tilde \xi^{(1)\alpha} = (A^0, A^i, \pi^i, \theta, \gamma, \lambda)$, agora com $\alpha = 1,2,...,10$, e usando a Eq. (\ref{psi_a0proca}), a seguinte matriz simplética é encontrada:

\noindent
$\tilde f^{(1)} = $
\be
\pmatrix{ 0 & 0 & 0 & -b\delta^{(3)} & 0 & \frac {\delta G_0(\vy)}{\delta A_0(\vx)} - m^2\delta^{(3)} \\[0.1in] \cr
	0 & 0 & -g_{ji} \delta^{(3)} & \frac{\delta \Psi(\vy)}{\delta A^i(\vx)} & 0 & \frac {\delta G_0(\vy)}{\delta A^i(\vx)} \\[0.1in] \cr
	0 & g_{ij} \delta^{(3)} & 0 & \frac{\delta \Psi(\vy)}{\delta \pi^i(\vx)} & 0 & \frac {\delta G_0(\vy)}{\delta \pi^i(\vx)} - {\partial}_i^y \delta^{(3)} \\[0.1in] \cr
	b\delta^{(3)} & - \frac{\delta \Psi(\vx)}{\delta A^j(\vy)} & - \frac{\delta \Psi(\vx)}{\delta \pi^j(\vy)} & \Theta_{xy} & - \delta^{(3)} & \frac {\delta G_0(\vy)}{\delta \theta (\vx)} \\[0.1in] \cr
	0 & 0 & 0 & \delta^{(3)} & 0 & bk \delta^{(3)} \\[0.1in] \cr
	- \frac {\delta G_0(\vx)}{\delta A_0(\vy)} + m^2 \delta^{(3)} & \frac {\delta G_0(\vx)}{\delta A^j(\vy)} & {\partial}_j^x \delta^{(3)} - \frac {\delta G_0(\vx)}{\delta \pi^j(\vy)} & - \frac {\delta G_0(\vx)}{\delta \theta(\vy)} & -bk \delta^{(3)} & 0}. \nonumber
\ee

\noindent
Acima, $\delta^{(3)}:= \dirac$.

Agora, em acordo com as Eqs.(\ref{intronugamma}, \ref{intronutheta}), selecionamos os seguintes modos-zero que serão responsáveis pela simetria de calibre:
\ba
	\tilde \nu_{(\theta)} & = & \pmatrix{a_0 & a \partial^i & c \partial^i & -kb & 0 & 1}, \nonumber \\
	\tilde \nu_{(\gamma)} & = & \pmatrix{1 & 0_{1 \times 3} & 0_{1 \times 3} & 0\;\; & b\; & 0} = \pmatrix{ \tilde \nu & 0}.
	\label{zmproca}
\ea
Esses modos-zero são  mais gerais do que os correspondentes à proposta inicial, segundo a qual $\tilde \cl$ deveria ser invariante por transformações do tipo $\delta_\vep A_\mu = \prt_\mu \vep$. Analisando a Eq. (\ref{introsymdotvep}), vê-se que de imediato pode-se tomar $a_0=0$ e $a = 1$ para essa proposta. Contudo, embora considerar esses valores neste momento reduza consideravelmente os cálculos futuros, carregaremos todos os termos acima, com a única restrição de serem constantes. 

Conforme apresentado no formalismo geral, $\tilde \nu_{(\gamma)}$ não gera novos vínculos. A imposição de que $\tilde \nu_{(\gamma)}$ seja modo-zero implica que
\be
	\frac{\delta G_0 (\vy)}{\delta A_0 (\vx)} = (m^2-b^2k)\dirac.
	\label{G0proca}
\ee

Da imposição de que  $\tilde \nu_{(\theta)}$ não deve gerar novos vínculos vem 
\ba
	\label{Gproca}
	&& 0 =  \int d^3y \; \tilde \nu^\alpha_{(\theta)} (\vx) \; \frac{\delta \tilde V (\vy)}{\delta \tilde \xi^{(1) \alpha}(\vx)}  \\[.2in]
	&& =  \int d^3y \left \{ a_0 \dirac ( -\partial_i \pi^i - m^2A_0 )  + a\partial^i_x \dirac ( \partial^j F_{ij} - m^2 A_i )  +     \right.  \nonumber \\[.2in]
	& & \left. + \; c\partial^i_x  \dirac ( -\pi_i + \partial_i A_0 ) + \rho^\mu_x \frac{\delta G(\vy)}{\delta A^\mu(\vx)} + c\partial^i_x \frac{\delta G(\vy)}{\delta \pi^i(\vx)} - kb \frac{\delta G(\vy)}{\delta \theta(\vx)} \right \}. \nonumber
\ea
O índice $x$ em $\partial^i$ significa que as derivadas devem ser avaliadas com respeito a $x$, em vez de $y$, e 
\be
	\rho^\mu_x := ( a_0, a\partial^i_x).
\ee

	Consideremos novamente a expansão de $G$ em série de potências. Sendo $\cg_n$ proporcional a $\theta^n$, escrevemos $G = \sum_n \cg_n$. A condição $G(\theta = 0) = 0$ implica que $n \ge 1$. A solução em primeira ordem da Eq.(\ref{Gproca}) é 
\be
	\cg_1 = \frac{\theta}{kb} ( -a_0 \partial_i\pi^i -m^2\rho^\mu A_\mu - c\partial^i\pi_i + c\partial^i\partial_i A_0).
\ee

A menos de termos de superfície, a contribuição de segunda ordem em $\theta$ é 
\be
	\cg_2 = - \frac 1{2(kb)^2} \{ c(2 a_0 + c) \partial_i \theta \partial^i \theta  + m^2 \rho^\mu \theta \rho_\mu \theta \}.
\ee 

A ausência de $A^\mu$ e $\pi^i$ em $\cg_2$ implica que $\cg_n = 0$ para qualquer $n \ge 3$. Conseqüentemente, a função $G$ é agora conhecida. Pode-se então calcular $G_0$, 
\be
	G_0(\vx) = \int d^3y \; \frac {\delta G(\vy)}{\delta A^0 (\vx)} = \frac 1 {kb} (c\partial^i \partial_i \theta - m^2 a_0 \theta).
	\label{G0proca2}
\ee
Aplicando esse resultado na Eq. (\ref{G0proca}), temos
\be
	k = \frac {m^2}{b^2}.
	\label{kproca}
\ee

Agora $G$ e $k$ já se encontram determinados em função dos parâmetros dos modos-zero $\tilde \nu$ e $\tilde \nu_\theta$. Falta impormos todas as condições necessárias sobre $\Psi$ a fim de que $\tilde \nu_\theta$ seja realmente um modo-zero. Dentre equações que levam à identidade trivial [usando as Eqs.(\ref{G0proca2}, \ref{kproca})], aparecem as seguintes:

\vspace{.2in}

\be
	c\partial_j^x \dirac + \frac{m^2}{b} \frac{\delta \Psi(\vx)}{\delta A^j(\vy)} = 0,
	\label{psiaproca}
\ee

\be
	- a\partial_j^x \dirac + \frac {m^2}b \frac {\delta \Psi (\vx)}{\delta \pi^j(\vy)} + \partial_j^x \dirac = 0,
	\label{psipiproca}
\ee

\be
	-b a_0 \dirac + a\partial_x^i \frac{\delta \Psi(\vy)}{\delta A^i(\vx)} + c \partial^i_x \frac {\delta \Psi (\vy)}{\delta \pi^i(\vx)} - \frac {m^2}b \Theta_{xy} - \frac{\delta G_0 (\vx)}{\delta \theta (\vy)} = 0.
	\label{psithetaproca}
\ee

\vspace{.2in}

	Por meio das Eqs. (\ref{psi_a0proca}, \ref{psiaproca}, \ref{psipiproca}, \ref{psithetaproca}), a menos de termos que envolvam $\theta$ exclusivamente, $\Psi$ pode ser encontrado como
\be
	\Psi = - \frac b {m^2} \{ m^2 A_0 + c \partial_i A^i + (1 - a)\partial^i \pi_i \}.
\ee

A Eq. (\ref{psithetaproca}) apenas fixa o valor de $\Theta_{xy}$ como nulo, a fim de termos consistência com as demais equações diferenciais.

\vspace{.4in}
Enfim, todas as incógnicas $\Psi$, $G$ e $k$ foram determinadas em função dos parâmetros dos modos-zero $\tilde \nu$  e $\tilde \nu_\theta$. Há porém certa restrição quanto ao valor de $a$ que iremos enocontrar agora. Primeiramente vamos remover os momentos $\pi$. Nota-se que $\Psi$ e $G$ possuem termos que dependem de $\pi$, portanto essas funções alteram a relação original entre $\pi$ e $A$ como segue 
\be
	\label{novopi}
	\pi_i = \partial_i A_0 - \dot A_i + \frac b {m^2} \{(1-a) \partial_i \dot \theta + (a_0 + c) \partial_i \theta \}.
\ee

Podemos também eliminar $\gamma$ usando
\be
	\gamma = \frac {b^2}{m^2} \dot \theta.
\ee

Assim como a eliminação de $\gamma$ não leva a uma Lagrangiana imediatamente invariante pelas transformações  geradas por $\tilde \nu_\theta$ e $\tilde \nu_\gamma$, pois, conforme visto na apresentação do formalismo geral, essa eliminação só tem sentido se $\vep_\gamma = - \dot \vep_\theta$, faz-se necessário analisar cuidadosamente essa eliminação de $\pi$. A fim de que a Eq.(\ref{novopi}) seja consistente com a aplicação de $\delta_\vep := \delta_{\vep_\theta} + \delta_{\vep_\gamma}$ devemos fixar $a=1$. Note que esse valor é exatamente aquele desejável para que recobremos a simetria $U(1)$, como comentado no início desta subseção. Uma eliminação de $\pi$ e $\gamma$ sem fixar o valor de $a$ leva a 
\ba
	\tilde \cl & = & - \frac 14 F_{\mu \nu} F^{\mu \nu} + \frac {m^2}2 A^\mu A_\mu + \frac b {m^2} \{ -m^2 A_0 \dot \theta + (1-a) \partial_i \dot \theta (\partial^i A_0 - \dot A^i) + \nonumber \\[.2in] 
	\label{procaembedL}
	&+& a_0 \theta (\partial^i \partial_i A_0 - \partial_i \dot A^i ) + \theta m^2 \rho^\mu A_\mu \}  + \frac {b^2}{2m^2}\dot \theta \dot \theta + \\[.2in] 
	&+&	\frac {b^2}{m^4} \left \{ \frac 32 (1-a)^2 \partial_i \dot \theta \partial^i \dot \theta - \frac 12 a_0^2 \partial_i \theta \partial^i \theta + (1-a) (a_0 + c) \partial_i \dot \theta \partial^i \theta + \frac{m^2}2 \rho^\mu \theta \rho_\mu \theta \right \}. \nonumber
\ea

	Alguns termos acima dependem de mais de uma derivada por campo, ademais há a constante $c$ que é arbitrária e não participa mais das transformações de calibre. Felizmente, todos esses problemas desaparecem ao considerarmos que $a=1$. Para esse valor de $a$, a Lagrangiana $\tilde \cl$ acima é invariante perante as transformações (\ref{introsymdotvep})
\ba
	\delta_\vep A_0 & = & \vep a_0 -  \dot \vep, \nonumber \\
	\delta_\vep A^i & = & - \partial^i \vep,  \\
	\delta_\vep \theta & = & - \frac {m^2}b \vep. \nonumber
\ea

	A constante $b$ é apenas um reescalonamento de $\theta$ e tem dimensão de massa ao quadrado. Assim escolher $b= m^2$ é equivalente a substituir $\frac b{m^2}\theta$ por $\theta$. Curiosamente, a constante $a_0$ permanece arbitrária. Pode-se checar diretamente que, para qualquer valor de $a_0$ e $a=1$, realmente temos $\delta_\vep \tilde \cl =0 $. O motivo disso é simples, um rearranjo da Eq. (\ref{procaembedL}) com $a=1$ mostra que $a_0$ e $A_0$ somente ocorrem dentro na seguinte combinação: $A_0 + a_0 \theta =: \tilde A_0$, logo escolher $a_0=0$ é equivalente a substituir $\tilde A_0$ por $A_0$.
	
	Equivalentemente, esses mesmos valores dos parâmetros dos modos-zero podem ser obtidos impondo-se invariância de Lorentz explícita, e a Lagrangiana $\tilde \cl$ torna-se manifestamente invariante por transformações $U(1)$:
\be
	\tilde \cl  =  - \frac 14 F_{\mu \nu} F^{\mu \nu} + \frac {m^2}2 A^\mu A_\mu - m^2A^\mu \partial_\mu \theta + \frac {m^2}2 \partial^\mu \theta \partial_\mu \theta.
\ee

	A Lagrangiana da Eq. (\ref{procaembedL}) não é a mais geral atingível através do método simplético de calibre, outras estruturas dos modos-zero $\tilde \nu_{(\theta)}$ e $\tilde \nu_{(\gamma)}$ são também possíveis e suas componentes podem também depender de campos.

\vspace{.4in}
\subsection{Comentários finais}
\label{comfin1}

Os dois exemplos anteriores ilustram o funcionamento do método simplético de calibre e apresentam os principais artifícios que são suficientes para sua aplicação em vários outros modelos. Os artifícios apresentados são eficientes também  em teorias não-comutativas, como é o caso do modelo de Proca não-comutativo \cite{symembproca} e do modelo autodual não-comutativo \cite{symemb}. Como mostrado em \cite{symembcor}, o método simplético pode ser usado de forma a obter resultados iguais ao do BFFT \cite{bfft}, embora isso não seja uma necessidade \cite{symembproca}. Assim como há vários métodos de imersão em calibre que se fundamentam nos princípios do método de Dirac, futuros avanços do método simplético de calibre provavelmente o irão subdividir em vários métodos\footnote{De certa forma, alguma subdivisão já há, pois o método com um campo de WZ é consideravelmente diferente do com dois campos de WZ.}. Dentre as investigações a serem feitas, encontra-se a formalização da implementação de simetrias de tipo não-Abeliana, isto é, de modos-zero que dependam dos próprios campos da teoria; ademais achamos que futuros desenvolvimentos do método podem o tornar consideravelmente mais sucinto em termos de trabalho algébrico. Uma quantidade significativa de reproduções de soluções de outros métodos já foi obtida através do método simplético \cite{symemb, symemboutros, symembcor, symembdual}, esperamos que futuro desenvolvimento desse método conduza a soluções de problemas físicos atuais. Essa esperança é justificável, o método simplético  trata de forma mais moderna e direta da estrutura do espaço de fase que o método de Dirac, sendo em particular mais econômico, é portanto natural considerar que seus princípios venham a se demonstrar consideravelmente mais convenientes, ou mesmo poderosos, que as abordagens predecessoras para resolução de problemas por meio de imersão em calibre.

\chapter{Não-comutatividade espaço-temporal}
\label{chapncet}

O espaço-tempo não-comutativo (NC) têm recebido muita atenção nos últimos anos, mas sua história é antiga, Heisenberg já teria proposto a inserção de certa não-comutatividade espacial a fim de resolver problemas relacionados com a auto-energia do elétron \cite{szabo, schaposniknc}. Essas considerações em parte levaram Snyder a publicar o primeiro artigo sobre o tema \cite{snyder}. Entretanto, o problema da auto-energia do elétron veio a ser resolvido através da teoria de renormalização, cujos resultados foram tão satisfatórios que eclipsaram as primeiras idéias de não-comutatividade espacial. Essa hipótese de inserir uma nova regra de não-comutatividade parece natural pois, assim como uma teoria quântica usual não tem seus estados físicos descritos por pontos no espaço de fase, mas sim por regiões de área proporcional a $\hbar$, um espaço-tempo NC também não possuiria pontos bem definidos (\asp fuzzy physics").

Em grande parte devido à formulação matematicamente rigorosa da geometria não-comutativa na década de 80 \cite{connes} e à conexão encontrada entre teorias de cordas e o espaço-tempo NC \cite{wittennc, connesnc, outrosdnc, sw}, a proposta de um espaço-tempo NC veio a ser revista em diversos contextos. Algumas de suas propriedades são bem vindas, como a eliminação do ponto espacial como estado físico e a conexão da Lagrangiana da teoria de Yang-Mills NC com gravitação, enquanto outras ainda precisam de melhor entendimento, em particular a renormalização dessas teorias. É perfeitamente possível que futuramente, analogamente ao que ocorreu com a Lagrangeana de Yang-Mills, muitas das interpretações atuais sobre não-comutatividade espaço-temporal sejam abandonadas, mas a estrutura formal dessas teorias venha a ser essencial para uma nova física, conectada ou não com as teorias de cordas.


Este capítulo é dedicado a uma apresentação geral das teorias NC's, com enfoque aos itens pertinentes à próxima seção. A estrutura matemática dessas teorias é muito rica, há muitos trabalhos sobre o assunto, contudo aqui não será dada ênfase ao lado matemático. Introduções sobre esse assunto podem ser encontradas nas Refs. \cite{connes, introncgeo}. Há novas abordagens recentes  a essas teorias que parecem promissoras que não serão abordadas aqui, veja por exemplo  \cite{balachandranrev} (e suas referências).

\vspace{.4in}
\section{Aspectos gerais}
Desde o estabelecimento da mecânica quântica no início do século XX, o emprego de operadores associados a observáveis físicos, em vez de variáveis reais, se tornou padrão na busca pelo entendimento das leis fundamentais da Natureza. A predição de resultados experimentais adquiriu uma natureza probabilística de caráter fundamental, consistentemente com a relação de incerteza de Heisenberg, a qual impõe um limite essencial ao conhecimento  dos estados dos observáveis físicos. Neste contexto, estados físicos deixam de ser descritos por pontos no espaço de fase e passam a ser descritos por regiões desse espaço de área mínima da ordem de $\hbar$. Essa relação de incerteza é modelada, em conjunto com a definição de valor esperado da medida de observáveis, pela imposição de que coordenada e seu momento conjugado não comutam entre si, isto é\footnote{Aqui o \asp chapéu" designa operador.},
\be
	[\hat x^i, \hat p_j] = i \hbar. \; \; \; \; \; i,j=1,2,...,N.
	\label{comutmq}
\ee

As teorias de espaço-tempo NC (ou, por simplicidade, teorias NC's) têm sido estudadas com o propósito de analisar a física de teorias com álgebras mais gerias que a anterior, a saber
\be
	[\hat \xi^\alpha, \hat \xi^\beta] = i \hat \Theta^{\alpha \beta}. \; \; \; \; \; \alpha, \beta=1,2,...,2N.
\ee
Em que $\hat \xi =  (\hat x^1,..., \hat x^N, \hat p_1,...,\hat p_N)$, $\hat \Theta^{ij} \not= 0$ para alguns valores de $i$ e $j$ e $\hat \Theta^{i \; (j+N)} = \delta^i_j \hbar + O(\hat \Theta^{ij}, \hat \Theta^{(i+ N) \; (j+N)})$.

\vspace{.4in}
Há várias formas de interpretar a alteração da álgebra quântica acima mencionada. A seguir, três abordagens a essa questão (não necessariamente populares) são comentadas:

\vspace{.1in}
\noindent
\emph{i)} Considerar que a existência do objeto $\hat \Theta^{ij} \not= 0$ seja tão fundamental quanto a de $\delta^i_j \hbar$. Esse novo objeto deveria introduzir pequenos desvios nos resultados teóricos que possuem boa concordância experimental e possibilitar a resolução de algum problema.\\

\noindent
\emph{ii)}  $\hat \Theta^{ij}$ é introduzido para modelar algum processo físico desconhecido ou seqüência de interações não controlada, não tendo portanto um status de grandeza fundamental tal qual $\delta^i_j \hbar$. Desta forma, introduz-se a não-comutatividade no espaço-tempo com o intúito de criar um modelo efetivo, o qual poderia ser consistente com muitos dos fenômenos conhecidos ou só com alguns muito particulares. \\

\noindent
\emph{iii)} Como visto no Capítulo \ref{capfs}, a relação (\ref{comutmq}) só é válida de forma geral em sistemas sem vínculos. Há teorias físicas que, considerando sua estrutura de vínculos e sob certos limites, tornam-se não-comutativas no espaço-tempo, embora originalmente tenham sido formuladas em um contexto comutativo. Ou seja, nesta abordagem, não se assume $\hat \Theta^{ij} \not= 0$ \emph{a priori}, mas a não-comutatividade espaço-temporal é obtida, sob certo limite, como uma nova descrição para a teoria original.\\

Essa classificação foi acima introduzida apenas para proporcionar uma visão ampla, porém vaga, de possíveis abordagens à não-comutatividade espaço-temporal. Não há na prática uma distinção bem definida entre essas abordagens.

O estudo da não-comutatividade espaço-temporal advinda da teoria de cordas sob o limite de Seiberg-Witten \cite{sw, lambda,  gmms}, assim como a análise do mais baixo nível de Landau em um contexto NC \cite{lllgamboa, lllhatsuda} e estudos gerais de sistemas vinculados cuja quantização leve a $\hat \Theta^{ij} \not=0$ \cite{deripoincare, girottirev, ncfromsimp}, são bons exemplos da abordagem \emph{iii}. Alguns exemplos de propostas fenomenológicas são a modificação da álgebra quântica usual a fim de modificar o limite GZK \cite{gzk, colemangzk, gamboagzk}, extensões do modelo padrão por meio da inserção de alguma não-comutatividade \cite{chainchianncsm, connes-lott, seankost}, modelagem da recentemente observada rotação da luz sob forte fundo magnético \cite{lightrot, nccp} e algumas propostas de gravitação quântica NC que não se reduzem, a priori, a conseqüências de teorias de cordas no limite de Seiberg-Witten \cite{nonstringnc}\footnote{As referências acima citadas são apenas alguns poucos exemplos, não são necessariamente as mais representativas. Além disso, embora a distinção entre as três interpretações expostas possa parecer clara, na prática comumente é difícil, ou mesmo impossível, classificar satisfatoriamente um artigo segundo esses critérios.}. Algumas referências sobre aspectos gerais fenomenológicos podem ser vistas em \cite{seankost, phenrevv}.

Ocasionalmente comenta-se sobre a possibilidade de uma não-comutatividade espaço-temporal fundamental nas linhas da abordagem \emph{i}, o que não está, em princípio, em acordo com a teoria de cordas. Como hoje se entende essas teorias NC's, elas não são teorias fundamentais da Natureza, pois, por menor que seja o parâmetro associado à não-comutatividade, várias inconsistências, ou sérias dificuldades, aparecem (especialmente no que concerne a quatização dessas teorias). Problemas dessa natureza ocorrem, por exemplo, no eletromagnetismo NC \cite{hayakawa}. Conforme veremos, a teoria $U(1)$ não-comutativa é, sob vários aspectos, mais próxima das teorias do tipo $SU(N)$ do que de uma $U(1)$. É perfeitamente possível que esses modelos não venham a encontrar aplicação em fenômenos eletromagnéticos, mas sejam úteis para outros fenômenos. Em particular, o eletromagnetismo NC, como atualmente o entendemos, possui liberdade assintótica e as únicas cargas possíveis são certa constante $e$, o negativo dessa constante e zero \cite{hayakawa}. Por outro lado, e gostaríamos de deixar isto claro, não é possível prever como as teorias NC's irão se desenvolver nos próximos anos. O que há de sólido são suas relações formais, mas novas têm sido descobertas, as quais talvez possibilitem novas interpretações e uma readequação à fenomenologia eletromagnética. Enfim, não há no momento condições de se considerar a não-comutatividade espaço-temporal como algum princípio fundamental, e nem há indícios claros nessa direção. Talvez, futuramente, estando alguns modelos NC's em bom acordo com a experiência, e dispondo esses de boa consistência interna, a questão da não-comutatividade espaço-temporal ser ou não ser fundamental se torne uma questão realmente relevante para a física; por enquanto ela fornece uma estrutura útil para propor novos modelos efetivos (de efeito Hall a gravitação quântica) e estudar outras teorias sob certos limites, como a teoria de cordas no limite de Seiberg-Witten.

\vspace{.4in}
No que segue, tal como grande parte da literatura sobre o assunto, iremos tratar somente da chamada não-comutatividade canônica, isto é, do caso em que $\hat \Theta^{ij}(\hat \xi)$ é uma constante, a qual será denotada por $\theta$ e assumiremos que seus índices assumem $D$ valores em um espaço-tempo $D$ dimensional, ou seja,
\be
	\label{nccan}
	[\hat x^\mu, \hat x^\nu] = i \theta^{\mu \nu}, 
\ee
com $\mu, \nu = 0,1,...,D-1.$ Estamos considerando uma deformação do espaço $\real^D$ em um $\theta$-deformado $\real_\theta^D$ \cite{deformacao} que satisfaz a álgebra acima e, em particular, 
\be
	\real^D_\theta \; \stackrel{\theta \rightarrow 0}\longrightarrow \; \real^D.
\ee
Essa deformação $\theta$ não altera a métrica, é verdadeiramente independente dela.

De forma análoga ao que ocorre na mecânica quântica usual, o comutador de $\hat x$ com uma função $f:\real^D \rightarrow \real \; \in C^\infty$ \asp aplicada"  no operador $\hat x$ [i.e., $f(\hat x)$], define uma derivada no espaço NC dada por
\be
	\label{ncopder}
	[ \hat x^\mu, \; f (\hat x)] = i \theta^{\mu \nu} \prt_\nu f(\hat x).
\ee

Se $\theta$ for matriz degenerada, existe em certo sistema de coordenadas $\mu_0$ tal que $\theta^{\mu_0 \nu}=0 \; \; \forall \nu$, logo $[ \hat x^{\mu_0}, \; f (\hat x)] = 0$. Neste último caso, a última relação não define derivada não-comutativa alguma, o que é natural, pois trata-se de uma componente que é verdadeiramente comutativa, logo sua derivada é a usual, independendo dos comutadores. Neste caso temos $\real_\theta^D = \real_{\theta_r}^{p} \times \real^{D-p}$, em que $p$ é o posto de $\theta$, $\real_{\theta_r}^{p}$ é o espaço $p$-dimensional $\theta_r$-deformado e $\theta_r$ é a matriz regular $p\times p$ obtida a partir de $\theta$.

Usando o lema de Baker-Hausdorff, a não-localidade de $\real^D_\theta$ pode ser observada de forma mais clara, pois
\be
	\label{transnc}
	\exp (i k_\mu \hat x^\mu) \; f(\hat x) \; \exp (- i k_\mu \hat x^\mu) = f(\hat x^\mu - \theta^{\mu \nu} k_\nu).
\ee

Esta relação já indica que há dificuldades no tratamento de observáveis em teorias de calibre no espaço-tempo NC. Em breve isto será visto.

\vspace{.4in}
\section{O símbolo de Weyl e o produto Moyal}
O produto Moyal \cite{moyal} (ou Weyl-Moyal, ou Groenewold-Moyal), embora atualmente freqüentemente associado à não-comutatividade espaço-temporal, foi originalmente proposto no contexto da mecânica quântica usual. Em particular, esse produto é uma deformação não-comutativa do produto usal entre funções reais \cite{deformacao}. Na mecânica quântica, em vez dos observáveis serem descritos por operadores em um espaço de Hilbert, esses podem ser descritos por  funções complexas obtidas a partir do símbolo de Weyl \cite{weyl}. Para preservar o mapa entre operadores e funções complexas dado pelo símbolo de Weyl, um novo produto entre funções é induzido, esse é chamado de produto Moyal e é associativo mas não-comutativo. Na mecânica quântica, esse produto é escrito como uma expansão em potências de\footnote{Para ser mais preciso, é uma expansão em termos de $\hbar$ vezes a estrutura de Poisson.} $\hbar$ e é especialmente útil para análises semiclássicas \cite{osborn}. No contexto do espaço-tempo NC, esse produto é escrito como uma expansão em termos da matriz $\theta$, como veremos.

A fim de evitar certas sutilezas matemáticas que estão além do objetivo desta tese, consideraremos apenas funções de Schwartz \cite{rieffel}, isto é, funções pertencentes a $C^\infty$ que vão para zero no infinito mais rapidamente que $1/x$, tal como todas suas derivadas\footnote{Alguns comentários simples e interessante sobre a extensão do produto Moyal para outras classes de funções por ser vista na Ref. \cite{wulkenhaar}.}. 

Para dada função real\footnote{De forma mais geral, a função $f$ pode ser complexa. Esta seção segue, em grande parte, a Ref. \cite{szabo}.} $f$ de Schwartz [i.e., $f: \real^D \rightarrow \real$, com $f \in {\cal S}(\real^D)$], podemos associá-la a um operador pelo símbolo de Weyl, que é dado por
\be
	\label{simbw}
	\hat W[f] = \int \frac{d^Dk}{(2 \pi)^D} \; \tilde f(k) \; \exp (i k \cdot \hat x),
\ee
com $k \cdot \hat x := k_\mu \hat x^\mu$ e 
\be
	\tilde f(k) := \int d^D x \; f(x) \; \exp (-i k \cdot  x).
\ee
	
Vários símbolos podem ser introduzidos de forma a associar operadores a funções complexas, esses dependem da escolha da ordenação dos operadores. O símbolo de Weyl segue a ordenação de Weyl, como pode ser conferido pela expansão da exponencial de $\hat x$. Entre outas propriedades, sendo $f$ uma função real, o operador de Weyl $\hat W$ é Hermiteano; e $\hat W[f]|_{\theta=0} = f(\hat x)$, em que $\hat x$ na última equação é interpretado como uma coordenada comutativa.

Temos liberdade de introduzir um traço no espaço dos operadores, o qual escolhemos de forma a satisfazer
\be
	\Tr \; e^{i k \cdot \hat x} = (2 \pi)^D \; \delta^D(k).
\ee
Portanto,
\be
	\Tr \; \hat W [f] = \int d^Dx \; f(x).
\ee
O traço no espaço dos operadores funciona como uma integral, e isso é condizente com o limite $\theta \rightarrow 0$ de $\hat W$ comentado anteriormente. Ocasionalmente, no lugar de $\Tr$ escreve-se $\int \Tr$ para deixar claro que $\Tr$ funciona como uma integral, mas não usaremos essa notação aqui. Isso é bem diferente do comportamento do traço que aparece na ação de teorias não-Abelianas, esse último não se confunde com a integral e o traço de um elemento da álgebra no ponto $x$ é um número nesse ponto. No caso NC não há um número em dado ponto, ou mesmo em dada pequena região, o único número que advém do traço depende de todo o espaço.
  
Seja 
\be
	\hat \Delta (x) := \int \frac{d^Dk}{(2 \pi)^D} \; e^{ik \cdot \hat x} \; e^{-ik \cdot  x}.
\ee

Usando o operador $\hat \Delta$ e o traço $\Tr$ pode-se inverter a Eq. (\ref{simbw}) de forma a obter uma função real a partir de um operador Hermiteano $\hat W$, isto é,
\be
	f(x) = \Tr \; \( \hat W[f] \; \hat \Delta (x) \),
\ee
pois
\be
	\Tr \; \( \hat \Delta(x) \; \hat \Delta (y) \) = \delta^D (x-y).
\ee

\vspace{.4in}
A multiplicação de operadores indica apenas a ordem em que devem ser aplicados, isto é, a multiplicação de $\hat W[f]$ por $\hat W[g]$ é simplesmente $\hat W[f] \; \hat W[g]$. Essa multiplicação define outro operador Hermiteano, o qual deve estar associado a outra função real pelo símbolo de Weyl. De forma geral, essa nova função não pode ser dada pelo produto usual (ou qualquer outro produto comutativo) entre $f$ e $g$, assim a nova função real se relaciona a $f$ e $g$ através de um novo produto, que denotaremos por $*$. De forma mais específica,
\ba
	(f * g)(x) &:=& \hat W^{-1}(\hat W[f] \; \hat W[g]), \\[.2in]
		& = & \Tr \( \hat W[f] \; \hat W[g] \; \hat \Delta(x) \), \nonumber \\[.2in]
		& = & \int \frac {d^Dk \; d^Dk'}{(2 \pi)^{2D}} \; \tilde f(k) \; \tilde g(k'-k) \; e^{- \frac i2 \theta^{\mu \nu} k_\mu k'_\nu} \; e^{ik'_\mu x^\mu},  \\[.2in]
		& = & f (x) \; \exp \( \frac i2 \theta^{\mu \nu} \stackrel{\leftarrow}{\prt}_\mu  \stackrel{\rightarrow}{\prt}_\nu \) \; g(x),
\ea
ou seja,
\be
	(f*g)(x) = f(x)\; g(x) + \frac i2 \theta^{\mu \nu} \prt_\mu f(x) \; \prt_\nu g(x) + O(\theta^2).
\ee

O produto *, como acima definido, é o produto Moyal. Algumas propriedades imediatas desse produto:
\ba
	\label{propmb1}
	[x^\mu, x^\nu]_* &=& i \theta^{\mu \nu}, \\[.2in]
	\label{propmb2}
	[x^\mu, f(x)]_* &=& i \theta^{\mu \nu} \prt_\nu f(x), \\[.2in]
	\label{propmb3}
	[f(x), g(x)]_* &=& 2i \; f(x) \; \mbox{sen} \( \frac 12 \theta^{\mu \nu} \stackrel{\leftarrow}\prt_\mu \stackrel{\rightarrow}\prt_\nu \) g(x), \\[.2in]
	\label{propmb4}
	\{f(x), g(x) \}_* &=& 2 \; f(x) \; \mbox{cos} \( \frac 12 \theta^{\mu \nu} \stackrel{\leftarrow}\prt_\mu \stackrel{\rightarrow}\prt_\nu \) g(x).
\ea
Acima, $[ f , g ]_* := f * g - g*f$ é o comutador Moyal e $\{ f , g \}_* := f * g + g*f$ é o anticomutador Moyal.

As Eqs. (\ref{propmb1}, \ref{propmb2}) devem ser comparadas com (\ref{nccan}) e (\ref{ncopder}) respectivamente. As Eqs. (\ref{propmb3}, \ref{propmb4}) serão úteis para o próximo capítulo. 

Pode-se verificar também que o produto Moyal é associativo, ou seja,
\be
	A * (B * C) = (A * B) * C.
\ee

Como somente estamos considerando funções de Schwartz, nota-se que
\be
	\label{moyal2}
	\int A * B \; d^Dx = \int A \; B \; d^Dx.
\ee
Em particular, termos quadráticos na ação em um espaço-tempo NC são iguais, a menos de termos de superfície, a seus correspondentes comutativos. Conseqüentemente, os propagadores das teorias NC's são os mesmos das teorias comutativas \cite{filk}.

Uma ação de uma teoria NC descrita por operadores deve usar o traço $\Tr$, e esse possui a propriedade cíclica. Devido às duas últimas propriedades apresentadas, é fácil notar que uma propriedade cíclica também está presente no quadro Moyal, pois
\be
	\label{moyalcic}
	\int A * B * C \; d^Dx= \int A \; B* C \; d^Dx = \int B * C * A \; d^Dx.
\ee

\vspace{.4in}
Uma forma interessante de analisar a não-localidade do produto Moyal é considerar sua descrição integral no espaço de configuração. Seja $K$ implicitamente definido por 
\be
	(f * g)(x) = \int d^Dy \; d^Dy' \; f(y) \; g(y') \; K(y, y', x).
\ee
Sendo
\be
	f(x) = \delta_z (x) := \delta^D(x-z) \;\;\;\;\; \mbox{ e } \;\;\;\;\; g(x) = \delta_{z'} (x) := \delta^D(x-z'),
\ee
vê-se que $K$ de forma geral é dado por
\be
	K(y,y',x) = (\delta_y * \delta_{y'} )(x) =: \delta^D(y-x) * \delta^D(y'-x).
\ee

Expressando as deltas de Dirac por meio de suas transformadas de Fourrier, vem
\ba
	\label{Kfourrier}
	K(y,y',x) &=& \frac 1 {(2 \pi)^{2D}} \int d^Dh \; d^D h' \; e^{i h \cdot  (x - y)} e^{-\frac i2 \; \theta^{\mu \nu} h_\mu h'_\nu} \; e^{i h' \cdot  (x - y')},  \\[.2in]
	\label{Kinv}
	& = & \frac 1{\pi^D \; |\det \theta|} \; e^{-2i \theta^{-1}_{\mu \nu} (x-y)^\mu \; (x-y')^\nu}.
\ea
A última equação é válida somente caso $\theta$ possua inversa, mas algo similar é válido para o subespaço em que $\theta$ for regular. A Eq. (\ref{Kfourrier}) mostra que o produto Moyal é uma modificação do produto ordinário tal que nenhum parâmetro além de $\theta$ é usado (nem mesmo a métrica) e a \asp interação" $\;$ ponto a ponto entre $f$ e $g$ é quebrada, pois não é possível integrar $h$ e $h'$ de forma a obter duas deltas se $\theta \not= 0$. Note que a não-localidade do produto Moyal é bem mais interessante que uma mera interação à distância entre pontos.

A Eq.(\ref{Kinv}) ilustra uma propriedade dificilmente esperada das teorias NC's. Para $y=y'$, temos
\be
	(\delta_y * \delta_y)(x) = \frac 1{\pi^D \; |\det \theta|}.
\ee
Ou seja, $\theta$ atua como um regularizador do quadrado de deltas de Dirac em $x=0$, o que \emph{a priori} pode ser visto como bem vindo. Considerando nossas expectativas iniciais, seria em princípio desejável, ou natural, que no caso NC existisse um $a_0$ cuja norma fosse da ordem de $\sqrt{||\theta||}$ tal que $\forall \; a \in \real^D \mbox{ com }|| a || > ||a_0 ||$ tivéssemos $(\delta * \delta) (a) = 0$. Mas isso não ocorre, o quadrado da delta em um mesmo ponto no espaço NC é uma constante para todo o espaço. Portanto temos uma não-localidade infinita associada ao quadrado de deltas de Dirac. Deltas de Dirac podem ser vistas como Gaussianas de largura infinitesimal, ou seja, um \asp campo" $\;$ formado por um pulso muito localizado. O produto de deltas no mesmo ponto (Gaussianas cocêntricas), contudo, faz com que esses pulsos extremamente localizados influenciem todo o espaço.  Essa propriedade pode ser vista como um indício da mistura ultravioleta/infravermelha, conforme mostrado na Ref. \cite{pertnc}, veja também \cite{szabo, wulkenhaar, dn, extlocbars, adriv}. 



\vspace{.4in}
\section{$D$-branas e o limite de Seiberg-Witten} 
\label{lsw}

A teoria exposta até aqui não determina como proceder com a quantização das teorias de espaço-tempo NC. Seria em princípio natural agora proceder com a dedução das regras de quantização canônica a partir do caso mecânico para, por fim, se possível, tratar das regras de Feynman. Mas esse não é o caminho padrão. Grande parte do interesse atual em teorias NC's vem de sua conexão com teorias de cordas. Como será visto de forma resumida, em certo limite de baixas energias e na presença de um fundo magnético (o limite de Seiberg-Witten), as $D$-branas \cite{dbrane, firstc} são efetivamente descritas através de uma geometria NC e o produto entre campos torna-se o Moyal\footnote{Em contextos diferentes, resultados similares foram anteriormente obtidos em \cite{wittennc, connesnc}.} \cite{sw}. Contudo, as medidas de integração permanecem as mesmas. Este último ponto não é trivial ou natural sob o ponto de vista de uma mecânica quântica não-comutativa, pois em princípio a quantização por integrais de Feynman poderia ser diferente, sua motivação original vem de deduções mecânicas de um espaço comutativo. Assim sendo conclui-se que a única novidade da quantização não-comutativa por integrais de Feynman no espaço dos momentos é a presença uma fase dependente de $\theta$ e dos momentos. As teorias livres, ou de ações com somente termos quadráticos, têm as mesmas propriedades quânticas da teoria comutativa equivalente devido à Eq. (\ref{moyal2}).

Seguindo a Ref. \cite{sw}, considere a ação de Polyakov para uma folha de mundo Euclideana $\Sigma$ de um espaço-alvo de métrica induzida $(g_{\mu \nu})$ na presença de um fundo magnético constante $B$, isto é,
\ba
	\label{stringb}
	S &=& \frac 1{4 \pi \alpha'} \int_\Sigma \( g_{\mu \nu} \prt_a x^\mu \prt^a x^\nu - 2 \pi i \alpha' B_{\mu \nu} \ep^{ab} \prt_a x^\mu \prt_b x^\nu \) \nonumber \\[.2in]
	&=& \frac 1{4 \pi \alpha'} \int_\Sigma g_{\mu \nu} \prt_a x^\mu \prt^a x^\nu - \frac i2 \int_{\prt \Sigma} B_{\mu \nu} x^\mu \prt_t x^\nu,
\ea
com $a=1,2$ e $\prt_t$ a derivada tangencial à borda da folha de mundo ( $\prt \Sigma$ ). O campo $B$ funciona como uma conveniente forma de selecionar as condições de contorno que as pontas das cordas estão sujeitas, pois igualando a variação de $S$ a zero obtemos
\be
	\label{stringcont}
	\(g_{\mu \nu} \prt_n x^\nu + 2 \pi i \alpha ' B_{\mu \nu} \prt_t x^\nu \)|_{\prt \Sigma} = 0,
\ee
sendo $\prt_n$ a derivada normal a $\prt \Sigma$. Portanto, para $B=0$ as pontas das cordas satisfazem condições do tipo Neumann, enquanto $B \rightarrow \infty$ (ou $g_{\mu \nu} \rightarrow 0$) leva a condições do tipo Dirichlet. Neste último caso, as pontas das cordas estão presas a $D$-branas \cite{dbrane}\footnote{Alternativamente, pode-se impor as condições de contorno como uma condição externa à ação, isto é, sem usar o campo de Neveu-Schwartz $B$.}.

Com as condições de contorno dadas por (\ref{stringcont}), o propagador em $\prt \Sigma$, sendo $\prt \Sigma$ parametrizado por $\tau \in \real$, é \cite{fradtse, callan}
\be
	\label{stringprop}
	\< x^\mu (\tau) x^\nu (\tau') \> = -\alpha' G^{\mu \nu} \; \mbox{log} (\tau - \tau')^2 + \frac i2 \theta^{ij} \ep(\tau - \tau'),
\ee
em que $\ep(\tau)$ vale 1 para $\tau$ positivo e $-1$ para $\tau$ negativo,
\ba
	\label{spop1}
	G_{\mu \nu} &:=& g_{\mu \nu} - (2 \pi \alpha')^2 \; (B g^{-1} B)_{\mu \nu}, \\[.2in]
	\label{spop2}
	G^{\mu \nu}  &=& \( \frac 1 {g + 2 \pi \alpha' B}g \frac 1 {g- 2\pi \alpha' B} \)^{\mu \nu}, \\[.2in]
	\label{spop3}
	\theta^{\mu \nu} &:= & - (2 \pi \alpha')^2 \( \frac 1 { g + 2 \pi \alpha' B} B \frac 1 {g- 2 \pi \alpha' B} \)^{\mu \nu}.
\ea

Devido à Eq.(\ref{stringprop}), $G$ é a métrica efetiva, enquanto $\theta$ é o parâmetro de não-comutatividade, pois
\be
	\label{stringnc}
	\< [ x^\mu (\tau), x^\nu	(\tau)] \> := \< x^\mu (\tau) x^\nu(\tau^-) - x^\nu (\tau^+) x^\mu(\tau) \> = i \theta^{\mu \nu},
\ee
com $\tau^+ := \tau + \vep$, $\tau^- := \tau - \vep$ e $\vep > 0$. Isto mostra que certo tipo de não-comutatividade já está presente. Veremos que, sob certo limite, o propagador acima estabelece um ordenamento normal que é descrito por certo produto associativo, e esse é justamente o produto Moyal anteriormente apresentado. Como será apresentado, o fato de o limite a seguir induzir o produto Moyal é apenas uma de suas características não-triviais.

Considere o seguinte limite de baixas energias:
\ba
	&&\alpha' \sim \ep^{1/2} \rightarrow 0, \\ 
	&&g_{\mu \nu} \sim \ep \rightarrow 0, \; \mbox{para $\mu, \nu = 1,2,...,r$},
\ea
sendo $r$ o posto da matriz $(B_{\mu \nu})$ (a qual satisfaz $B_{\mu \nu} = 0 \; \; \forall \mu > r$, o que pode ser sempre atingido por transformações de coordenadas). Através desse limite, que neste contexto é referido por limite de Seiberg-Witten, explora-se o limite de baixas energias de teorias de cordas sem alterar os parâmetros que caracterízam as cordas abertas ($G$, $\theta$ e $G_s$), enquanto elimina-se a influência das cordas fechadas (descritas por $g$, $B$ e $g_s$). As contantes $g_s$ e $G_s$ serão em breve introduzidas. No limite de Seiberg-Witten, as Eqs. (\ref{spop1}-\ref{spop3}) são dadas por

\ba
	G_{\mu \nu} & =& \left \{ \begin{array}{ll}  - (2 \pi \alpha')^2 \; ( B g^{-1} B )_{\mu \nu}  \; \; \; & \mbox{para } \mu, \nu = 1,2,...,r \\ 
			g_{\mu \nu} & \mbox{para } \mu,\nu > r \end{array} \right. ,\\[.2in]
	G^{\mu \nu}& = & \left \{ \begin{array}{ll} - \frac 1 {(2 \pi \alpha')^2} \( \frac 1B g \frac 1B \)^{\mu \nu}  \; \; \; & \mbox{para } \mu, \nu = 1,...,r \\[.1in] 
			g^{\mu \nu} & \mbox{para } \mu,\nu > r  \end{array} \right. ,\\[.2in]
	\theta^{\mu \nu} &=& \left \{ \begin{array}{ll} \( \frac 1B \)^{\mu \nu} \; \; \; &\mbox{para } \mu, \nu = 1,2,...,r \\[.1in]
			0 & \mbox{para } \mu, \nu > r \end{array} \right. .
\ea

Agora o propagador perde a singularidade associada com $\tau \rightarrow \tau'$, pois
\be
	\label{stringpropsw}
	\< x^\mu (\tau)\; x^\nu (\tau') \> = \frac i2 \theta^{ij} \ep(\tau - \tau').
\ee
Para quaisquer duas funções do operador $x$, o ordenamento normal associado a esse propagador satisfaz  \cite{pol}
\be
	: f(x(\tau)): \; : g(x(0)): \; = \; e^{\frac i2 \; \ep(\tau) \; \theta^{\mu \nu} \frac \prt {\prt x^\mu(\tau)} \frac \prt {\prt x^\nu(0)}}\; :f(x(\tau))\; g(x(0)):
\ee 
e
\be
	\mbox{lim}_{\tau \rightarrow 0^+}\; : f(x(\tau)): \; : g(x(0)): \; = \; :f(x(0))* g(x(0)):.
\ee 

Assim, temos 
\be
	\< \prod_{n=1}^m \; f_n(x(\tau)) \> = \int d^Dx \; (f_1*f_2*...*f_m)(x(\tau)).
\ee

A relação com o operador de Weyl pode agora ser facilmente estabelecida, 
\be
	\< \prod_{n=1}^m \; f_n(x(\tau)) \>  =   \Tr \hat W[f_1*...*f_m] = \Tr \prod_{n=1}^m \hat W[f_n].
\ee
 
\vspace{.4in}
É interessante analisar como teorias de calibre se comportam nesse espaço não-comutativo. Campos de calibre se acoplam à borda da folha de mundo $\prt \Sigma$ ($Dp$-brana) por meio do termo
\be
	\label{stringgaugecoup}
	- i \int d \tau A_\mu(x) \prt_\tau x^\mu,
\ee
$\mu = 1,2,...,p+1$, que é adicionado à ação (\ref{stringb}). 

Considerando que $A$ esteja associado a uma simetria de calibre ordinária,
\be
	\label{varord}
	\delta A_\mu = \prt_\mu \lambda,
\ee
a variação clássica de (\ref{stringgaugecoup}) é nula, pois
\be
	\delta \int d \tau A_\mu \prt_\tau x^\mu = \int d \tau \prt_\tau \lambda = 0.
\ee

De forma geral, porém, é necessário avaliar a variação da função partição, a qual em primeira ordem em $A$ se transforma por
\be
	\int d \tau \; : A_\mu \prt_\tau x^\mu: \;  \int d\tau' \; :\prt_{\tau'} \lambda:.
\ee

Esse produto de operadores pode ser regularizado de várias formas. Em particular, usando a regularização de Pauli-Villars, o produto acima coincide com a variação clássica e é nulo. Uma outra regularização de interesse é a de separação de pontos (\asp point splitting"). Nesta, a região $|\tau - \tau'| < \delta$ é eliminada da integração, em seguida o limite $\delta \rightarrow 0$ é tomado. Assim procedendo,
\ba
	& \int d \tau \; : A_\mu (x(\tau)) \prt_\tau x^\mu: \; : (\lambda(x(\tau^-)) - \lambda(x(\tau^+))): \;  =  \nonumber \\[.2in]
	\label{firstvarf}
	& = \int d \tau \; :\( A_\mu * \lambda - \lambda * A_\mu \) \; \prt_\tau x^\mu :.
\ea
Devido à não-comutatividade do produto Moyal, essa regularização quebra a simetria $U(1)$ presente no caso clássico. Por outro lado, a regularização de separação de pontos está associada a uma outra simetria de calibre. Em vez da variação (\ref{varord}), deve-se cosiderar a seguinte\footnote{Embora só a primeira ordem tenha sido avaliada em (\ref{firstvarf}), essa transformação de calibre é válida para todas as ordens da expansão da função partição; os detalhes se encontram em \cite{sw}.}
\be
	\label{calnc}
	\hat \delta \hat A_\mu = \prt_\mu \hat \lambda + i[ \hat \lambda, \hat A_\mu]_*.
\ee
O símbolo \asp  $\; \hat{} \;$" acima foi empregado para diferenciar os campos regularizados por Pauli-Villars dos regularizados por separação de pontos, ou seja, esse \asp chapéu" não designa operadores, serve apenas para diferenciar os campos que pertencem à álgebra $u(1)$ dos que pertencem a $u_*(1)$ (i.e., álgebra $u(1)$ NC). Os grupos NC's serão introduzidos em outra seção.

A ação mais simples possível capaz de conferir dinâmica ao campo de calibre $\hat A \in Dp$-brana de métrica $G$ e que seja invariante por transformações do tipo $U_*(1)$ é
\be
	 \label{acaoncu1}
	 S_{\hat A} = k_p \; \int \hat F \wedge_* \str \hat F.
\ee
Acima, $k_p$ é uma constante que depende da dimensão da brana associada, $\wedge_*$ é o produto exterior $*$-deformado (e.g., $A \wedge_* B = A_\mu * B_\nu \; dx^\mu \wedge dx^\nu$, para duas quaisquer 1-formas) e $\hat F$ é dado por
\be
	\hat F := d \hat A - i \hat A \wedge_* \hat A.
\ee
Essa construção segue de perto a contrução usual de teorias de calibre não-Abelianas, embora no presente caso tenhamos uma teoria Abeliana. 

Assim sendo, a descrição efetiva NC depende de $\theta$ somente através do produto Moyal $*$. Conforme minuciosa e originalmente argumentado na Ref. \cite{sw}, ação (\ref{acaoncu1}) é a ação efetiva de uma $D$-brana com campos de calibre de posto um, regularizada por separação de pontos e no limite de Seiberg-Witten. Para o caso não-Abeliano, o aspecto da ação (\ref{acaoncu1}) é o mesmo, exceto pela presença de um traço no espaço da álgebra.

\vspace{.4 in}
Por outro lado, desde 1985 \cite{dbifromstrings} sabe-se que $D$-branas no limite de campos lentamente variantes ($ \sqrt{ 2 \pi \alpha '} |\frac{\prt F}{F}| \ll 1$) podem ser descritas por ações de Dirac-Born-Infeld (DBI) (para revisões veja por exemplo Refs. \cite{revdbi, pol}). Usando esse resultado e nesse limite, a Lagrangiana efetiva é 
\be
	\label{dbic}
	\cl_{DBI} = \frac 1 {g_s (2 \pi)^p (\alpha')^{\frac {p+1}2}} \sqrt{\det [g + 2 \pi \alpha'(B+F)]},
\ee
sendo $g_s$ uma contante que não depende da dimensão da $D$-brana no espaço, isto é, do valor de $p$.

É importante notar que nessa Lagrangiana $B$ e $F$ ocorrem exclusivamente na soma $B+F$, o que leva à chamada simetria $\Lambda$: as transformações $A \rightarrow A + \Lambda$ e $B \rightarrow B - d\Lambda$ não alteram a Lagrangiana. É natural que essa simetria esteja presente na Lagrangiana efetiva acima, pois ela se encontra na ação (\ref{stringb}) adicionada do termo de calibre (\ref{stringgaugecoup}), para $B$ constante\footnote{A presença de $B$ pode ser interpretada como uma modificação no potencial, $A \rightarrow A^B$, com $A^B_\mu := A_\mu + \frac 12 B_{\nu \mu} x^\nu$. Sendo $F^B := dA^B$, vem $F^B = F + B$.}.

A Lagrangiana (\ref{dbic}) é dada em um espaço comutativo, sendo compatível com a regularização de Pauli-Villars. Consideremos o problema de determinar a Lagrangiana efetiva análoga para o caso NC. Como visto, no espaço NC os produtos são substituídos por produtos Moyais e esses carregam toda a influência do campo magnético B. Além disso a métrica efetiva é dada por $G$, portanto\footnote{Essa expressão é independente do limite de Seiberg-Witten.}
\be
	\label{dbinc}
	\cl_{DBI*} = \frac 1 {G_s (2 \pi)^p (\alpha')^{\frac {p+1}2}} \sqrt{\det (G + 2 \pi \alpha'\hat F)}.
\ee

Dado que as Lagrangianas acima só tem sentido como teorias efetivas no limite de campos lentamente variantes, os produtos Moyais que aparecem na expansão do determinante da Lagrangiana NC podem ser ignorados, assim toda a influênia de $B$ encontra-se no produto Moyal interno a $\hat F$. A análise da simetria $\Lambda$ no quadro NC é menos óbvia, mas pode ser encontrada nas Refs. \cite{sw, backgroundseiberg, lambda}.

A relação entre as constantes $g_s$ e $G_s$ é encontrada mediante a igualdade 
\be
	\cl_{DBI}(F = 0) = \cl_{DBI*}(\hat F=0),
\ee
ou seja,
\be
	G_s = g_s \( \frac{ \det G}{ \det (g + 2 \pi \alpha' B)} \)^{1/2} = g_s \( \frac{ \det (g + 2 \pi \alpha' B)}{\det g } \)^{1/2}.
\ee 

E a relação de $G_s$ com a constante de acoplamento de calibre, aqui denotada por $g_{YM}$, é obtida  a partir da análise do termo de segunda ordem em $\hat F$ da expansão de $\cl_{DBI*}$, ou seja,
\be
	\label{gymcordas}
	\frac 1 {g_{YM}^2} = \frac{(\alpha')^{\frac {3-p}2}}{(2 \pi)^{p-2} G_s}.
\ee

Em particular, nota-se que $g_{YM}^2$ só é adimensional em D3-branas, em conformidade com a invariância de escala de Yang-Mills em 4D. Em uma D2-brana, $g_{YM}^2$ tem dimensão de massa, tal qual é exigido por uma análise dimensional de uma teoria de calibre $U(1)$ [ou $U(N)$] em 3D, se os campos de calibre tiverem dimensão de massa.

A Eq. (\ref{gymcordas}) fixa o valor de $k_p$ em (\ref{acaoncu1}) como sendo [veja Eq. (\ref{symc})]
\be
	k_p = \frac 1{2 g_{YM}^2} =  \frac{(\alpha')^{\frac {3-p}2}}{2(2 \pi)^{p-2} G_s}.
\ee
 
Esse limite de campos lentamente variantes (CLV) será importante para algumas considerações sobre o mapa de Seiberg-Witten e certos resultados apresentados no próximo capítulo.

\vspace{.4in}
\textbf{Resumo.} i) Não-comutatividade espaço-temporal ocorre naturalmente em teorias de cordas na presença de um fundo magnético (\ref{stringnc}); ii) o limite de Seiberg-Witten é, em particular, um limite de baixas energias que elimina a física das cordas fechadas, as cordas abertas têm condições de contorno do tipo Dirichlet, o produto Moyal é diretamente induzido pelo ordenamento normal e a ação de calibre efetiva é quadrática em\footnote{Em particular a ação efetiva pode depender de ordens superiores de $\hat F$. No limite de campos lentamente variantes, sem considerar o limite de Seiberg-Witten, ordens superiores de $\hat F$ ocorrem, veja a Eq. (\ref{dbinc}).} $\hat F$; iii) dependendo da técnica de regularização, a ação efetiva em uma $Dp$-brana pode ser tanto descrita por uma teoria de calibre ordinária quanto por uma NC.

\vspace{.4in}	
\section{Tópicos sobre quantização}
\label{topq}

Segundo o procedimento padrão de quantização de campos no espaço-tempo NC, as expectativas inicias de eliminar divergências ultra-violetas (UV) por meio de não-comutatividade espaço-temporal são ingênuas. Teorias NC's de campos sofrem das mesmas divergências das teorias ordinárias e ainda possuem outras divergências, essas associadas a outros gráficos de Feynman que não ocorrem no caso ordinário, os gráficos não-planares \cite{filk}. Os gráficos não-planares estão associados à mistura ultra-violeta/infra-vermelho (UV/IV) \cite{pertnc} e uma abordagem direta usando as ferramentas usuais da teoria de campos não é capaz de tratar esse dipo de divergência. Há propostas referentes a essa questão, a abordagem ao problema pode variar bastante, veja \cite{uvirdiv}. Não é claro se essa mistura UV/IV é um problema a ser resolvido através de novas técnicas, ou se é uma \asp patologia" que deve ser eliminada por meio de outras abordagens à quantização.

Se a mistura UV/IV seria dificilmente prevista nos primórdios da  não-comutatividade espaço-temporal, dificuldades referentes à causalidade ou unitariedade de teorias NC's com $\theta^{0i} \not= 0$ poderiam ser esperadas \cite{uni, cas}. Essas teorias são às vezes referidas por teorias de tempo/espaço NC e contém não-localidade temporal. Há essencialmente três abordagens a essa questão: i) qualquer teoria NC realista deve sempre satisfazer $\theta^{0i} = 0$ \cite{uni}; ii) qualquer teoria realista deve satisfazer $\theta^{\mu \nu} \theta_{\mu \nu} > 0$ ($\theta$ deve ser do tipo espaço), assim pode-se sempre mudar de referencial para um em que $\theta^{0i}=0$ \cite{seankost}; iii) supõe-se que o problema encontra-se na forma de quantização, assim outras abordagens poderiam ser realistas e evitar o probleman de unitariedade \cite{altqua}. Pouco depois do aparecimento do artigo \cite{uni}, foi visto que há certa classe de teorias de tempo/espaço NC que não tem problemas de unitariedade, mesmo usando a quantização padrão, trata-se das teorias em que $\theta$ é do tipo luz $\theta^{\mu \nu} \theta_{\mu \nu} = 0$ \cite{lightlikenc}. Sob uma perspectiva de teoria de cordas, há realmente dificuldades de se produzir uma teoria NC de campos com não-comutatividade temporal \cite{notimespace}.

\vspace{.4in}
Segundo os preceitos gerais apresentados nas seções anteriores, a versão NC de $\lambda \phi^4$ deve ser dada por
\ba 
	S_{\phi *} &=& \Tr  \( \hat W [\prt_\mu \phi] \hat W [\prt^\mu \phi] + \frac {m^2}2 \hat W^2[\phi] + \frac \lambda {4!} \hat W^4[\phi] \), \nonumber \\[.2in]
	&=& \Tr \( \frac 12 \hat W[ \prt_\mu \phi *  \prt^\mu \phi] + \frac {m^2}2 \hat W[\phi*\phi] + \frac \lambda {4!} \hat W[\phi*\phi*\phi*\phi] \), \nonumber \\[.2in]
	&=& \int \( \prt_\mu \phi \; \prt^\mu \phi + \frac {m^2}2 \phi \phi + \frac \lambda {4!} \phi*\phi*\phi*\phi \)d^Dx.
\ea
Não há produtos Moyais nos dois primeiros termos devido à propriedade (\ref{moyal2}). Assim, o propagador \asp livre" da teoria $\lambda \phi^4$ NC é igual ao da versão NC. A diferença do caso NC advém do termo
\be
	\Tr \hat W^4 [\phi] = \prod_{a = 1}^4 \int d^Dk \; \tilde \phi (k_a) \; \delta^D\( \sum_{a=1}^4 k_a \) \; V(k_1, k_2, k_3, k_4),
\ee
com
\be
	V(k_1, k_2, k_3, k_4) := \prod_{a <b}^4 e^{- \frac i2 k_a \wedge k_b}
\ee
e
\be
	k_a \wedge k_b := k_{ai} \theta^{ij}k_{bj}.
\ee

Devido à conservação do momento, ou seja, à delta que aparece acima, $V$ é invariante por permutações cíclicas, embora, em geral,
\be
	V(k_1, k_2, k_3, k_4) \not= V(k_2, k_1, k_3, k_4).
\ee
Isso faz com que seja necessário prestar atenção ao ordenamento dos propagadores associados a cada vértice. Como regra adota-se o ordenamento cíclico para todos os vértices. Como veremos, dependendo da ordem de associação dos propagadores em relação ao ordenamento dos vértices, a contribuição de $V$ pode ser uma fase global ou uma não-trivial fase  interna ao gráfico. O primeiro caso é representado por diagramas planares, o segundo por não-planares, conforme será ilustrado a seguir.

As regras de Feynman para o propagador e para o vértice são
\pagebreak

\vspace{.2in}

\begin{figure*}[htbp]
	\centering
	\includegraphics[width=0.30\textwidth]{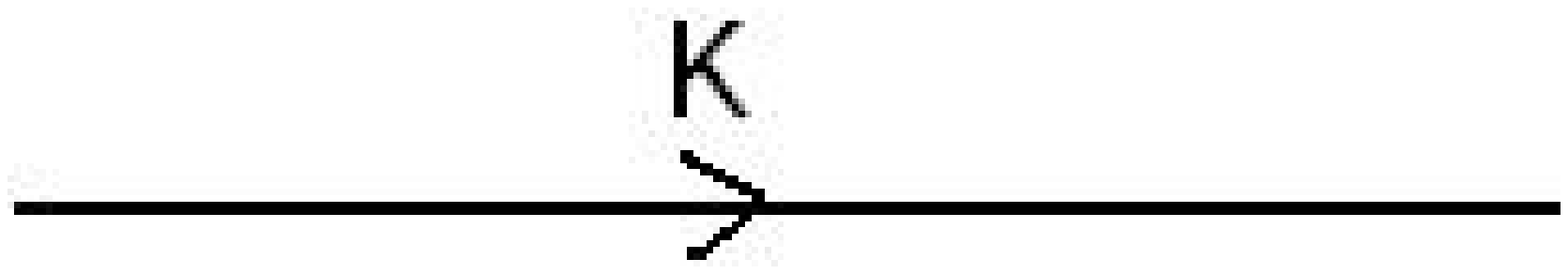}
	\hspace{1.1in} \includegraphics[width=0.20\textwidth]{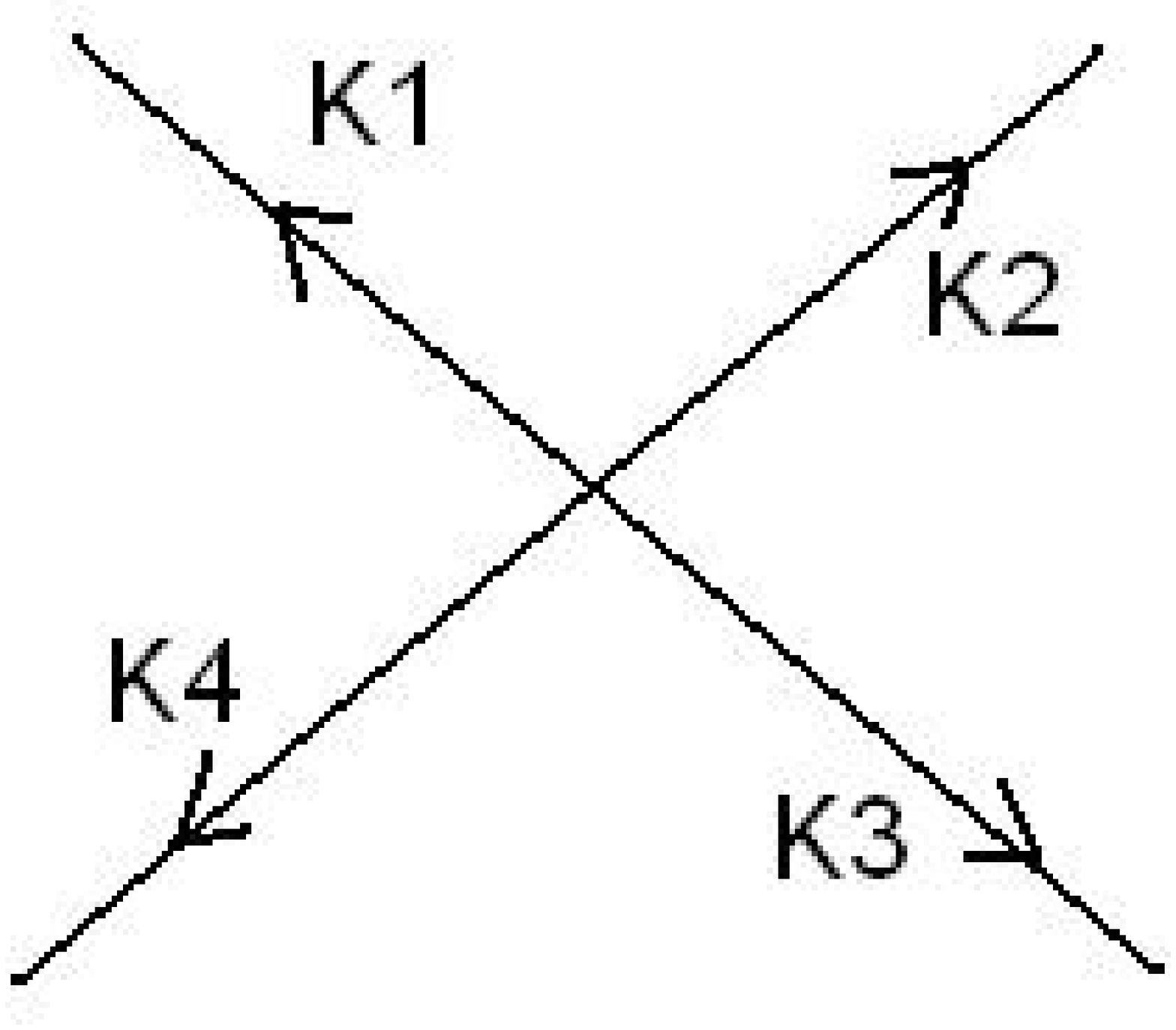}
\end{figure*}
\vspace{-.5in}

$$\mbox{\hspace*{.4in}Fig.3: }\frac 1 {k^2 + m^2} \mbox{\hspace{1.1in} Fig.4: } \frac \lambda {4!} \; \prod e^{-\frac i2 k_a \wedge k_b}\; \delta(\sum k_a)$$

\vspace{.3in} 

Considere os seguintes diagramas

\vspace{.2in}

\begin{figure*}[h]
	\centering
	\includegraphics[width=0.30\textwidth]{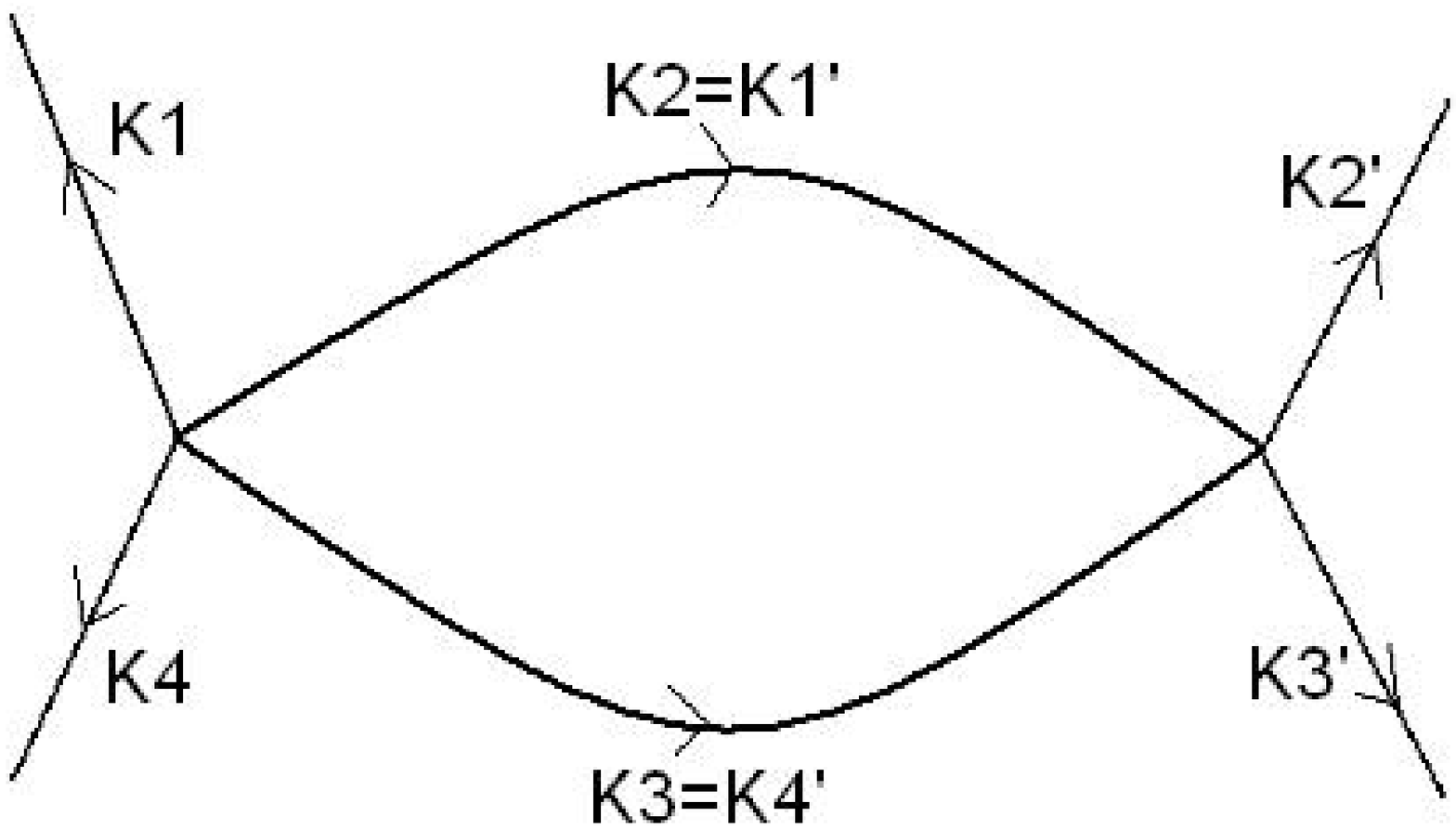}
	\label{fig:graf3}
	\hspace{1in} \includegraphics[width=0.40\textwidth]{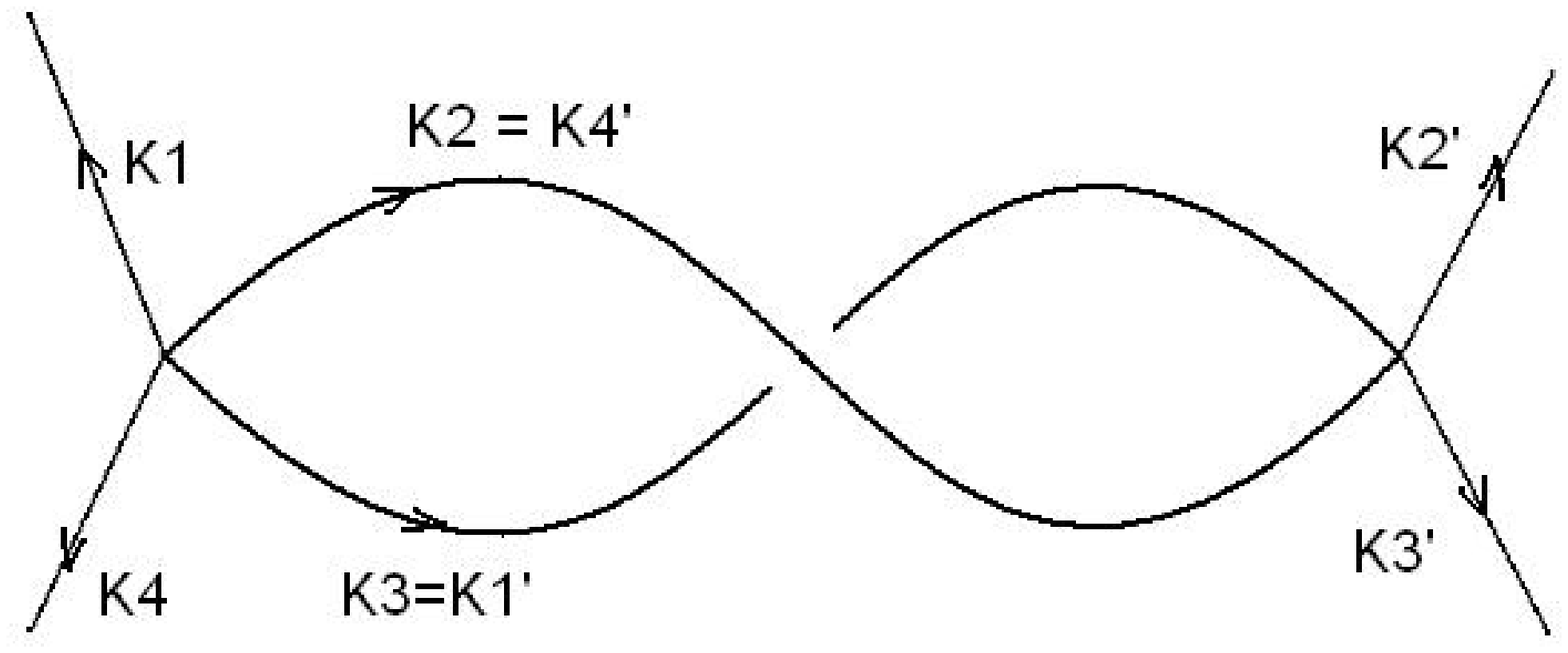}
\end{figure*}
\vspace{-.2in}

\begin{center}
	Fig.5: Diagrama Planar \hspace{1.1in} Fig.6: Diagrama Não-Planar
\end{center}

\vspace{.2in}

A contribuição dos vértices do diagrama não-planar é 
\ba
	&V(k_1, k_2, k_3, k_4) \; V(-k_3, k_2', k_3', -k_2) \; \stackrel{ \mbox{ \tiny on  shell}}= \nonumber \\
	& \; \; \; \; \; \; \; \; \; \; \; \; \; \; \; \; \; \; V(k_2, k_3, k_4, k_1) \; V(-k_2, -k_3, k_2', k_3').
\ea

Seja $k := k_1 + k_4$, logo $k = -k_2' - k_3'$ e $k_3 = -k - k_2$. Assim,
\be
	V(k_2, k_3, k_4, k_1) \; V(-k_2, -k_3, k_2', k_3') = e^{-\frac i2 k \wedge (2 k_2 + k_1 + k_2')}.
\ee

Contrariamente ao caso acima, a contribuição dos vértices para o diagrama planar é tal que os momentos internos se cancelam, ou seja, a única dependência associada à não-comutatividade reside nos momentos externos, a saber:
\ba
	& V(k_1, k_2, k_3, k_4) \; V(-k_2, k_2', k_3', -k_3)  \stackrel{ \mbox{ \tiny on  shell}}= \nonumber \\ 
	& V(k_2, k_3, k_4, k_1) \; V(-k_3, k_2, k_2', k_3') = e^{- \frac i2 k \wedge(k_1 + k_2')}.
\ea

Conseqüentemente, os gráficos planar e não-planar acima representam respectivamente as seguintes expressões:
\be
	\frac {\lambda^2}{4!^2} e^{- \frac i2 k \wedge (k_1 + k_2')} \int \frac 1 {k_2^2 + m^2} \frac 1{(k+k_2)^2 + m^2} d^Dk_2,
\ee

\be
	\frac {\lambda^2}{4!^2} e^{- \frac i2 k \wedge (k_1 + k_2')} \int \frac 1 {k_2^2 + m^2} \frac {e^{-i \; k \wedge k_2}}{(k+k_2)^2 + m^2} d^Dk_2.
\ee

O que foi aqui apresentado para esses dois gráficos pode ser, sem maiores dificuldades, generalizado para quaisquer gráficos de teorias do tipo $\phi^n$ \cite{filk, dn, szabo}.

Na teoria $\phi^4$ comutativa há 6 gráficos de auto-energia em 1 \emph{loop} todos equivalentes entre si. No caso NC, dois desses são do tipo não-planares e as auto-energias em 1 \emph{loop} são dadas por
\be
	\Pi_p^{(1)}(p) = \frac 13 \int \frac {d^Dk}{(2 \pi)^D} \frac 1{k^2 + m^2},
\ee
\be
	\Pi_{np}^{(1)}(p) = \frac 16 \int \frac {d^Dk}{(2 \pi)^D} \frac {e^{i k \wedge p}}{k^2 + m^2}.
\ee	
Acima, $p$, em vez do $k$ usado anteriormente, é o momento total externo.

A primeira expressão é exatamente a que ocorre no caso ordinário, estando portanto sujeita a todas as complicações usuais. A fase que aparece em $\Pi_{np}^{(1)}$ funciona como um cutoff desde que o momento externo $p$ seja suficientemente grande. Não sendo esse o caso, $\Pi_{np}$ adquire a divergência UV de $\Pi_p$, porém agora sob a forma de uma divergência de longa distância (\asp UV/IR mixing") \cite{pertnc}.

Para melhor quantificar essa e outras propriedades quânticas, costuma-se usar a parametrização de Schwinger
\be
	\frac 1 {k^2 + m^2} = \int^\infty_0 d \alpha \; e^{- \alpha (k^2 + m^2)},
\ee
assim
\be
	\Pi_{np} = \frac 1 {6 (4 \pi)^{D/2}} \int^\infty_0 \frac {d \alpha}{\alpha^{D/2}} e^{- \alpha m^2 - \frac {p \circ p}{4 \alpha} - \frac 1 {\Lambda^2 \alpha}},
\ee
em que $\Lambda$ é o cutoff empregado no caso usual e 
\be
	p \circ p := p_\mu \theta^{\mu \nu} \delta_{\nu \lambda} \theta^{\lambda \kappa} p_\kappa.
\ee

Esse não convencional produto aparece naturalmente ao se avaliar a Gaussiana que aparece na integração de $k$ em $\Pi_{np}$. $p \circ p $ é uma notação padrão no contexto de teorias NC's de campos.

Nas Refs. \cite{uni, cas, lightlikenc} foi demonstrado que uma condição necessária para que as integrais de Feynman convirjam de forma a preservar unitariedade é $p \circ p \ge 0$, o que em geral não ocorre no espaço de Minkovisky se $\theta^{0i} \not=0$.

\vspace{.4in}
\section{Teorias $U_*(N)$ e o mapa de Seiberg-Witten}
\label{teoriasunnc}
A fim de se construir extensões NC's de teorias de calibre, o primeiro passo é definir os grupos apropriados. O grupo $U(N)$ possui uma extensão NC natural que denotaremos por $U_*(N)$ [grupo $U(N)$ NC]. Entretando, embora $U(N) = U(1) \times SU(N)$, não existe em princípio uma versão NC do grupo $SU(N)$. Isso se deve a $\det (A * B) \not= \det A * \det B$ em geral. Há porém formas de contornar essa dificuldade e assim tratar de teorias NC's com grupos de calibre diferentes do $U_*(N)$ \cite{outrosgruposnc, algenvuniveoutrosgrupos}. Após a definição do grupo $U_*(N)$, definiremos a teoria de calibre correspondente e veremos que não será possível definir os observáveis tão diretamente como ocorre nas teorias $U(1)$ ou $SU(N)$, pois transformações de calibre em particular envolvem translações espaciais (\ref{transnc}). Veremos que há certo mapa capaz de quebrar a interpolação entre translações espaciais e transformações de calibre, e esse mapa é exatamente o mesmo que associa a descrição ordinária com a não-comutativa de $D$-branas, conforme indicado na Seção \ref{lsw}.   

Diremos que uma matrix $N \times N$ complexa $g$ é elemento de $U_*(N)$ se satisfizer 
\be
	g * g^\dagger = \id.
\ee

O conjunto $U_*(N)$ é um grupo, pois, para dados $g,h,j \in U_*(N)$,
\ba
	&& i)\; g*(h*j) = (g*h)*j,  \\
	&& ii) \; \id * g = g * \id = g, \\
	&& iii) \; \exists \; g^{-1}= g^\dagger, \\
	&& iv) \; (g*h) \in U_*(N).
\ea  

Assim como no caso $U(N)$, os geradores são dados por matrizes Hermiteanas. Pela definição do produto Moyal, é imediato conferir que, dados $A$ e $B$ Hermiteanos, 
\be
	(i[A, B]_*)^\dagger = i[A,B]_*.
\ee

Pode-se checar que, além da constante de estrutura anti-simétrica $f$, a fim de expressar o comutador acima em função de certa base Hermiteana $\{ t^a \}$, é necessário também uma constante de estrutura simétrica $d$. Outros detalhes sobre a álgebra $u_*(N)$ não serão importantes para o objetivo desta tese, mas podem ser encontrados na Ref. \cite{algenvuniveoutrosgrupos} (veja \cite{franz} para uma revisão).

Analogamente ao que foi apresentado na Seção \ref{tu2}, mas substituindo o produto usual pelo Moyal, constrói-se a seguinte ação para a teoria $U_*(N)$ de calibre:
\be
	 \label{acaoncun}
	 S_{\hat A} = \frac 1 {2g^2} \; \int \tr \; \hat F \wedge_* \str \hat F.
\ee
Acima $\wedge_*$ é o produto exterior $*$-deformado (e.g., $A \wedge_* B = A_\mu * B_\nu \; dx^\mu \wedge dx^\nu$, para duas quaisquer 1-formas), o $\tr$ acima é um traço na álgebra e $\hat F$ é dado por
\be
	\hat F := d \hat A - i \hat A \wedge_* \hat A.
\ee
O \asp $ \; \; \hat{} \; $ " acima designa apenas campos que pertencem a uma álgebra NC, $\hat A$ e $\hat F$ não são operadores. $\hat W[\hat A]$ e $\hat W [\hat F]$, logo abaixo novamente utilizados, são operadores.

Sendo $S \in U_*(N)$, $\hat A$ e $\hat F$ se transformam segundo
\be
	\hat A \rightarrow \hat A' = S*\hat A*S^\dagger +i S * dS^\dagger
\ee
\be
	\hat F \rightarrow \hat F' = S*\hat F*S^\dagger
\ee

É importante notar que mesmo no caso $U_*(1)$ as transfomações de calibe têm uma estrutura não-Abeliana. Em particular, $\hat F$ não é um observável, pois se transforma por mudanças de calibre. Algo similar ocorre nas teorias $SU(N)$, conforme visto na Seção \ref{tu2}. Naquela seção foi possível construir um observável a partir do traço de $F$. O mesmo ocorre no caso $U_*(1)$, porém precisa-se do traço no espaço dos operadores, ou seja,
\be
	\Tr \hat W [\hat F] \rightarrow \Tr (\hat W[S] \; \hat W[\hat F] \; \hat W^\dagger[S]) = \Tr \hat W [\hat F].
\ee
Equivalentemente, usando a propriedade (\ref{moyalcic}),
\be
	\int \hat F \; d^Dx \rightarrow \int \; S * \hat F * S^\dagger \; d^Dx = \int \hat F \; d^Dx.
\ee
Para $U_*(N)$ precisa-se dos dois traços, $\tr$ e $\Tr$.
	
Como o traço $\Tr$ precisa ser empregado, a abordagem acima não constrói observáveis locais. É natural que observáveis locais em uma teoria NC não possam ser facilmente construídos, pois, sendo $S \in U_*(N)$, em particular podemos escrever $S = \exp_*(i \lambda(x))$, o qual está, por meio do símbolo de Weyl, diretamente associado ao operador de translações espaciais, veja a Eq. (\ref{transnc}).

Observáveis locais podem ser determinados em uma teoria NC através de caminhos menos diretos. Uma das possibilidades é usar uma variação do formalismo de \emph{loops} de Wilson para espaços NC's \cite{wilsonloopsnc}. A outra é utilizar certo mapa entre teorias de calibre ordinárias com as NC's, chamado de mapa de Seiberg-Witten \cite{sw}. A conexão entre essas duas abordagens foi explorada nas Refs. \cite{liustartrek}. Nesta tese, somente a segunda possibilidade será explorada.

\vspace{.4in}
Os resultados da Seção \ref{lsw} indicam que deve haver algum mapa que relacione uma teoria de calibre ordinária com uma NC. Existindo esse mapa, será possível definirmos observáveis locais usando o mesmo procedimento de uma teoria ordinária. Necessariamente esse mapa deve satisfazer
\be
	\label{eqorbitas}
	\hat \delta_{\hat \lambda} \hat A (A) = \hat A(A + \delta_\lambda A) - \hat A(A);
\ee
ou seja, qualquer variação infinitesimal da conexão $A$, que seja uma transformação de calibre, induz, necessariamente, uma transformação infinitesimal de calibre em $\hat A$. A recíproca também procede. É importante notar que não estamos exigindo que os parâmetros $\lambda$ e $\hat \lambda$ sejam iguais, $\hat \lambda$ é em princípio uma função qualquer de $\lambda, \theta$ e $A$.  

Como órbitas de calibre são topológicas (pois são conseqüências imediatas da identidade de Bianchi) e a equação acima descreve certo mapa local entre órbitas de calibre, vamos usar a condição adicional de que o mapa dado por $\hat A (A)$ somente pode depender de $A$ e de $\theta$; qualquer parâmetro adicional, o que inclui a métrica, não deve aparecer\footnote{Essa será uma condição útil para o mapa ser construído. Contudo, embora fugindo da \asp filosofia" de um mapa entre estruturas topológicas, pode-se afrouxar essa condição e permitir que outros parâmetros apareçam no mapa, dentre eles a métrica. Em $4D$, curiosamente, todas as contribuições adicionais em primeira ordem em $\theta$ ou são nulas ou são termos de superfície, quando aplicados na ação de YM NC; mas em $3D$ aparecem contribuições não triviais \cite{hr}. Neste trabalho vamos sempre aderir à \asp filosofia" de que o mapa $\hat A(A)$ só deve depender de $A$, $\prt A$ e $\theta$ (esses são os únicos fatores que as identidades de Bianchi ordinária e NC dependem).}. Sob essas condições, para $\theta$ suficientemente pequeno e com a condição de contorno $\hat A|_{\theta = 0} = A$, o mapa mais geral entre $\hat A$ e $A$ é (para $N=1$)
\be
	\label{ansatzsw1}
	\hat A_\mu (A) = A_\mu + \theta^{\alpha \beta} A_\alpha (k_1 \; \prt_\mu A_\beta + k_2 \; \prt_\beta A_\mu) + k_3 \theta^{\alpha \beta} \prt_\alpha \prt_\mu A_\beta.
\ee
As constantes $k_1$, $k_2$ e $k_3$ serão avaliadas a seguir.

Como
\ba
	\label{varcalinfo}
	\delta_\lambda A_\mu &=& \prt_\mu \lambda, \\
	\label{varcalinfnc}
	\hat \delta_{\hat \lambda} \hat A_\mu &=& \prt_\mu \hat \lambda + i[ \hat \lambda, \hat A_\mu]_*,
\ea
usando a Eq. (\ref{eqorbitas}), vem
\ba
	\prt_\mu \hat \lambda - \theta^{\alpha \beta} \prt_\alpha \lambda \; \prt_\beta A_\mu &=& \tilde A_\mu + \theta^{\alpha \beta} \tilde A_\alpha (k_1 \; \prt_\mu \tilde A_\beta + k_2 \; \prt_\beta \tilde A_\mu) + \nonumber \\
	\label{expswc}
	&&+ k_3 \; \theta^{\alpha \beta} \prt_\alpha \prt_\mu \tilde A_\beta - \hat A(A),
\ea

\ba
	\prt_\mu f &=& \theta^{\alpha \beta} \prt_\alpha \lambda \prt_\beta A_\mu (k_2 +1) + k_1 \theta^{\alpha \beta}(\prt_\alpha \lambda \prt_\mu A_\beta + A_\alpha \prt_\mu \prt_\beta \lambda + \prt_\alpha \lambda \prt_\mu \prt_\beta \lambda ) + \nonumber \\
	\label{prtf}
	&& + k_2 \theta^{\alpha \beta} (A_\alpha \prt_\beta \prt_\mu \lambda + \prt_\alpha \lambda \prt_\beta \prt_\mu \lambda).
\ea
Em que $\tilde A := A + \prt \lambda$ e $f$ é dado por
\be
	\hat \lambda (\lambda, A) = \lambda + f(\lambda, A).
\ee

A fim de que possa-se integrar $f$ em (\ref{prtf}), vem
\be
	k_2 = -1.
\ee
Além disso, os termos de segunda ordem nas derivadas do parâmetro de calibre $\lambda$ devem ser desprezados. Já seria de se esperar que certas restrições ao comportamento do parâmetro de calibre seriam necessárias, dada a grande diferença entre teorias $U(1)$ e $U_*(1)$. A última, embora Abeliana, tem uma estrutura não-Abeliana e ainda mistura transformações de calibre com transformações espaciais. 

Conseqüentemente, 
\be
	\prt_\mu f = \theta^{\alpha \beta} [k_1(\prt_\alpha \lambda \prt_\mu A_\beta + A_\alpha \prt_\mu \prt_\beta \lambda) - (A_\alpha \prt_\beta \prt_\mu \lambda)],
\ee	
cuja solução é
\be
	k_1 = \frac 12
\ee
e
\be
	f = \frac 12 \theta^{\alpha \beta} \; \prt_\alpha \lambda \; A_\beta.
\ee
	
Curiosamente, a constante $k_3$ ficou livre, pois a Eq. (\ref{eqorbitas}) independe dela. Por outro lado, a ação do eletromagnetismo NC mapeada por Seiberg-Witten também independe dela, pois $k_3$ está associada a um termo de superfície \cite{asakawa, hr}. Em acordo com a Ref. \cite{sw}, vamos escolher $k_3=0$. 

Por fim, para $N=1$ e em primeira ordem em $\theta$, o mapa de Seiberg-Witten é dado por
\ba
	\label{swm11}
	\hat A_\mu &=& A_\mu + \theta^{\alpha \beta} A_\alpha \( \frac 12 \; \prt_\mu A_\beta -  \prt_\beta A_\mu \), \\[.2in] 
	\label{swm11lambda}
	\hat \lambda &=& \lambda + \frac 12 \theta^{\alpha \beta} \; \prt_\alpha \lambda \; A_\beta.
\ea

\vspace{.4in}
Para $N$ qualquer, deve-se usar
\ba
	\label{varcalinfoN}
	\delta_\lambda A_\mu &=& \prt_\mu \lambda + i[ \lambda, A_\mu], \\
	\label{varcalinfncN}
	\hat \delta_{\hat \lambda} \hat A_\mu &=& \prt_\mu \hat \lambda + i[ \hat \lambda, \hat A_\mu]_* \\
	& = & \prt_\mu \hat \lambda + i [ \hat \lambda, \hat A_\mu] - \frac 12 \theta^{\alpha \beta} \{\prt_\alpha \hat \lambda, \prt_\beta \hat A_\mu \} + O(\theta^2). \nonumber
\ea
A presença do anti-comutador acima não é tão inesperada quanto pode parecer a princípio. No início desta seção, foi mencionado que em teorias não-Abelianas NC's precisa-se de duas constantes de estruturas, uma anti-simétria e outra simétrica. O produto $\hat \lambda  \hat A$ não pode ocorrer acima diretamente [como ocorre na expansão em primeira ordem da Eq. (\ref{varcalinfnc})], pois esse produto é proporcional ao quadrado dos geradores da álgebra, enquanto os demais termos da expressão são lineares nos geradores. Assim, só restam duas alternativas: ou emprega-se o anti-comutador desse produto ou o seu comutador. O último caso não é condizente com $N=1$ (\ref{varcalinfnc}, \ref{expswc}), mas o primeiro é.

Considerando a discussão sobre mapeamento de órbitas de calibre apresentada, no lugar de (\ref{ansatzsw1}) temos\footnote{Além do termo associado a $k_3$ que aparece em (\ref{ansatzsw1}), outras possibilidades coerentes, e em particular compatíveis com o caso $N=1$, são $k'_3\theta^{\alpha \beta} \prt_\mu [A_\alpha, A_\beta]$, $k_4 \theta^{\alpha \beta} \prt_\alpha [A_\mu, A_\beta]$ e $k'_4\theta^{\alpha \beta} [A_\alpha, [A_\mu, A_\beta]]$. Nenhum desses termos é essencial para a existência do mapa, e não são considerados na Ref. \cite{sw}. Maiores detalhes sobre as conseqüências desses termos encontram-se em \cite{asakawa}.}
\be
	\label{ansatzswN}
	\hat A_\mu (A) = A_\mu + \theta^{\alpha \beta} \{ A_\alpha, k_1 \; \prt_\mu A_\beta + k_2 \; \prt_\beta A_\mu  + k'_2 [A_\beta, A_\mu]\}.
\ee
Conseqüentemente,
\ba
	\prt_\mu \hat \lambda & \!+ & \! i[ \hat \lambda, \hat  A_\mu] - \frac 12 \theta^{\alpha \beta} \{\prt_\alpha  \lambda, \prt_\beta  A_\mu \} = \tilde A_\mu + \nonumber \\[.2in]
	\label{avswN}
	&& + \; \theta^{\alpha \beta} \{ \tilde A_\alpha, k_1 \; \prt_\mu \tilde A_\beta +   k_2 \; \prt_\beta \tilde A_\mu  + k'_2 [\tilde A_\beta, \tilde A_\mu]\} - \hat A_\mu,
\ea
com $\tilde A := A + \prt \lambda + i[\lambda, A]$.

Usando artifícios análogos\footnote{Para $N>1$, a seguinte identidade pode ser útil: $\{A,[B,C]\} + \{[A,B],C\} + [\{C,A\},B]=0$.} aos já apresentados, descobre-se que o mapa de Seiberg-Witten para qualquer $N$ e em primeira ordem em $\theta$ é dado por
\ba
	\label{swmN1}
	\hat A_\mu &=& A_\mu - \frac 14 \theta^{\alpha \beta} \{ A_\alpha,  \prt_\mu A_\beta +  F_{\beta \mu}\}, \\[.2in] 
	\label{swmN1lambda}
	\hat \lambda &=& \lambda + \frac 14 \theta^{\alpha \beta} \{ \prt_\alpha \lambda, A_\beta\}.
\ea

\vspace{.4in}
Segundo a análise feita na Seção \ref{lsw}, $\theta$ não é necessariamente muito pequeno, portanto deve haver alguma forma de generalizar o mapa apresentado em primeira ordem para qualquer ordem. Uma forma elegante de extensão do mapa apresentado é reinterpretar o $\theta$ anteriormente utilizado como uma variação infinitesimal $\delta \theta$ entre duas órbitas de calibre, uma de parâmetro $\theta$ e a outra com $\theta + \delta \theta$ \cite{sw}. Desta forma as Eqs.(\ref{swmN1}, \ref{swmN1lambda}) se tornam equações diferenciais cuja solução é um mapa $\hat A(A)$ para qualquer ordem em $\theta$. Seguindo esse princípio, as Eqs. (\ref{swmN1}, \ref{swmN1lambda}) são promovidas para
\ba
	\label{swmNg}
	\delta \hat A_\mu &=&  - \frac 14 \delta \theta^{\alpha \beta} \{ \hat A_\alpha,  \prt_\mu \hat A_\beta +  \hat F_{\beta \mu}\}_*, \\[.2in] 
	\label{swmNglambda}
	\delta \hat \lambda &=&  \frac 14 \delta \theta^{\alpha \beta} \{ \prt_\alpha \hat \lambda, \hat A_\beta\}_*.
\ea
As deltas que aparecem acima no lado esquerdo das igualdades são iguais a $\delta \theta \frac \prt {\prt \theta}$.  Para resolver as equações acima para dada ordem em $\theta$, expande-se $\hat A$ e $\hat \lambda$ em potências de $\theta$, assim, com a condição de contorno $\hat A|_{\theta = 0} = A$, encontra-se uma relação recursiva cujos termos de primeira ordem em $\theta$ são dados por (\ref{swmN1}, \ref{swmN1lambda}). Não é difícil conferir diretamente que as Eqs. (\ref{swmNg}, \ref{swmNglambda}) são condizentes com (\ref{eqorbitas}). Note que para esse caso a equação análoga a (\ref{avswN}) é obtida pela substituição dos $\theta$'s da expansão por $\delta \theta$ e de todos os comutadores e anti-comutadores por suas versões Moyais. Como não é necessário expressar esses (anti-)comutadores em função das constantes de estrutura para obter o mapa de Seiberg-Witten de $N$ arbitrário, toda a demonstração segue sem maiores alterações para o caso da expansão em $\delta \theta$.


\chapter{Dualidades eletromagnéticas no espaço-tempo não-comutativo}
\label{cap6}

Extensões das dualidades vistas na Seção \ref{tu4} para o caso NC não são triviais e suas conseqüências dificilmente podem ser consideradas como esperadas. As teorias de calibre $U_*(1)$ têm uma estrutura similar às teorias não-Abelianas, assim, em particular, mesmo sem fontes são teorias com interação. Entretanto, devido ao mapa de Seiberg-Witten, teorias $U_*(1)$ podem ser mapeadas em teorias ordinárias de calibre. Embora a teoria mapeada seja também não-linear, por ser uma teoria $U(1)$, vale a identidade de Bianchi ordinária $dF=0$, o que será de grande ajuda para empregar os métodos apresentados na Seção \ref{tu4}. 

Neste capítulo trataremos de duas dualidades de natureza eletromagnética. Primeiramente veremos o caso em 3D com massa topológica até a primeira ordem em $\theta$ com enfoque à Ref. \cite{nossoNCMCS}, depois o caso sem massa topológica tanto em 4D quanto em 3D, seguindo a Ref. \cite{issues}. Mesmo na época da publicação do artigo \cite{nossoNCMCS}, já havia outros que tratavam dessa dualidade \cite{ghosh, ghoshm, cm, cm2, dayi, rob}, porém com conclusões distintas e, em certos casos, incompatíveis. Curiosamente, a teoria dual encontrada em \cite{nossoNCMCS} não é idêntica a nenhuma anteriormente apresentada, embora compatível com as Refs. \cite{dayi, rob}. Posteriormente sua resposta foi explicitamente confirmada em \cite{hr}.

É importante notar que a dualidade eletromagnética não-massiva não é um caso particular da com massa topológica, deve-se lembrar que a primeira, em 3D, relaciona vetores com escalares, enquanto a segunda relaciona vetores com vetores. Em \cite{issues}, motivados especialmente pelos resultados em 4D das Refs. \cite{ganor, aschieri}, resolvemos estudar essa dualidade no contexto tridimensional e avaliar a necessidade do limite de campos lentamente variantes (CLV). Entender o funcionamento das dualidades eletromagnéticas é essencial para o entendimento das teorias $U_*(1)$, pois, em particular, essas dualidades conectam teorias sem não-comutatividade temporal a teorias com não-comutatividade temporal. Esse artigo \cite{issues} foi o primeiro a tratar da dualidade NC tridimensional não-massiva e o primeiro a avaliar a partir de que ordem o limite de CLV se faz necessário a fim de preservar certa regra de dualidade ($\theta \rightarrow g^2 \str \theta$) \cite{ganor, gmms} e manter o caráter $S$ da dualidade.

\vspace{.4in}
\section{Sobre as extensões não-comutativas do modelo autodual induzidas por dualidade}
\label{sec6ad}

O termo de Chern-Simons não-comutativo (CS NC) tem sido estudado em vários contextos desde o final da década de 90 \cite{nccsantigo, susskindnccs, inducedcschu, gs} e é dado por
\be
	S_{CS*} = \frac m{2g^2} \int \tr \(A \wedge_* dA - \frac {2i}3 A \wedge_* A \wedge_* A \).
\ee
Esta é uma extensão natural para o modelo de CS [veja a Eq.(\ref{CSnA})], ademais ela pode ser também obtida por meio de bosonização do modelo de Thirring NC \cite{gs, inducedcschu} e  é motivada por resultados de teoria de cordas \cite{nccsdbranes}.

Dado o termo de Chern-Simons acima, constrói-se imediatamente o modelo NC de Maxwell-Chern-Simons (NC MCS). No contexto de dualidades, dado que a teoria MCS é equivalente à autodual, uma questão natural é se há um modelo autodual NC e se esse é equivalente ao modelo MCS NC. Considerando a Eq. (\ref{acaoadf}), tem-se a seguinte generalização (veja também \cite{gs}): 
\be
	\label{ad*}
	S_{ad*} =  a_f  \int \tr \[ f \wedge_* \str f + \frac 1{m}\( f \wedge df - \frac {2i}3 f \wedge_* f \wedge_* f \)\].
\ee 

Na Ref. \cite{cm2} mostra-se a equivalência entre as ações $S_{ad*}$ e $S_{MCS*}$ em primeira ordem em $\theta$, contudo os autores usam uma hipótese não ortodoxa, também usada nas Refs. \cite{ghoshm}, referente à existência de um mapa análogo ao mapa de Seiberg-Witten para teorias sem simetria de calibre ($\hat f = f + O(\theta)$). Mais recentemente \cite{rob, hr} demonstrou-se de forma direta que as ações $S_{ad*}$ e $S_{MCS*}$ não são equivalentes entre si. Em particular, a Ref. \cite{hr} mostra por meio de uma aplicação do mapa de Seiberg-Witten\footnote{Para empregar um mapa de Seiberg-Witten em $S_{ad*}$, campos auxiliares foram inseridos de forma a implementar uma simetria de calibre, seguindo o resultado de \cite{abf}.} que, mesmo em primeira ordem em $\theta$, as ações $S_{ad*}$ e $S_{MCS*}$ mapeadas não são equivalentes.

Embora não se possa empregar diretamente um mapa de Seiberg-Witten para a ação (\ref{ad*}), a teoria MCS NC induz por dualidade uma espécie de teoria autodual NC mapeada, como veremos. Essa questão foi explorada nas Refs. \cite{dayi, ghosh, nossoNCMCS, hr}, as quais usam em linhas gerais a seguinte técnica: a partir da versão mapeada por Seiberg-Witten de MCS NC, certa técnica de dualidade é empregada para encontrar a teoria dual, a qual, no limite $\theta \rightarrow 0$ torna-se a teoria autodual ordinária. Note que essa teoria autodual NC assim induzida não é necessariamente autodual e não há indícios de relações com algum modelo de Thirring NC; contudo, é uma questão relevante tanto para um melhor entendimento das teorias $U_*(1)$ quanto para possíveis futuras aplicações. 

As teorias autoduais obtidas nas Refs. \cite{nossoNCMCS, hr} são idênticas entre si, o que inclui o mapa entre os campos do modelo MCS NC com os do autodual NC. Todavia os modelos autoduais obtidos anteriormente nas Refs. \cite{dayi, ghosh} diferem entre si e dos dois útilmos citados. Em particular, em \cite{ghosh} afirma-se que a teoria autodual induzida por dualidade é, ao menos em primeira ordem em $\theta$, igual à autodual ordinária; tese essa que é expressa de forma contundente no título \asp Maxwell-Chern-Simons Theory is Free for Marginally Noncommutative Spacetimes". Nosso resultado, obtido por meio da técnica projeção dual \cite{dual}, embora diferente, é compatível com o da Ref. \cite{dayi}, mas descredita a tese de \cite{ghosh}. Devido à existência na época de dois resultados diferentes \cite{dayi, ghosh}, os quais também diferiam do nosso, depois de estabelecida a dualidade, checamos explicita e diretamente a correspondência entre os parênteses generalizados e a correspondência das equações de movimento. Nas próximas subseções detalharemos os passos e as conseqüências da Ref. \cite{nossoNCMCS}.

\vspace{.4in}
\subsection{A projeção dual do modelo MCS não-comutativo}

A Lagrangiana do modelo MCS NC é dada por 

\be
	\label{ncmcs}
	{\cal L}_{MCS*}= - \frac 1 {4g} \hfu * \hfd + \frac m {2g} \epu \( \hamd * \prt_\nu \hald - \frac {2i}3 \hamd * \hand * \hald \),
\ee
com
\be
	\hfd := \prt_\mu \hand - \prt_\nu \hamd - i \hamd * \hand + i \hand * \hamd,
\ee
$\mu, \nu, \lambda = 0,1,2$ e métrica $(g_{\mu \nu}) = \diag \pmatrix { + & - & -}$. Na notação anteriormente usada, acima deveria aparecer $g^2$ no lugar de $g$, aqui estamos seguindo a notação de \cite{nossoNCMCS}.

Usando o mapa de Seiberg-Witten em primeira ordem em $\theta$ (\ref{swm11}), a Lagrangiana ${\cal L}_{MCS*}$ torna-se
\be
	g \lnm = -\frac 14 \[ \fmnu \fmnd + 2 \theta^{\alpha \beta} \( F_{\mu \alpha} F_{\nu \beta} \fmnu - \frac 14 F_{\alpha \beta} \fmnu \fmnd \) \] + \frac m2 \epu \amd \prt_\nu \ald\, .
\ee
O termo de CS NC é mapeado no termo usual de CS, não só na primeira ordem, como um cáculo direto demonstra, mas também em todas as ordens em $\theta$ \cite{gs}.

No que segue, tal como feito em \cite{dayi, nossoNCMCS}, vamos nos restringir ao caso natural em que o tempo é uma coordenada comutante $\theta^{0i}=0$. Essa restrição não é realmente necessária para a obtenção dos próximos resultados, como comentado em \cite{ghosh} e provado de forma mais geral em \cite{issues}, comentaremos sobre essa questão na próxima subseção. Assim, $\lnm$ pode ser reescrita como
\ba
	g\lnm &=& -\frac 14 \fmnu \fmnd + \frac m2 \epu \amd \prt_\nu \ald - \frac 18 \theta^{\alpha \beta} F_{\alpha \beta} \fmnu \fmnd \nonumber \\[0.2in]
	\label{e1}
	&=& - \frac 12 \( 1+ \theta \tilde F_0 \) \tilde F^\mu \tilde F_\mu + \frac m2 A^\mu \tilde F_\mu,
\ea
em que $\theta := \theta^{12}$ e $\tfmu := \frac 12 \epu F_{\nu \lambda} = \epu \prt_\nu A_\lambda$.

A fim de proceder com a técnica da projeção dual \cite{dual}, primeiramente introduzimos um campo auxiliar $\pi^\mu$ como segue
\be
	\label{e2}
	g \lnm = \pi^\mu \tfmd + \frac 12 \( 1 - \theta \tilde F_0 \) \pi^\mu \pi_\mu + \frac m2 A_\mu \tfmu.
\ee
Como comentado na Seção \ref{tu4}, esse procedimento é apenas uma transformada de Legendre de parte da Lagrangiana. A equivalência entre as Eqs. (\ref{e1}) e (\ref{e2}) é verificada por meio da substituição das equações de movimento de $\pi^\mu$ em (\ref{e2}), ou seja,
\be
	\label{dp1}
	\pi_\mu  = - \( 1 + \theta \tilde F_0 \) \tfmd.
\ee

Nosso objetivo agora é realizar redefinições de campos\footnote{Por redefinições de campos subentende-se que os novos campos introduzidos se relacionam com os originais de forma bijetiva, ou seja, trata-se do análogo de transformações de coordenadas para campos. Em particular, essas redefinições são transformações canônicas.} a fim de separar a Lagrangiana $\lnm$ em uma parte de puro calibre e outra sem simetria de calibre e que carrega toda a dinâmica. E isso realmente é possível, conforme veremos. Introduzindo
\be
	\label{dp2}
	\chi_\mu := \pi_\mu - \frac 12 \theta \delta^0_\mu \pi^\alpha \pi_\alpha,
\ee
temos
\be
	\label{e4}
	g \lnm = \( \chi^\mu + \frac m2 A^\mu \) \tfmd + \frac 12 ( 1 +  \theta \chi_0 ) \chi^\mu \chi_\mu.
\ee

Agora façamos uma \asp translação" em $\chi^\mu$, definindo o campo $p^\mu$,
\be
	\label{dp3}
	p^\mu := \chi^\mu + \frac m2 A^\mu,
\ee
assim
\be
	g \lnm = p^\mu \tfmd + \frac 12 \( p - \frac m2 A \)^\nu \( p - \frac m2 A \)_\nu \(1 + \theta \(p_0 - \frac m2 A_0 \) \).
\ee

A equação acima está pronta para a aplicação do último passo da projeção dual. Sejam
\ba
	A_\mu &=& A^+_\mu + A^-_\mu, \nonumber \\
	\label{dp4}
	p_\mu &=& \frac m2 (A^+_\mu - A^-_\mu).
\ea
Essa escolha desacopla os campos, como pode ser visto a seguir: 
\be
	g \lnm = \frac m2 \epu A^+_\mu \partial_\nu A^+_\lambda + \frac {m^2}2
(1 - m\theta A^-_0)A^-_\mu A^{-\mu} - \frac m2 \epu  A^-_\mu \partial_\nu A^-_\lambda.
\ee
O termo puro de CS é puro calibre, não propaga grau de liberdade algum.  Entretanto, ele é responsável pela simetria de calibre observada no modelo original. A parte da Lagrangiana acima dependente de $A^-$ field possui um grau de liberdade e nenhuma simetria de calibre. Essa carrega o mesmo conteúdo dinâmico da teoria dada por $\lnm$. Portanto, chamamos essa parte da Lagrangiana como modelo autodual NC induzido por dualidade, a qual pode ser escrita como
\be
	\label{ncsd}
	\lns =  \frac 1{2g} \( 1 -   \theta f_0 \) f_\mu f^{\mu} - \frac 1{2mg} \epu  f_\mu \partial_\nu f_\lambda,
\ee
após a substituição 
\be
	\label{dp5}
	m A^-_\mu \rightarrow f_\mu.
\ee

O mapa entre os campos $A$ e $f$ é encontrado ao se \asp desfazer" as passagens (\ref{dp1}, \ref{dp2}, \ref{dp3}, \ref{dp4}, \ref{dp5}), sendo dado por
\be
	\label{e10}
	f^\mu = \tf^\mu +  \tf^\mu \tf^\alpha \tth_\alpha + \frac 12 \tth^\mu \tf^\alpha  \tf_\alpha
\ee
e, conseqüentemente,
\be
	\label{e11}
	\tf^\mu = f^\mu -  f^\mu f^\alpha \tth_\alpha -\frac 12 \tth^\mu f^\alpha f_\alpha,
\ee
em que  $\tth_\mu := \frac 12 \epd \theta^{\nu \lambda}$, logo $\tth_0 = \theta$ e $\tth_i = 0$.

Com a última passagem o método da projeção dual foi concluído e obtivemos êxito em determinar a teoria dual. Na próxima subseção apresentamos maiores detalhes sobre esse resultado e comparação com os demais existentes. 

A Ref. \cite{ghosh} critica o uso de Lagrangianas mestras se não acompanhadas de uma verificação da álgebra (i.e., dos parênteses generalizados, do setor simplético). Esse procedimento nos parece redundante, contudo, devido aos conflitos existentes, e embora a projeção dual posssa ser vista como um procedimento mais transparente que o da Lagrangiana mestra, em \cite{nossoNCMCS} consideramos conveniente checar explicitamente a correspondência das álgebras e das equações de movimento perante o mapa (\ref{e10}). Apresentaremos aqui também essa parte.

\vspace{0.4in}
As equações de movimento dos modelos dados por (\ref{ncmcs}) e  (\ref{ncsd}) são respectivamente
\be
	\ep_{\mu \nu \lambda} \partial^\nu \(-\tf^\lambda - \frac 12 \tth^\lambda \tf^\alpha  \tf_\alpha - \tth^\alpha  \tf_\alpha \tf^\lambda + m A^\lambda \) = 0,
\ee
\be
	f_\mu - \frac 1m \epd \partial^\nu f^\lambda - \frac 12 \tth_\mu f^\alpha  f_\alpha -  \tth^\alpha f_\alpha f_\mu = 0.
\ee
A existência de uma correspondência entre elas segue diretamente da correspondência entre seus campos encontrada em (\ref{e10}), como pode ser diretamente verificado, o que prova que ambos os modelos possuem a mesma dinâmica clássica, apenas a expressam através de diferentes campos. Veremos na próxima subseção que a dinâmica quântica desses modelos também é equivalente.

A álgebra do modelo MCS NC mapeado em primeira ordem em $\theta$ foi calculada na Ref. \cite{ghosh} [veja sua Eq. (14)]. Para encontrar a álgebra do modelo autodual NC usaremos o método simplético \cite{mfj, mbw}. Logo após a primeira iteração, observa-se que a matriz (pré-)simplética é singular. Seu modo-zero leva ao seguinte vínculo
\be
	- f_0 + \frac 1{m} \ep_{ij} \prt^i f^j + \frac 12 \theta f^i  f_i + \frac 32 \theta f_0 f_0 = 0.
\ee

Neste modelo, a forma mais rápida de encontrar os parênteses generalizados é considerar $f_0$ como função de $f_i$, resolvendo o vínculo acima. Paralelamente, pode-se também dar continuidade ao algoritmo do método simplético e inserir esse vínculo no setor cinético. Ver-se-á que a iteração seguinte produzirá uma matriz simplética não-degenerada e os parênteses generalizados (obtidos a partir das componentes de sua inversa) serão também dados por
\ba
	\{ f_0(\vx), f_0(\vy) \}_* & = & g \theta \( f_i(\vx) + f_i(\vy) \) \prt_{(x)}^i \dirac, \nonumber \\
	\{ f_0(\vx), f^i(\vy) \}_* & = & g\( 1 + 3 \theta f_0(\vx) \) \prt_{(x)}^i \dirac + mg\theta \ep^{ij} f_j (\vx) \dirac,  \\
	\{ f^i(\vx), f^j(\vy) \}_* & = & - mg \, \ep^{ij} \dirac. \nonumber
\ea
O asterisco acima associado aos colchetes indica apenas que se trata de parênteses generalizados (nenhuma relação neste caso com o produto Moyal).

Usando a prescrição dada pela Eq. (\ref{e10}), encontramos a seguinte álgebra para nosso modelo autodual NC:
\ba
	\{ B(\vx), B(\vy) \}_* & = & 0, \nonumber \\
	\{ E^i(\vx), B(\vy) \}_* & = & g\ep^{ij}(1 + \theta B(\vx))\prt_j^{(x)} \dirac, \\
	\{ E^i(\vx), E^j(\vy) \}_* & = & -gm\ep^{ij}(1 + 2\theta B)\dirac - g\theta ( \ep^{kj} E^i(\vx)+ \ep^{ki} E^j(\vy)) \prt^{(x)}_k \dirac, \nonumber
\ea
em que $E^i := - \ep^{ij} \tf_j$ e $B := - \tf_0$. Como esperado, vê-se que há coincidência com a álgebra da teoria MCS NC mapeada por Seiberg-Witten, originalmente calculada na Ref\footnote{A menos do coeficiente $g$, que se encontra nessa referência ausente desde o início, e de um erro de digitação de um sinal.}. \cite{ghosh} [veja sua Eq. (14)].

\vspace{.4in}
\subsection{Comentários finais}
\label{comfin2}

Como inicialmente anunciado, foi possível estabelecer um modelo autodual NC induzido por dualidade. Em \cite{hr}, através de outros métodos, a mesma dualidade foi encontrada com o mesmo mapa, embora usando a função partição e sem a particularização $\theta^{0i}=0$. Essa nossa particularização é desnecessária para obter a dualidade, isso mais tarde ficou claro ao obtermos certa relação mais geral em \cite{issues} [veja a Eq. (\ref{rel}) na próxima seção]. Como  mencionado em \cite{ganor} em um contexto similar, em primeira ordem em $\theta$ essa dualidade é automaticamente válida na função partição. Essa propriedade foi apresentada de forma mais detalhada em \cite{issues}, justificando o motivo de \cite{nossoNCMCS} e \cite{hr} terem obtido a mesma resposta, embora o primeiro tenha usado somente argumentos de ordem clássica. Na próxima seção, que é dedicada à Ref. \cite{issues}, esse ponto será retomado.

Nota-se que a Lagrangiana (\ref{ncsd}) é diferente da apresentada em \cite{dayi} [sua Eq. (17)]. Infelizmente, nessa última referência, não se encontra explícita a resposta final, só há uma indicação. Todavia, dando continuidade aos seus cálculos, nota-se que sua Lagrangiana do modelo autodual que aparece em sua Eq. (17) não depende de um termo cúbico em $f$ sem derivadas como na nossa Eq.(\ref{ncsd}), mas sim de um termo cúbico em $\prt f$. Por esse motivo, o autor sugeriu que essa dualidade induziria uma modificação no termo de CS, como diz o título \asp \emph{(...) a new noncommutative Chern-Simons theory in d=3}". Embora não tenhamos obtido essa  modificação no termo de CS, nossa resposta não nos parece ser incompatível com a de \cite{dayi}. Há ambigüidades, mesmo em primeira ordem em $\theta$, na definição do modelo autodual induzido por dualidade. O mapa entre $f$ e $A$ induzido pelo procedimento da Ref. \cite{dayi} difere do nosso mapa\footnote{Infelizmente, esse mapa não se encontra explícito nessa referência, mas pode ser deduzido seguindo suas indicações.} (\ref{e10}). Como ambos resultados parecem perfeitamente corretos, as Lagrangianas autoduais induzidas pela dualidade obtidas em \cite{dayi, nossoNCMCS} são equivalentes entre si em primeira ordem em $\theta$, a despeito de suas diferenças aparentes. Esse ponto parece merecer uma análise mais geral.

Como antes comentado, a Ref. \cite{ghosh} também apresenta um modelo autodual induzido por dualidade diferente do nosso. Essa referência não usa nenhuma técnica de dualidade específica. Após certas manipulações, conclui que a teoria autodual induzida é igual à autodual ordinária, isto é, não depende de $\theta$. Um mapa explícito dessa dualidade não é fornecido nessa referência, ademais não conseguimos encontrar qualquer mapa consistente com essa proposta. Se tal mapa existisse, em particular teríamos que o modelo MCS NC mapeado por Seiberg-Witten em primeira ordem em $\theta$ seria equivalente ao modelo MCS ordinário. Como sabe-se que o termo de CS NC é mapeado no termo de CS ordinário \cite{gs}, isso implicaria dizer que a Lagrangiana do eletromagnetismo NC mapeada por Seiberg-Witten é equivalente à Lagrangiana do eletromagnetismo ordinário, o que não pode ser correto por vários motivos; em particular, na versão mapeada do eletromagnetismo, mesmo em primeira ordem em $\theta$, há quebra de isotropia \cite{seankost}.

\vspace{.4in}
\section{Dualidades eletromagnéticas não-comutativas em 3D, 4D e o limite de CLV}
\label{sec6clv}

Conforme visto na Seção \ref{tu4}, a dualidade eletromagnética em 4D associa duas teorias de calibre do tipo $U(1)$ cujas Lagrangianas são idênticas, a menos da inversão da constante de acoplamento $g^2$. No espaço-tempo NC 4D em primeira ordem em $\theta$, como originalmente apresentado na Ref. \cite{ganor} através do emprego do mapa de Seiberg-Witten, dualidades eletromagnéticas relacionam duas teorias $U_*(1)$ que diferem  pela troca de $\theta$ por $g^2 \; \str \theta$. Esse resultado é mais que uma mera curiosidade, pois sugere um sério problema de consistência. Em 4D, sendo $\theta$ do tipo espaço, isto é, $\theta^{\mu \nu}\theta_{\mu \nu} > 0$, necessariamente $\str \theta$ é do tipo tempo (pois $\str\str \theta \wedge \str \theta = - \str \theta \wedge \theta$). Devido aos resultados das Refs. \cite{uni, cas}, como comentado na Seção \ref{topq}, em princípio estaria-se obtendo uma correspondência física entre uma teoria sem problemas de unitariedade com uma com problemas de unitariedade. O resultado da Ref. \cite{ganor} é apenas um indicativo da presença das dificuldades mencionadas, pois a dualidade foi obtida apenas na expansão em primeira ordem em $\theta$. Mesmo assim, as conseqüências desse resultado tecnicamente simples não são triviais, outros trabalhos, por esse motivados e se atendo aos mesmos limites, apareceram depois \cite{4Dswduality}.

Pouco depois da publicação da Ref. \cite{ganor}, ainda no ano 2000, foi publicado um estudo exato em $\theta$ (sem usar o mapa de Seiberg-Witten) sobre dualidade $S$ em espaços NC's num contexto de teoria de cordas\cite{gmms}. A dualidade $S$ nesse contexto induz uma espécie de extensão NC da dualidade de Montonen-Olive para Super-Yang-Mills em 4D com ${\cal N}=4$. Os autores obtiveram evidências para a conjectura da existência da extensão dessa dualidade, a qual, em particular, impõe para todas as ordens que $\theta$ se transforma em $g^2 \str \theta$ via dualidade, como originalmente indicado pela Ref. \cite{ganor}. Nesse contexto os autores verificam que essa dualidade $S$ não conecta duas teorias de campos NC's, pois teorias de campos com não-comutatividade temporal não são obtidas como teorias efetivas de cordas. 

Nas Refs. \cite{aschieri}, a dualidade eletromagnética NC em 4D é analisada sob o limite de campos lentamente variantes (CLV) \cite{sw, esw} num contexto de teoria de campos. Os autores usam o fato de, nesse limite, o mapa de Seiberg-Witten associar duas Lagrangianas de DBI (\ref{dbic}), uma comutativa e outra NC, assim verificam que a regra $\theta \rightarrow g^2\; \str \theta$ é exata em $\theta$ (no limite de CLV e usando o mapa de Seiberg-Witten). Considerando os resultados das Refs. \cite{ganor, gmms, uni, cas, lightlikenc} é proposto em \cite{aschieri} que o parâmetro $\theta$ de teorias $U_*(1)$ fisicamente consistentes tem de ser do tipo luz ($\theta^{\mu \nu}\theta_{\mu \nu} =0$). Essa restrição, contudo, pode ser eliminada se outras abordagens à quantização de teorias NC's forem usadas \cite{altqua}.

A extensão dessa dualidade NC para o caso 3D e uma análise da necessidade do limite de CLV, em um contexto de teoria de campos e por meio do emprego do mapa de Seiberg-Witten, é feita em \cite{issues}. Deve ser notado que os argumentos de \cite{ganor} e outros dependem da dimensão do espaço-tempo, em particuar, somente em 4D $\theta$ e seu dual $\str \theta$ são ambos 2-formas. Em \cite{issues}, a teoria escalar dual a $U_*(1)$ é obtida até segunda ordem em $\theta$ e é mostrado que até essa ordem a regra $\theta \rightarrow \tilde \theta := g^2 \str \theta$ pode ser naturalmente estendida para 3D. A partir da terceira ordem, de forma geral, para qualquer dimensão $D \ge 2$, não é possível escrever a ação dual como função de $\tilde \theta$ sem recorrer a $\theta$ ($ \propto \str \tilde \theta$), o que quebra a regra anterior e a versão \asp clássica" de dualidade $S$, pois a constante de acoplamento da teoria dual não é uma simples inversão da teoria original. Contudo, no limite de CLV, para todas as ordens em $\theta$, seja em 4D ou 3D, a teoria dual é expressa por meio de $\tilde \theta$ somente. Em \cite{issues} é também deduzida uma fórmula que simplifica consideravelmente o mapa de Seiberg-Witten em 3D, essa foi usada para estabelecer alguns dos resultados acima mencionados. A seguir, apresentaremos os resultados da Ref. \cite{issues}.

Não se encontra na Seção \ref{tu4}, mas a dualidade eletromagnética ordinária em 3D pode ser vista em detalhes na Ref. \cite{issues}. Não há novidades técnicas, basta repetir os passos da Seção \ref{tu4} para o caso 3D.

\vspace{.4in}
\subsection{Dualidade 3D em primeira ordem em $\theta$}

A ação da teoria de calibre $U_*(1)$, como visto nas Seções \ref{lsw} e \ref{teoriasunnc}, é dada por
\be
	\label{sm}
	S_{\hat A*} = a \int \hat F \wedge_*  {^\star} \hat F ,
\ee
em que $a = -1/(2 g^2)$ é constante.

Como $d \hat F \not= 0$, a técnica da ação mestra \cite{master}, como apresentada na Seção \ref{tu4}, não pode ser diretamente aplicada. Todavia, como visto na Seção \ref{teoriasunnc}, teorias de calibre NC's podem ser mapeadas em teorias de calibre ordinárias; assim, em particular, o caso $U_*(1)$ é mapeado em uma teoria cujo tensor eletromagnético satisfaz $d F = 0$ \cite{sw}. Além disso, esse mapa possibilita um tratamento direto dos observáveis, devido a desacoplar transformações de calibre das transformações espaciais. Para a primeira ordem em $\theta$, o mapa de Seiberg-Witten para o caso $U(1)$ é dado por
\be
	\label{sw01}
	\hat A = \[ A_\mu - \theta^{\alpha \nu} A_\alpha \( \prt_\nu A_\mu - \frac 12 \prt_\mu A_\nu \) \] dx^\mu,
\ee
com $\delta_{\hat \lambda} \hat A = d \hat \lambda - 2 i \hat A \wedge_* \hat \lambda$ e $\delta_\lambda A = d \lambda$. 

Aplicando (\ref{sw01}) em $S_{\hat A*}$, em primeira ordem em $\theta$, 
\ba
	\label{a*}
	S_{A_\theta} 	&& =  a \int F \wedge \fdu \; \( 1 + \langle \theta, F \rangle \)  \\[.1in]
	&& = - a  \int  \( \vec E^2 - B^2 \) \(1 - \vec \theta \cdot \vec E - \theta B \) d^3x, \nonumber
\ea
em que $F:=dA$, $(\vec \theta)^i := \theta^{i0}$, $\theta := \theta^{12}$ e $ \langle \; , \, \rangle$ é o produto escalar entre formas diferenciais, em particular $\langle F, \theta \rangle = {^\star}( \fdu \wedge \theta) = \frac 12 \theta^{\mu \nu} F_{\mu \nu}$. Acima já usamos que o termo $F^{\mu \nu} F_{\nu \lambda} \theta^{\lambda \kappa} F_{ \kappa \mu} \, d^3x$ é proporcional a $\fdu \wedge F \,  \langle F, \theta \rangle $ (veja Seção \ref{sec6ad}). As equações de movimento são
\be
	\vec \nabla \cdot  \vec D  = 0,
\ee
\be
	\vec \nabla \times H = \dot{\vec D},
\ee
\be
	\label{eqt3}
	\vec \nabla \times \vec E = - \dot B.
\ee
Com $\vec D = \vec E (1 - \vec \theta \cdot \vec E - \theta B) - \frac 12 \vec \theta (\vec E^2 - B^2)$ e $H = B (1 - \vec \theta \cdot \vec E - \theta B) + \frac 12 \theta (\vec E^2 - B^2)$ (essas definições são análogas às usadas na Ref. \cite{testing}). A Eq. (\ref{eqt3}) não é modificada pela não-comutatividade do espaço-tempo, pois advém diretamente da identidade de Bianchi. Claramente, $\vec \theta$ (e não $\theta^{12}$) é responsável pela violação de isotropia espacial.

Explorando a identidade de Bianchi, a seguinte ação mestra é proposta:
\be
	\label{amt}
	S_{M_\theta}[F, \phi] = \int \[ a \, \fdu \wedge F \, \( 1 + \langle \theta, F \rangle \)  - d \phi \wedge F \].
\ee
A ação acima será usada para encontrar a dualidade em primeira ordem em $\theta$, enquanto uma natural generalização dela será empregada para revelar dualidade em ordens mais altas. Essa não é a única ação mestra possível, as seguintes também levam a dualidades entre as mesmas descrições vetorial e escalar:
\be
	S_{M_{\theta,c}} [G, \phi] = \int \[ a \; G \wedge {^\star G} \; ( 1 + c \langle \theta, G \rangle ) - \( 1 + \frac 12 (c-1) \langle \theta , G \rangle \) d \phi \wedge G \],
\ee

\be
	\label{aml}
	S_{M'_\theta}[B, A] =  \int \[ - \frac 1{4a} B \wedge {^\star B} \( 1 - \frac 1{2a} \langle \theta, {^\star B} \rangle \)  - B \wedge dA \].
\ee

A primeira é uma generalização da ação mestra de (\ref{amt}) por um parâmetro arbitrário e contínuo $c$, sendo a última reobtida para $c=1$. A ação $S_{M_{\theta,c}}$ possui a interessante propriedade de balancear a contribuição da não-comutatividade entre seus dois termos. Entretanto, para qualquer $c$, os modelos que ela conecta são os mesmos modelos vetorial e escalar que são encontrados por $S_{M_{\theta}}$. Na Eq.(\ref{aml}), $A$ e $B$ são 1-formas. Essa outra ação mestra equivalente parece ser mais adequada ao problema inverso ao proposto, isto é, o de encontrar o quadro vetorial a partir do quadro escalar.

Dando continuidade à analise de (\ref{amt}): de sua variação com respeito a $\phi$, obtemos $dF=0$, o que implica  $F=dA$; inserindo esse resultado em $S_{M*}$, $S_{A*}$ é obtida.	Para determinar o outro lado da igualdade, a variação de $F$ será avaliada, assim procedendo,
\be
	\label{phiF}
	\frac 1 {2a} d \phi =   \fdu \( 1 + \langle \theta , F \rangle \) + \frac 12 \langle F,  F \rangle \, \tdu.
\ee
Acima, a propriedade $ F \wedge \fdu \, \langle F, \theta \rangle = \langle F, F \rangle \; \tdu \wedge F$ foi usada. Com respeito aos campos $\vec D$ e $H$, temos,
\be
	- \frac 1{2a} \vec \nabla \times \phi = \vec D,
\ee
\be
	- \frac 1 {2a} \dot \phi = H.
\ee

Para primeira ordem em $\theta$, o inverso das relações acima é

\be
	\label{Fphi}
	\fdu = \frac 1 {2a} d \phi \( 1 - \frac 1 {2a} \langle \ddu \phi, \theta \rangle \) - \frac 1  {8a^2} \langle d \phi, d \phi \rangle \, \tdu ,
\ee

\be
	\vec E = - \frac 1 {2a} \( 1 - \frac 1 {2a} \vec \theta \cdot \vec \nabla \times \phi - \frac 1 {2a} \dot \phi \theta \) \, \vec \nabla \times \phi + \frac 1 {8a^2} (\vec \nabla \phi \cdot \vec \nabla \phi - \dot \phi^2 ) \, \vec \theta,
\ee

\be
	B = - \frac {\dot \phi}{2a} \( 1 - \frac 1 {2a} \vec \theta \cdot \vec \nabla \times \phi - \frac 1 {2a} \dot \phi \theta \) - \frac 1 {8a^2} (\vec \nabla \phi \cdot \vec \nabla \phi - \dot \phi^2 ) \, \theta.
\ee

A inserção da expressão para $F$ em $S_{M_\theta}$ leva a uma extensão NC da ação do campo escalar, a saber,
\be
	\label{phi*}
	S_{M_\theta} \lra - \frac {1}{4a}\int d \phi \wedge \ddu \phi \, \( 1 - \frac 1 {2a} \langle \ddu \phi, \theta \rangle \) = S_{\phi_\theta}.
\ee
	
A correspondência das equações de movimento entre os modelos vetorial e escalar, como esperado, é dado por $F=dA$ junto com a Eq.(\ref{phiF}) (e sua inversa). Realmente, se $d$ for aplicado em ambos os lados de (\ref{phiF}), com $F=dA$, a equação de movimento de  $S_{A_\theta}$ é obtida; enquanto a aplicação de $d^\star$ em (\ref{Fphi}) produz a equação de movimento de $S_{\phi_\theta}$.

Verifica-se diretamente que o mapa (\ref{phiF}) corretamente relaciona as Hamiltonianas e colchetes de ambas as representações.

\vspace{.4in}

Com o último resultado, foi definido um novo modelo escalar cuja ação é, em primeira ordem, classicalmente equivalente à teoria de calibre $U_*(1)$ em 3D. Embora haja termos cúbicos na Lagrangiana, essa dualidade não é apenas clássica, ela também é válida quanticamente. Isso também foi afirmado em \cite{ganor}. Uma computação explícita com integrais de caminho de Feynman, no contexto da dualidade NC entre os modelos Maxwell-Chern-Simons e autodual, foi feita na Ref. \cite{hr}, a qual apresenta a mesma dualidade de \cite{nossoNCMCS}, que não usa argumentos quânticos. Esse resultado pode ser generalizado. Esquematicamente, sejam $\cl_1 (A)$ e $\cl_2(B)$ duas Lagrangianas classicalmente equivalentes que estão relacionadas pela Lagrangiana mestra $\cl_m(A,B)$ cuja função partição é 
\be
	Z = \int \cd A \; \cd B \; \exp \[ -i \int \( a_1 A^2 + \theta A^3 + a_2 B A + f(B) \) d^Dx \].
\ee

Integrações em $A$ podemser conertidas em intrgrais Gaussianas por meio da introdução de mais dois campos, como segue,
\ba
	 Z & = & \mbox{cte} \int \cd A \; \cd B \; \cd C  \; \cd D \; \exp \[ -i \int \( a_1 A^2 + \theta ACC + D(C-A) + \right. \right . \nonumber \\
	&& \left. \left. + \; a_2 B A + f(B) \) d^Dx \].
\ea

Agora a integral em $A$ pode ser diretamente computada, deve-se substituir $A$ por $\frac 1 {2 a_1} (- \theta CC + D - a_2B)$.  Portanto, na teoria acima, se a dualidade clássica for válida para qualquer $\theta$ e dualidade quântica for válida para $\theta = 0$, a dualidade quântica também é válida. Esses argumentos se aplicam à dualidade NC escalar/vetorial aqui apresentada.

\vspace{.4in}
\subsection{Dualidade em ordens mais altas e o limite de CLV}

Na segunda ordem em $\theta$, (\ref{a*}) é escrita como\footnote{Note que a Ref.\cite{ganor} usa uma convenção diferente nos fatores constantes das formas diferenciais.} \cite{lambda, ganor, dayi},
\be
	\label{a*2}
	S_{A_\theta} = \frac a2 \int \[F^{\mu \nu} F_{\mu \nu} \( 1 + \frac 12 \theta^{\mu \nu} F_{\mu \nu} \) + L_{\theta^2} \] d^3x,
\ee
com
\ba
	L_{\theta^2} =&& -2 \; \tr (\theta F \theta F^3 ) + \tr (\theta F^2 \theta F^2) + \tr (\theta F) \; \tr(\theta F^3) - \frac 18 \tr (\theta F)^2 \; \tr (F^2) + \nonumber \\
	&& + \frac 14 \tr(\theta F \theta F) \; \tr (F^2)
\ea
e $\tr (AB) := A_{\mu \nu} B^{\nu \mu}$, $\tr (ABCD) := A_{\mu \nu} B^{\nu \lambda} C_{\lambda \kappa} D^{\kappa \mu} \;$ \emph{etc}. Este traço aqui usado não é o mesmo $\tr$ que aparece como traço na álgebra, embora o símbolo seja o mesmo. 

Felizmente, no espaço-tempo 3D, a expressão acima pode ser consideravelmente simplificada. Já foi usado na Eq. (\ref{a*}) que $\tr (FF \theta F) = \frac 12  \tr (FF) \; \tr (F \theta)$; com certa reflexão, essa relação pode ser generalizada para a seguinte: 
\be
	\label{rel}
	\tr (A B_1 A B_2 \; ... \; A B_n) = \( \frac 12 \)^{n-1} \prod^n_{k=1} \tr (A B_k),
\ee
para quaisquer tensores anti-simétricos de \emph{rank} 2 $A, \{ B_k \}$. Portanto,
\be
	\label{ls}
	L_{\theta^2} = \frac 14 \tr (FF) \; \tr( \theta F)^2.
\ee

A ação mestra $S_{M_\theta}$ (\ref{amt}) pode agora ser estendida para segunda ordem em  $\theta$, isso é obtido pela adição de $- a \int {^\star F} \wedge F \; \langle F, \theta \rangle {^2} $ à expressão em primeira ordem. Assim,

\ba
	^\star F = && \frac {d \phi} {2a} \( 1 - \frac {\langle \theta, {^\star d \phi} \rangle}{2a} - 3 \frac {\langle \theta, { ^\star d \phi} \rangle{^2}}{ 4a^2} + \langle \theta, \theta \rangle \frac {\langle d \phi, d \phi \rangle }{ 8a^2} \) - \nonumber \\
	\label{fm}
	&& - {^\star} \theta \frac {\langle d \phi,  d \phi \rangle }{8 a^2} \( 1 -  5 \frac {\langle \theta, {^\star} d \phi \rangle} {2a} \)
\ea
e
\be
	\label{p*2}
	S_{\phi_\theta} = - \frac 1 {4a} \int d \phi \wedge {^\star} d \phi \( 1 - \langle  \tht, d \phi \rangle + 3 \langle  \tht, d \phi \rangle^2 + \frac 14 \langle \tht, \tht \rangle \; \langle d \phi, d \phi \rangle \),
\ee
em que  $\tht = {^\star} \theta / 2a$. Portanto, no quadro escalar, ao menos até a segunda ordem, $\tht$ é o parâmetro de violação de Lorentz e $\theta$ é desnecessário. Note que apenas através do emprego de $\tilde \theta$ a constante de acoplamento $a$ da teoria de calibre original aparece no quadro dual como um fator global $a^{-1}$. \emph{A priori}, pode-se conjecturar que $\tht$ é o parâmetro fundamental do quadro escalar, enquanto $\theta$ é inferido pela dualidade.Entretanto, a menos que o limite de campos lentamente variantes (CLV) seja empregado, essa é apenas uma ilusão de uma simetria não exata. 

A partir da expansão em terceira ordem em $\theta$, termos com mais derivadas que campos aparecem no mapa de  Seiberg-Witten de $\hat F$ e estão presentes em $L_{\theta^3}$, como será mostrado (qualquer $L_{\theta^n}$ pode apenas depender de $A$ através de $F$, mas pode possuir mais derivadas que $A$'s). Esses fatores frustram a simetria acima sugerida. Para os inferir, usaremos a seguinte equação diferencial de Seiberg-Witten\cite{sw} [é obtida a partir da definição de $\hat F$ e da equação diferencial que define o mapa $\hat A(A)$ (\ref{swmNg})]
\ba
	 \label{dsw}
	 \delta \hat \fmnd	(\theta) &=& \frac 14 \delta \theta^{\alpha \beta} \[ 2 \hat F_{\mu \alpha} * \hat F_{\nu \beta} + 	2 \hat  F_{\nu \beta} * \hat F_{\mu \alpha}  - \hat A_\alpha * (\hat D_\beta \hat F_{\mu \nu} + \partial_\beta \hat F_{\mu \nu})  - \right. \\ \nonumber
	&&  \left. - (\hat D_\beta \hat F_{\mu \nu} + 	\partial_\beta \hat F_{\mu \nu}) * \hat A_\alpha \].
\ea

Expandindo $\hat F$ e $\hat A$ em potências de $\theta$ até a terceira ordem, temos
\be
	 \label{dsw3}
	 \delta \hat \fmnd^{(3)}	(\theta) = - \frac 14 \delta \theta^{\alpha \beta} \theta^{\alpha' \beta'}\theta^{\alpha'' \beta''} \( \prt_{\alpha'}\prt_{\alpha''}F_{\mu \alpha} \prt_{\beta'}\prt_{\beta''}F_{\nu \beta} - \prt_{\alpha'}\prt_{\alpha''} A _\alpha \prt_{\beta'}\prt_{\beta''} \prt_\beta F_{\mu \nu} \) +...
\ee
Em que $F_{\mu \nu} = \hat F_{\mu \nu}^{(0)}$ e $A_\mu = A^{(0)}_\mu$. Apenas os termos com mais derivadas que campos foram escritos na expressão acima. Inserindo esse resultado na Eq.(\ref{sm}), os únicos termos de $L_{\theta^3}$ que possuem mais derivadas que campos são os que estão na seguinte expressão\footnote{Esta solução pode também ser obtida através dos resultados na Ref. \cite{fidanza}, Seção 3.2; nela o mapa de Seiberg-Witten é expandido em potências de  $A$.}
\be
	\label{df}
	\theta^{\alpha \beta} \theta^{\alpha' \beta'} \tr(\prt_{\alpha} \prt_{\alpha'} F \; \theta \; \prt_{\beta} \prt_{\beta'} F \; F) - \frac 14 \theta^{\alpha \beta} \theta^{\alpha' \beta'} \; \tr(F \theta) \; \tr (\prt_{\alpha} \prt_{\alpha'} F \; \prt_{\beta} \prt_{\beta'} F). 
\ee

A contribuição desses termos para as equações de movimento é dada por
\be
	\label{ec}
	\theta^{\alpha \beta} \theta^{\alpha' \beta'} \prt_\mu \[ F_{\alpha \alpha'}^\mu \; \theta \; F_{\beta \beta'}^\kappa + \frac 12 \tr(F_{\alpha \alpha'} \theta) \; F^{\kappa \mu}_{\beta \beta'} +   F_{\alpha \alpha'}^{[\mu}  \; F_{\beta \beta'} \; \theta^{\kappa]} + \frac 14 \tr(F_{\alpha \alpha'} F_{\beta \beta'}) \; \theta^{\kappa \mu} \].
\ee
Acima introduzimos uma notação compacta: índices não explícitos são contraídos como na notação matricial, índices extras em $F$ são derivadas e $F_{\alpha \alpha'}^{[\mu}  \; F_{\beta \beta'} \; \theta^{\kappa]} = F_{\alpha \alpha'}^{\mu}  \; F_{\beta \beta'} \; \theta^{\kappa} -  F_{\alpha \alpha'}^{\kappa}  \; F_{\beta \beta'} \; \theta^{\mu}$. Por exemplo, escreve-se o primeiro termo em (\ref{ec}) como $\prt_\alpha \prt_{\alpha'} F^\mu_{\;\; \nu} \; \theta^{\nu \lambda} \; \prt_\beta \prt_{\beta'} F_\lambda^{\;\; \kappa}$.

Uma cuidadosa análise das simetrias e anti-simetrias de cada termo de (\ref{ec}) e de sua independência linear para $\theta$ arbitrário com $D \ge 4$ mostra que (\ref{ec}) não é nulo. Para diretamente garantir que, em qualquer dimensão($D \ge 2$), (\ref{ec}) não é a identidade trivial [ou, equivalentemente, que (\ref{df}) não é nem nulo e nem um termo de superfície] pode-se avaliar um caso particular de (\ref{ec}); por exemplo, para $D\ge 3$, sejam $\kappa = 2$ e $\theta$ nulo exceto pelas componentes $\theta^{01}$ and $\theta^{10}$, assim (\ref{ec}) equivale a 
\ba
	&& \( \theta^{10} \)^3 \[ \prt_\mu \( \stackrel{..}{F}^{\mu \, [0} F''^{\, 1]\, 2} + F''^{\, \mu \, [0} \stackrel{..}F^{\, 1]\, 2} - 2 \, \dot F'^{\mu \, [0} \dot F'^{\, 1] \,2} +  \stackrel{..}{F}^{\, 1 0} F''^{\, 2 \mu} + \right. \right.  \\  
	&& \left. \left. + F''^{\, 1 0} \stackrel{..}F^{\, 2 \mu} - 2 \, \dot F'^{\, 1 0} \dot F'^{\, 2 \mu}\) + \prt_{[0} \( \stackrel{..}{F}^{\, 2 \nu} F''_{\, 1] \, \nu} + F''^{\, 2 \nu} \stackrel{..}F_{\, 1] \, \nu} - 2 \, \dot F'^{\, 2 \nu} \dot F'_{\, 1] \, \nu} \) \], \nonumber 
\ea
em que cada ponto e cada \asp linha" significa, respectivamente, $\prt_0$ e $\prt_1$. A expressão acima não é identicamente nula em nenhuma dimensão (maior que dois). Esse resultado está em conflito com certa proposição da Ref.\cite{bc}, veja a próxima subseção para mais detalhes.

As expressões (\ref{dsw} - \ref{ec}) são váidas para dimensões espaço-temporais arbitrárias. Novamente, no espaço-tempo 3D, uma simplificação considerável é possível. Embora a propriedade (\ref{rel}), naquela forma, não posssa ser usada diretamente em (\ref{df}), uma computação direta mostra que um resultado análogo é válido. No espaço-tempo 3D, a expressão (\ref{df}) é igual a
\be
	\label{df3d}
	\frac 14 \; \theta^{\alpha \beta} \theta^{\alpha' \beta'} \; \tr( F_{\alpha \alpha'} F_{\beta \beta'}) \; \tr(F \theta).
\ee 

Aderindo à expansão em terceira ordem, a contribuição da expressão acima a $S_{\phi_\theta}$ (\ref{p*2}) é obtida pela substituição $F \rightarrow {^\star} d \phi / (2a)$. Conseqüentemente, até a terceira ordem em $\theta$,  $S_{\phi_\theta}$ não pode ser expresso exclusivamente através de $\tilde \theta$, $\theta$ também é necessário\footnote{Pode-se artificialmente inserir $\ep$'s a fim de mudar $\theta^{\alpha \beta} \prt_\alpha \prt_\beta$ para $\propto \tilde \theta_{\mu} \ep^{\mu \alpha \beta} \prt_\alpha \prt_\beta$, esse procedimento é inócuo pois ${^\star} \tilde \theta \propto \theta$; mas estamos adotando a regra de sempre escrever ou $\theta$, ou $\tilde \theta$, nunca ${^\star} \tilde \theta$. Ademais esse procedimento não evita as dificuldades com a dualidade $S$, pois $\tilde \theta$ não ocorrerá proporcionalmente a $\phi$ no quadro dual.}. Isso viola a simetria entre  $\theta$ e $\tilde \theta$ presente na dualidade eletromagnética até a segunda ordem em $\theta$. Conseqüentemente, no quadro escalar, a constante $a$ não aparece como um fator global $a^{-1}$ e a estrutura esperada de uma dualidade $S$ é quebrada.

No limite de CLV, os termos na ação mapeada por Seiberg-Witten que dependem das derivadas de $F$ são ignorados, portanto $S_{\phi_\theta}$, na terceira ordem em $\theta$, pode ser exclusivamente expresso em termos de  $\tilde \theta$. Neste limite, como a Lagrangiana mapeada por Seiberg-Witten é uma função somente de $F$ (sem derivadas) \cite{sw}, a Lagrangiana é expressada como uma função de $\tr (FF)$ e $\tr(F \theta)$ somente [devido à Eq.(\ref{rel})); portanto, a ação escalar dual $S_{\phi_\theta}$ para todas as ordens em  $\theta$ pode ser expressa por $\tilde \theta$, sem referência a  $\theta$ (ou ${^\star} \tilde \theta$). Embora a propriedade (\ref{rel}) seja em geral falsa no espaço-tempo 4D, a ação dual pode também ser expressa exclusivamente por $\tilde \theta$ no espaço-tempo 4D, em todas as ordens em $\theta$, se o limite de CLV for usado \cite{aschieri}. A relação (\ref{rel}) simplifica consideravelmente a análise no espaço-tempo 3D.

\vspace{.4in}
\subsection{Comentários finais}
\label{comfin3}

Na última subseção demonstramos diretamente, por meio de cáculos válidos em qualquer dimensão maior que dois, que  a Lagrangiana mapeada por Seiberg-Witten da teoria NC eletromagnética ($L_{A_\theta}$) depende de $F$ e suas derivadas\footnote{Embora não tenha sido explicitamente mostrado na subseção anterior, não é difícil avaliar que uma particularização de (\ref{df}) para $D=2$ é também diferente de zero.}. Até a segunda ordem em $\theta$, as derivadas em  $F$ podem ser combinadas com os campos $A$ de forma a produzirem outro $F$ (eliminando toda a dependência explícita nos $A$'s, pois há um número igual de derivadas e $A$'s). Entretanto, a equação diferencial de Seiberg-Witten (\ref{dsw}) leva ao aparecimento de termos com mais derivadas que campos na expansão em terceira ordem. Esses termos foram aplicados na Lagrangiana do eletromagnetismo NC ($L_{\hat A_*})$ e os termos resultates foram expressos em (\ref{df}). Talvez surpreendentemente, esses termos não são nem nulos e nem são termos de superfície, como mostrado subseqüentemente. Esse resultado não está em acordo com a primeira parte de uma proposição na Ref.\cite{bc}. Achamos que esse nosso resultado deve ser visto como um contra-exemplo a essa proposição. Realmente, a primeira parte da Proposição 3.1 não parece ser correta de forma geral \cite{comment}. Todavia, deve ser enfatizado que essa proposição claramente é válida no limite de campos lentamente variantes (CLV) e que, nesse limite, ela é compatível com nossos resultados; ademais, quaisquer resultados que dependam dessa proposição são perfeitamente válidos naquele limite. Há outras interessantes conseqüências que estamos agora avaliando \cite{dm}. 

Correntes podem ser facilmente inseridas na dualidade tratada, seguindo os passos da Seção \ref{tu4}, se for assumido um acomplamento independente de $\theta$ como $A \wedge {^\star J}$ na ação mapeada. Contudo, isso viola a correspondência com a teoria $U_*(1)$, que impõe o acoplamento $\hat A \wedge_* {^\star} \hat J$, cujo mapa foi encontrado na  Ref.\cite{banerjeecurrents}.

É interessante notar que a dualidade eletromagnética não viola isotropia espacial e, caso uma das teorias a viole, a direção privilegiada é rodada por $\pi/2$. Na teoria $U_*(1)$, a direção privilegiada no espaço é dada por 
$\theta^{0i}$ enquanto na teoria dual, até segunda ordem ou no limite de CLV, a direção privilegiada espacial é dada por 
\be
	\tilde \theta^i = \frac 1 {4a} \ep^{i \mu \nu} \theta_{\mu \nu} = \frac 1 {2a} \ep^{0 i j } \theta_{j 0},
\ee
donde nota-se que a direção dada por $\vec \theta$ é ortogonal à de $\vec{\tilde \theta}$.

\chapter{Conclusão}

Os resultados por nós obtidos e apresentados nesta tese se dividem em três partes: $i)$ o desenvolvimento do formalismo simplético de calibre (Seção \ref{secformsympcal}); $ii)$ a extensão não-comutativa (NC) da dualidade entre os modelos autodual e Maxwell-Chern-Simons (MCS) (Seção \ref{sec6ad}); $iii)$ o estabelecimento da dualidade eletromagnética NC em 3D sem massa topológica e avaliação da relevância do limite de campos lentamente variantes (CLV) para essa dualidade tanto em 3D quanto em 4D (Seção \ref{sec6clv}). Os principais resultados obtidos em cada uma dessas partes foram comentados nas Subseções \ref{comfin1}, \ref{comfin2} e \ref{comfin3} (veja também as Refs. \cite{symemb, symembjf, nossoNCMCS, issues}).


Acreditamos que os resultados apresentados nas partes $ii$ e $iii$ acima são de valor para o entendimento das teorias eletromagnéticas não-comutativas, as quais, como mencionado, possuem relações com diversas áreas da física, embora o significado físico delas ainda não se encontre plenamente estabelecido; o que, achamos nós, não tardará a ocorrer. É possível que resultados do \emph{Large Hadron Collider} (LHC), cujas atividades terão início no próximo ano, nos dêem pistas de como implementar a não-comutatividade espaço-temporal ou revelem novos fenômenos (não necessariamente relacionados ao eletromagnetismo) que possam ser modelados pela Lagrangiana $U_*(1)$. Deve-se ressaltar a conexão das teorias NC's com as teorias de cordas, pois alguns aspectos das $D$-branas são melhor entendidos através da descrição NC \cite{szaborev}. Além disso, novos avanços continuam a surgir \cite{connesrec}\footnote{Para citar apenas um dos vários artigos que se poderiam ser citados aqui.}. 

O formalismo simplético de calibre (parte $i$ acima) não deve ser visto como um método plenamente acabado, mas já se demonstrou capaz de reproduzir alguns resultados de outros métodos e nos parece que poderá se revelar útil para a resolução de problemas físicos atuais, dentre eles os relacionados com quantização, dualidade ou mesmo com o mapa de Seiberg-Witten.

Estamos no momento interessados em algumas interessantes conseqüências da Ref. \cite{issues} ainda não exploradas. Dentre elas, gostaríamos de analisar minuciosamente os resultados da Ref. \cite{bc} que se fundamentam na parte problemática de sua Proposição 3.1 e tentar extrair novos resultados referentes a aspectos mais físicos do eletromagnetismo NC em 4D. A Eq. (\ref{rel}) simplifica consideravelmente a expansão em Seiberg-Witten em 3D, esperamos usá-la em uma análise do eletromagnetismo NC nessa dimensão. O seu quadro dual escalar pode se demonstrar conveniente para essa análise. Avaliar as conseqüências do resultado acima num contexto de teoria de cordas é também tema de interesse.

\appendix

\chapter{Formas diferenciais no espaço $\hat M$ e equações de Maxwell}
\label{apformasm}

Aqui introduzimos formas diferenciais no espaço $\hat M$, seguindo as idéias da Seção \ref{tu1}, para expressar as Eqs. (\ref{MDa} - \ref{MDd}). Retomemos a Eq.(\ref{ebf}) introduzindo o operador $\hat d$ como segue 
\be
	\hat d E \wedge dt + d_0 B + \hat d B = 0,
	\label{d343}
\ee
com $d_0 := dt \; \prt_0$ e $\hat d := dx^i \; \prt_i$. Nota-se que $\hat d$ \'e a derivada exterior de $\hat M$. Usando essa nota\c{c}\~ao, a equa\c{c}\~ao de movimento de $F$ (\ref{maxmov}), ou equivalentemente de $E$ e $B$, pode ser reescrita como
\be
	d_0 \str (E \wedge dt) + \hat d \str (E \wedge dt) + \hat d \str B = - e(J_0 \ddu t + J_i \ddu x^i).
	\label{d344}
\ee
O termo $d_0 \str B$ n\~ao aparece acima pois \'e identicamente nulo devido \`a condi\c{c}\~ao de ortogonalidade (\ref{ort}).

Na equa\c{c}\~ao anterior, a separa\c{c}\~ao entre $dt$ e $\{dx^i\}$ torna-se patente mediante a introdu\c{c}\~ao de um operador de dualidade em $T_x^\star \hat M$. Seja $C(x)$ uma $k$-forma em $T^\star_x \hat M$, define-se o operador $\hat \str$ por
\be
	\hat \str C = \frac 1 {k!} \; \frac 1 {(d-k)!} \; C_{i_1 ...i_k} \hat \omega^{i_1... i_k j_1 ... j_{d-k}} \; dx_{j_1} ...  dx_{j_{d-k}},
\ee
com $\hat \omega_{i_1... i_d} = \ep_{i_1... i_d} \hat f$, $\hat f = \sqrt{| \hat g|}$ e $\hat g = \det (g_{ij})$. Usando a Eq. (\ref{ort}) algumas rela\c{c}\~oes \'uteis podem ser deduzidas:
\ba
	\label{hstra}
	\str dt &=& (-1)^{a_0} \; \sqrt{|g^{00}|} \; \hat f \; d^dx, \\
	\label{hstrb}
	\str dx^i &=& - \sqrt{ |g_{00}|} \; dt \wedge  \hat \str dx^i, \\
	\label{hstrc}
	\str C &=& (-1)^k \sqrt{|g_{00}|} \; dt \wedge \hat \str C, \\
	\label{hstrd}
	\str(C \wedge dt) &=& (-1)^{a_0 +k} \; \sqrt{|g^{00}|} \; \hat \str C,
\ea
em que $a_0 = 0$ se $g^{00} > 0$ e $a_0 = 1$ se $g^{00} < 0$. Nota: $g_{00} = 1/g^{00}$. 

Portanto, a Eq. (\ref{d344}) pode ser reescrita como
\ba
	&& (-1)^{1+a_0} \, \sqrt{|g^{00}|} \,  d_0 \hat \str E + (-1)^{1+a_0} \, \sqrt{|g^{00}|} \, \hat d \hat \str E  - \sqrt{|g_{00}|} \; dt \wedge \hat d \hat \str B = \nonumber \\[.2in]
	&& = 	- e \( (-1)^{a_0} \sqrt {|g^{00}|} \; J_0 \; \hat f \; d^dx - \sqrt{|g_{00}|} \; J_i \; dt \wedge \hat \str dx^i \).
	\label{d353}
\ea

Cada uma das Eqs. (\ref{d343}) e (\ref{d353}) pode ser quebrada em duas, como segue\footnote{Pode-se conferir diretamente que essas equações são invariantes pela transformação $g \; \rightarrow \; -g$.}:
\ba
	\label{d354a}
	&& \hat d E  + \dot B = 0, \\
	\label{d354b}
	&& \hat d B = 0,  \\
	\label{d354c}
	&& g^{00} \; \hat \str \dot E + \; \hat d \hat \str B = - e \; J_i \; \hat \str dx^i \\
	\label{d354d}
	&& \hat d \hat \str E = e \; J_0 \; \hat f \; d^dx,
\ea
em que o ponto sobre um campo \'e uma derivada temporal parcial, i.e., $\dot B := \prt_0 B$. 

As equações acima são equivalentes às Eqs. (\ref{MDa} - \ref{MDd}). Alguns textos apresentam o eletromagnetismo em formas diferenciais tendo como ponto de partida as equações acima. As últimas duas equações dependem da métrica, essa dependência possibilita a introdução das formas $H$ e $D$ que são iguais a $B$ e $E$ somente para certo valor da métrica (métrica do vácuo) \cite{azul}.






\setlength{\baselineskip} {20 pt}

\end{document}